\documentclass[twocolumn,secnumarabic,amssymb, nobibnotes, superscriptaddress, nofootinbib, aps, prd]{revtex4-2}

\newcounter{inlinefig}

\usepackage[pdfencoding=auto,psdextra]{hyperref}
\hypersetup{
    colorlinks=true,
    linkcolor=blue,
    filecolor=magenta,
    urlcolor=blue,
    citecolor=blue
}

\usepackage[T1]{fontenc}

\usepackage{multirow}
\usepackage{colortbl}
\usepackage{array}

\usepackage{lipsum}
\usepackage{aas_macros}
\usepackage{orcidlink}

\usepackage{natbib}
\usepackage{hyperref}
\usepackage{amsmath}
\usepackage{enumitem}
\setlist{wide, labelwidth=!, labelindent=0pt}
\usepackage{multirow}
\hypersetup{    
  colorlinks      = true,
  linkcolor       = {blue},
  linkbordercolor = {white},
  citecolor       = {blue},
  citebordercolor = {white},
  urlcolor        = {blue},
  urlbordercolor  = {white},
}

\usepackage{tabularx}
\usepackage{xspace}
\usepackage{listings}
\usepackage{lineno}
\usepackage{graphicx,booktabs}
\usepackage{blindtext}
\usepackage{soul}
\sethlcolor{yellow}
\usepackage{xcolor}
\usepackage{placeins}
\usepackage{float}
\usepackage{amssymb}
\usepackage{ragged2e}
\usepackage{amsmath, wasysym}
\usepackage{tikz}

\usetikzlibrary{backgrounds}
\usetikzlibrary{arrows}
\usetikzlibrary{shapes}
\usepackage{txfonts}
\usepackage{hyperref}
\usepackage{newtxmath}
\usepackage{bm}

\usepackage[utf8]{inputenc}
\usepackage[T1]{fontenc}
\usepackage{amsmath,amssymb}
\makeatletter
\def\widetext@rule{}  
\makeatother
\usepackage{graphicx}
\usepackage{hyperref}

\begin{document}

\title{Probing gravity with non-linear clustering in redshift space}

\author{Cristian~Viglione\orcidlink{0000-0002-3391-3469}}
\email[]{crisneryvi@gmail.com}
\email[]{viglione@ice.csic.es}
\affiliation{Institute of Space Sciences (ICE, CSIC), Campus UAB, Carrer de Can Magrans, s/n, 08193 Barcelona, Spain}
\affiliation{Institut d'Estudis Espacials de Catalunya (IEEC), Edifici RDIT, Campus UPC, 08860 Castelldefels, Barcelona, Spain}

\author{Pablo~Fosalba\orcidlink{0000-0002-1510-5214}}
\affiliation{Institute of Space Sciences (ICE, CSIC), Campus UAB, Carrer de Can Magrans, s/n, 08193 Barcelona, Spain}
\affiliation{Institut d'Estudis Espacials de Catalunya (IEEC), Edifici RDIT, Campus UPC, 08860 Castelldefels, Barcelona, Spain}

\author{Isaac~Tutusaus\orcidlink{0000-0002-3199-0399}}
\affiliation{Institut de Recherche en Astrophysique et Planétologie (IRAP), Université de Toulouse, CNRS, UPS, CNES, 14 Av. Edouard Belin, 31400 Toulouse, France}

\author{Linda~Blot\orcidlink{0000-0002-9622-7167}}
\affiliation{Center for Data-Driven Discovery, Kavli IPMU (WPI), UTIAS, The University of Tokyo, Kashiwa, Chiba 277-8583, Japan}
\affiliation{Laboratoire Univers et Théorie, Observatoire de Paris, Université PSL, Université Paris Cité, CNRS, 92190 Meudon, France}

\author{Jorge~Carretero\orcidlink{0000-0002-3130-0204}}
\affiliation{Centro de Investigaciones Energéticas, Medioambientales y Tecnológicas (CIEMAT), Avenida Complutense 40, E-28040 Madrid, Spain}
\affiliation{Port d'Informació Científica (PIC), Campus UAB, C. Albareda s/n, 08193 Bellaterra (Barcelona), Spain}

\author{Pau~Tallada\orcidlink{0000-0002-1336-8328}}
\affiliation{Centro de Investigaciones Energéticas, Medioambientales y Tecnológicas (CIEMAT), Avenida Complutense 40, E-28040 Madrid, Spain}
\affiliation{Port d'Informació Científica (PIC), Campus UAB, C. Albareda s/n, 08193 Bellaterra (Barcelona), Spain}

\author{Francisco~Castander\orcidlink{0000-0001-7316-4573}}
\affiliation{Institute of Space Sciences (ICE, CSIC), Campus UAB, Carrer de Can Magrans, s/n, 08193 Barcelona, Spain}
\affiliation{Institut d'Estudis Espacials de Catalunya (IEEC), Edifici RDIT, Campus UPC, 08860 Castelldefels, Barcelona, Spain}


 
\begin{abstract}
We compute the gravity model testing parameter ($E_G$) on realistic simulated modified gravity galaxy mocks adopting the more accurate estimator described in Ref.~\cite{Wenzl2024ConstrainingBOSS}. The analysis is conducted using two twin simulations presented in Ref.~\cite{Arnold2018TheDistributions}: one based on general relativity (GR) and the other on the ($f(R)$) Hu $\&$ Sawicki model with ($f=10^{-5}$) (F5). This study aims to measure the ($E_G$) estimator in GR and ($f(R)$) models using high-fidelity simulated galaxy catalogs, with the goal of assessing how future galaxy surveys can detect deviations from standard gravity. Deriving this estimator requires precise, unbiased measurements of the growth rate of structure and the linear galaxy bias. We achieve this by implementing an end-to-end cosmological analysis pipeline in configuration space, using the multipoles of the 2-point correlation function. In our analysis we estimate the scale-dependent growth rate predicted by non-standard gravity models using COMET-VDG fits. We split the estimation of the RSD ($\beta$) parameter over distinct scale ranges, separating large (quasi-linear) and small (non-linear) scales. We show that this estimator can be accurately measured using mock galaxies in low redshift bins ($z < 1$), although its discriminating power between competing theories is limited. We find that, for an all-sky galaxy survey and neglecting observational systematics, accurate and largely unbiased estimations of ($E_G$) can be obtained across all redshifts. However, the error bars are too large to clearly distinguish between the theories. When measuring the scale-dependence of the ($E_G$) estimator, we note that state-of-the-art theory modeling limitations and intrinsic "prior volume effects" prevent high-accuracy constraints. Alternatively, we propose a null test of gravity using RSD clustering, which, if small scales are modeled accurately in future surveys, could detect  departures from GR.
\end{abstract}

\maketitle



\section{Introduction}

The need of a theoretical explanation for the observed accelerated expansion of the universe has forced the inclusion of the cosmological constant ($\Lambda$) as a dark energy component that acts as a negative pressure or effectively "repulsive" gravity on large cosmological scales. Little advances have been achieved in determining the nature of the cosmological constant since its introduction, despite massive recent observational efforts (SDSS \cite{Margony_1999}, DES \cite{DES}, BOSS \cite{Ivanov2019CosmologicalSpectrum}, DESI \cite{DESI_Collaboration_2022, desicollaboration2025desidr2resultsii}). Several alternative gravity models \cite{ishak2019modifiedgravitydarkenergy} are also able to explain this accelerated expansion without the need of a cosmological constant. So far, the validity of General Relativity (GR) has been mainly tested on relatively small scales \cite{Jain2013ASTROPHYSICALUNIVERSE, Koyama2016CosmologicalGravity}. However modified gravity models apply corrections to GR that only become important at cosmological scales much larger than the Solar System where screening effects make deviations from standard gravity vanish. In particular, current galaxy surveys are trying to break the degeneracy between modified gravity models and dark energy models in observations by sampling the largest accessible scales. 

The large number of alternative theories of gravity that have been proposed in recent years have motivated the need to develop methods to probe the validity of these models. Of particular interest are those approaches that focus on observables directly related to the underlying theory of gravity. One of the first observables that was put forward, presented in Ref.~\cite{Zhang2007AScales}, provides a direct test for gravity on large scales. The $E_G$ estimator corresponds to the ratio between curvature, related to the $\Phi$ and $\Psi$ gravitational potentials, and the velocity field, which is tied to the growth rate of structures $f$. In GR, these relations are governed by the Poisson equation, which links the gravitational potentials to the matter density, and the Euler equation, which describes the evolution of the velocity field under the influence of gravity. The resulting $E_G$ prediction in GR takes a scale-independent form that depends only on the matter density parameter and the growth rate at a given redshift. This property makes $E_G$ particularly useful to test the validity of GR, as any observed deviation from its predicted value could indicate that GR breaks down on the largest cosmological scales.

Observationally, the velocity field, connected to redshift space distortions, can be  inferred from galaxy density auto-correlations, while the curvature field can be derived from the correlation between weak gravitational lensing (shear) of background galaxies and the positions of foreground galaxies. In this study, we analyze these fields in Fourier space, specifically using the angular power spectrum—a curved-sky generalization of the power spectrum. We employ a pseudo-$C_\ell$ estimator to achieve improved separation between large and small scales. Previous analysis \cite{Ghosh2018TheStatistics, Yang2018CalibratingRelativity, Pullen2014ProbingLensing, Pullen2016ConstrainingVelocities, Abidi2022ModelIndependentClustering, Wenzl2024ConstrainingBOSS} have used CMB lensing in order to estimate the convergence field, since this estimator is not affected by systematics related to intrinsic alignments and it has a broad kernel that samples dark-matter clustering at higher redshifts. We instead opt for using galaxy-galaxy lensing since this allows us to be self-consistent with the data as we can extract the galaxy source information directly from the same lightcone simulation. This also gives us the opportunity to select different source sample populations to optimize the gravity estimator. In this context, we leave for future work the potential impact of intrinsic alignments in our analysis.

Estimating the value of the growth rate can be difficult since it suffers from a degeneracy with galaxy clustering bias, \( b_1 \), which describes the relationship between the distribution of galaxies and the underlying matter density field, and the scalar amplitude \( A_s \) or \(\sigma_8\). The scalar amplitude \( A_s \) quantifies the initial amplitude of scalar perturbations in the early universe, while \(\sigma_8\) is the root-mean-square (RMS) fluctuation of the matter density field on a scale of \( 8 \, h^{-1} \, \mathrm{Mpc} \), often used as a proxy for the overall matter clustering strength. This degeneracy arises when these parameters are constrained solely from the power spectrum. So working with multipoles of the correlation function have become a common approach to break this degeneracy as each multipole exhibits a different dependence on these clustering amplitude parameters. In this context, several emulators have been proposed to produce fast and accurate predictions of the clustering multipoles for a given cosmology that are also able to reproduce different sources of non linear effects. For this analysis the public code COMET-EMU \cite{Eggemeier2023COMETTheory} is used to predict the multipoles of the correlation function which allows to emulate non-linear galaxy clustering in redshift space using different perturbation theory approaches (EFT, VDG). We note that, currently,  COMET-EMU does not allow predictions for MG cosmologies so we will incorporate the MG boost directly on the multipoles.

In this study, we incorporate both the monopole and quadrupole in configuration space to adopt a more robust approach \cite{Cabr1}. Ref.~\cite{Crocce_2011} uses the angular galaxy-galaxy autocorrelation to estimate $f$, which is further expanded in Ref.~\cite{Gazta_aga_2012} and Ref.~\cite{Asorey_2014} by using additional weak lensing cross-correlations between
redshift bins. While Ref.~\cite{Hoffmann_2014} uses the shape of the reduced
three-point correlation and a second method with a combination of third-order one- and two-point cumulants to estimate the linear growth factor D.

Modified gravity models often introduce an additional scalar degree of freedom, which generates a fifth force, characterized by $f_{R0}$, \cite{euclidcollaboration2024euclidpreparationsimulationsnonlinearities}. In this study, we examine the $f(R)$ gravity model, where the fifth force has a finite range, denoted by $\lambda_c$. Within this range, the additional force strength in the linear regime is equivalent to one-third of the standard gravitational force, effectively modifying $G$ to $\frac{4}{3}G$ on small scales (r << $\lambda_c$), while preserving the standard $G$ on larger scales. Such a significant modification would be incompatible with observations if not for the presence of a screening mechanism, which suppresses these changes in regions of high density.

In this paper, we calculate the $E_G$ estimator from a synthetic galaxy mock that follows an $f(R)$ Hu $\&$ Sawicki modified gravity model \cite{Arnold2018TheDistributions}. Potential deviations from GR are estimated by comparing measurements of the gravity estimator in this mock with respect to a reference galaxy mock that uses the same galaxy assignment pipeline applied to a LCDM simulation (\emph{i.e.}, same cosmological parameters and initial conditions, but standard gravity force).  The fact that we use simulated data allows us to ignore systematics that affect the estimator in observations, namely the effect of lensing and magnification that can produce errors on the $E_G$ estimation of up to $40\%$ on high redshift photometric samples \cite{Yang2018CalibratingRelativity, Ghosh2018TheStatistics}. Moreover, since we want to assess whether it is possible at all to distinguish gravity models that are consistent with the set of current observational data, we focus as a working case on an ideal all-sky survey, and neglect sources of astrophysical systematics such as intrinsic alignments or photometric errors. We adopt the refined $E_G$ estimator introduced by Ref.~\cite{Wenzl2024ConstrainingBOSS}, which builds upon the traditional method of using the angular power spectrum, as presented by Ref.~\cite{Pullen2014ProbingLensing}. Additionally, Ref.~\cite{Grimm_2024} proposed a novel, entirely model-independent $E_G$ estimator that combines galaxy velocity measurements from surveys with the Weyl potential. However, this new approach is not explored in this study.

Due to the nature of $f(R)$ in the Hu $\&$ Sawicki model the value of the growth rate $f$ is scale dependent. Consequently, we will estimate this parameter across different scales. In GR, this parameter is typically calculated by fitting a model to the multipoles of the correlation function over a broad range of scales, under the assumption of scale independence. Estimating this parameter over limited scales, linear and nonlinear, introduces several complexities, which we will address throughout this paper. A substantial body of work has explored the impact of modified gravity on nonlinear clustering and redshift-space observables, both theoretically and using N-body simulations. Early studies demonstrated the scale dependence of the growth rate and redshift-space distortions in 
$f(R)$ gravity using simulations \cite{Jennings_2012}. Subsequent work developed perturbative redshift-space power spectrum models explicitly designed for modified gravity and general Horndeski theories, and quantified the biases incurred when GR-based templates are applied to MG data \cite{Bose_2016,Bose_2017,Valogiannis_2020}.

Beyond two-point statistics, a wide range of nonlinear and velocity-sensitive probes have been shown to enhance sensitivity to modified gravity effects on small scales, including higher-order and anisotropic counts-in-cells \cite{Hellwing_2013, Drozda_2022, drozda2025(2)anisotropiccountsincellsredshiftspace, drozda2025skewnessprobegravityreal}, marked correlation functions and environment-dependent clustering \cite{Armijo_2024}, clustering ratios \cite{Arnalte_Mur_2017}, and pairwise velocity statistics \cite{Li_2012,Jaber_2024}. Nonlinear matter power spectrum modelling in modified gravity, including halo-reaction and emulator-based approaches, has also been extensively investigated \cite{Winther_2015,Gupta_2023, ruan2023emulatorbasedhalomodelmodified}.

In Ref.~\cite{Garc_a_Farieta_2019}, a similar approach to the one adopted in this work is followed by estimating the growth rate in $\Lambda$CDM and $f(R)$ at small scales ($s < 50,h^{-1}\mathrm{Mpc}$), although they use dark matter (halo) simulations incorporating massive neutrinos. They are not able to distinguish between the models at the 1$\sigma$ level at redshifts $z < 1.0$, but they only consider linear theory unlike this work. They build upon the previous work on Ref.~\cite{Garc_a_Farieta_2021} by also employing a galaxy sample. However, they are similarly unable to distinguish the growth rate between GR and $f(R)$ gravity when relying solely on linear theory. They estimate that a 10$\%$ difference between GR and F5 at the 2$\sigma$ level could be achieved at low redshifts with proper modeling of the multipoles. In this work, we aim to improve the accuracy of growth rate estimates by using nonlinear modeling, particularly at smaller scales where the differences between GR and $f(R)$ are expected to be more pronounced.

This paper is organized as follows: in section \ref{sec:fR} we introduce the $f(R)$ Hu $\&$ Sawicki model and emulators to calculate the dark matter and growth-rate boost, in section \ref{sec:EG estimator} we explain the theoretical predictions for the $E_G$ estimator for F5 and GR, while in section \ref{sec:Comet}, we introduce the COMET emulator and the VDG model that we use to perform the fits to the growth rate. Then we present the GR and F5 simulated catalogs that we use to obtain the data vectors in Section \ref{sec:Data}. A description of the different ingredients that enter in the $E_G$ estimator and how they can be accurately computed is presented in section \ref{sec:methodology}. Our main results are discussed in section \ref{sec:Results}, and we propose a null test of gravity in section \ref{sec:nulltest}, as a simple alternative test to distinguish between GR and F5 using the 2-point clustering in redshift space. Finally, in sections \ref{sec:Discussion} and \ref{sec:Conclusions} we discuss our main findings in detail, and present our conclusions and future work.

\section{The $\lowercase{f}(R)$ Hu $\&$ Sawicki model}
\label{sec:fR}

The \(f(R)\) gravity model, a widely studied modified gravity (MG) framework, extends GR by introducing a scalar function \(f(R)\), where \(R\) is the Ricci scalar, into the gravitational action:

\begin{equation}
S = \int \mathrm{d}^4 x \sqrt{-g}\left[\frac{R + f(R)}{16 \pi G} + \mathcal{L}_m\right],
\end{equation}
where \(g\) is the determinant of the spacetime metric, \(\mathcal{L}_m\) represents the matter field's Lagrangian, and \(G\) is the gravitational constant. In this model, \(f(R)\) serves as a generalization of the cosmological constant, or, when constant, it represents the cosmological constant itself.

By varying the action with respect to the metric, one derives the field equations, commonly referred to as the Modified Einstein Equations \cite{Arnold2018TheDistributions}:

\begin{equation}
G_{\mu\nu} + f_R R_{\mu\nu} - \left( \frac{f}{2} - \Box f_R \right) g_{\mu\nu} - \nabla_\mu \nabla_\nu f_R = 8\pi G T_{\mu\nu}, 
\label{Eequn}
\end{equation}

where \(\nabla\) represents the covariant derivative with respect to the metric, \(\Box \equiv \nabla_\nu \nabla^\nu\) is the d'Alembert operator, and \(T_{\mu\nu}\) is the energy-momentum tensor for the matter fields. \(R_{\mu\nu}\) is the Ricci tensor, and \(f_R \equiv \frac{d f(R)}{d R}\) is the derivative of the scalar function with respect to the Ricci scalar \(R\).

The form of the \(f(R)\) function depends on the specific model chosen. To simulate the observed structure formation, a functional form for \(f(R)\) must be selected. According to \cite{Hu2007ModelsTests}, an appropriate \(f(R)\) function should satisfy the following conditions: (1) it should reproduce the \(\Lambda\)CDM model at high redshifts (consistent with CMB observations), (2) at low redshifts, it must behave similarly to a cosmological constant, driving accelerated expansion, (3) it should include free parameters to model various low-redshift phenomena, and (4) it must recover GR results at small scales (e.g, solar system scales) to be consistent with observational constraints. The Hu-Sawicki (HS) model satisfies these criteria and takes the following form:

\begin{equation}
f(R) = -m^2\frac{C_1 \left( \frac{R}{m^2} \right)^n}{C_2 \left( \frac{R}{m^2} \right)^n + 1},
\end{equation}
where \(m^2 \equiv \Omega_m H_0^2\), and \(C_1\), \(C_2\), and \(n\) are model parameters. For this work, \(n = 1\) is used.

Additionally, the derivative of \(f(R)\) respect the Ricci scalar is given by:

\begin{equation}
f_R = -n \frac{C_1 \left( \frac{R}{m^2} \right)^{n-1}}{\left[C_2 \left( \frac{R}{m^2} \right)^n + 1\right]^2}
\label{eq:fR}
\end{equation}

In the high curvature regime (\(R \gg m^{2}\)), as shown by \cite{Oyaizu2008NonlinearMethodology}, Eq.~\eqref{eq:fR} becomes:

\begin{equation}
f_{R} \approx -n \frac{C_{1}}{C_{2}^{2}} \left( \frac{m^{2}}{R} \right)^{n+1}, 
\label{f_R approx}
\end{equation}

In Ref.~\cite{Hu2007ModelsTests} was demonstrated that a background resembling the standard \(\Lambda\)CDM model can be recovered by enforcing the condition:

\begin{equation}
\frac{C_{1}}{C_{2}} = 6 \frac{\Omega_{\Lambda,0}}{\Omega_{m,0}}, 
\end{equation}

where \(\Omega_{\Lambda,0}\) and \(\Omega_{m,0}\) represent the present-day densities of dark energy and matter, normalized by the critical density. This condition reduces the number of free parameters in the equation to one: either \(C_{1}\) or \(C_{2}\), since, as established earlier, n=1 for this work.

The remaining free parameter is described by the scalar field's background value at redshift \(z=0\), denoted \(\bar{f}_{R0}\), which is treated as a free parameter to constrain the HS \(f(R)\) model:

\begin{equation}
\frac{C_{1}}{C_{2}^{2}} = -\frac{1}{n} \bar{f}_{R_{0}} \left( \frac{R_{0}}{m^{2}} \right)^{n+1}.
\end{equation}

By appropriately selecting this parameter, the \(f(R)\) model can recover GR in high-density regions, ensuring consistency with solar system tests via the chameleon mechanism \citep{Hu2007ModelsTests}.

In cosmological simulations based on standard gravity, it is common to use the Newtonian limit of GR, which assumes weak gravitational fields and a quasi-static evolution of matter fields. This approximation is also applied in most modified gravity simulations, including those in this work. The limitations of this approach, specifically in the context of \(f(R)\) gravity, are explored in Ref.~\cite{Sawicki2015LimitsCosmologies}. As stated in this paper, the quasi-static approximation in modified gravity and dark energy cosmologies is valid only within the dark energy sound horizon making its applicability limited to specific conditions. For the quasi-static approximation to hold, the sound speed of dark energy must exceed certain thresholds, such as 1$\%$ of the speed of light for current galaxy surveys and up to 10$\%$ for future wide-field surveys like Euclid. The approximation also fails near the sound horizon due to corrections from friction and oscillatory terms.

Under the Newtonian limit, the complex 16-component field equation (Eq.~\eqref{Eequn}) simplifies to two key equations. The first is the Modified Poisson Equation:

\begin{equation}
\nabla^2 \Phi = \frac{16\pi G}{3}\delta\rho_{m} a^{2} - \frac{1}{6} \delta R, \label{poisson}
\end{equation}

where \(\Phi\) represents the perturbation to the time-time component of the metric, \(\delta \rho_{m} = \rho_{m} - \bar{\rho}_{m}\) is the perturbation from the background matter density \(\bar{\rho}_{m}\), and \(\delta R\) is the perturbation from the background value of the Ricci scalar, \emph{i.e.}, the background curvature. The second equation describes the scalar degree of freedom \(f_R\):

\begin{equation}
\nabla^2 f_R =  \frac{1}{3}\left(\delta R - 8\pi G\delta\rho_{m}\right). \label{fRequn}
\end{equation}

Combining Eq.~\eqref{poisson} and Eq.~\eqref{fRequn}, the Modified Poisson Equation is expressed as:

\begin{equation}
\nabla^{2} \Phi = \frac{8 \pi G}{2} \delta \rho_{m} a^{2} - \frac{1}{2} \nabla^{2} f_{R}, 
\end{equation}

where it is more clear that \(f_{R}/2\) acts as the potential for the modified gravity force. The equation approaches the standard GR expression within the Solar System, thanks to the chameleon mechanism \cite{Khoury2003ChameleonCosmology, Hu2007ModelsTests}. In scenarios with small values of \(\bar{f}_{R_{0}}\), the background expansion remains indistinguishable from that in \(\Lambda\)CDM \cite{Kou2023ConstrainingLensing}. In observations, numerous constraints on HS \(f(R)\) gravity focus on \(f_{R_{0}}\). On cosmological scales, constraints have been derived from cluster number counts, CMB, cosmic shear, supernovae, and BAO data, with Ref.~\cite{Cataneo2014NewConstraints} placing \( \log | f_{R_{0}} | < -4.79 \), Ref.~\cite{Hu2016TestingApproach} finding \( \log | f_{R_{0}} | < -4.5 \), Ref.~\cite{Hojjati2015SearchingData} setting \( \log | f_{R_{0}} | < -4.15 \),  Ref.~\cite{Kou2023ConstrainingLensing} finds \( \log | f_{R_{0}} | < -4.61 \) and Ref.~\cite{bai2024testingfrgravitycosmic} obtains \( \log | f_{R_{0}} | < -4.79 \). On more local scales constraints arise from galactic studies, with Ref.~\cite{Naik2019ConstraintsSample} setting \( \log | f_{R_{0}} | < -6.1 \) through galaxy rotation curves, and Ref.~\cite{Desmond2020GalaxyGravity} obtained \( \log | f_{R_{0}} | < -7.85 \) based on galaxy morphology. This means that small-scale measurements force practically all astrophysical objects to be screened, \emph{i.e.} behave like GR. In Ref.~\cite{euclidcollaboration2024euclidpreparationsimulationsnonlinearities} an upper limit of \( \log | f_{R_{0}} | < -5.6 \) is constrained using dark matter simulations with baryonic effects, but neglecting other systematic effects.  While \cite{Liu_2016} establishes \( \log | f_{R_{0}} | < -4.82 \) and \( \log | f_{R_{0}} | < -5.16 \) with WMAP9 and Planck15 priors, respectively, using WL peak abundance.

Despite these tight bounds, \(f(R)\) gravity remains a valuable framework for exploring deviations from GR on cosmological scales. In this study, we explore the \(f(R)\) gravity model using a \(|\bar{f}_{R_{0}}| = 10^{-5}\) (also known as F5), which, while slightly conflicting with local observational constraints, 
is within the HS constraints on cosmological scales \cite{Arnold2018TheDistributions}. F5, with its more significant deviation from GR compared to other studied deviations like F6, provides key insights into the effects of gravity modifications on large-scale phenomena such as weak lensing and clustering statistics. Understanding these effects is crucial for upcoming large-scale structure surveys like DESI, Euclid and LSST, which aim at testing GR as one of their main scientific goals.

\begin{figure*}[t!]
    \centering
    \includegraphics[width=\linewidth]{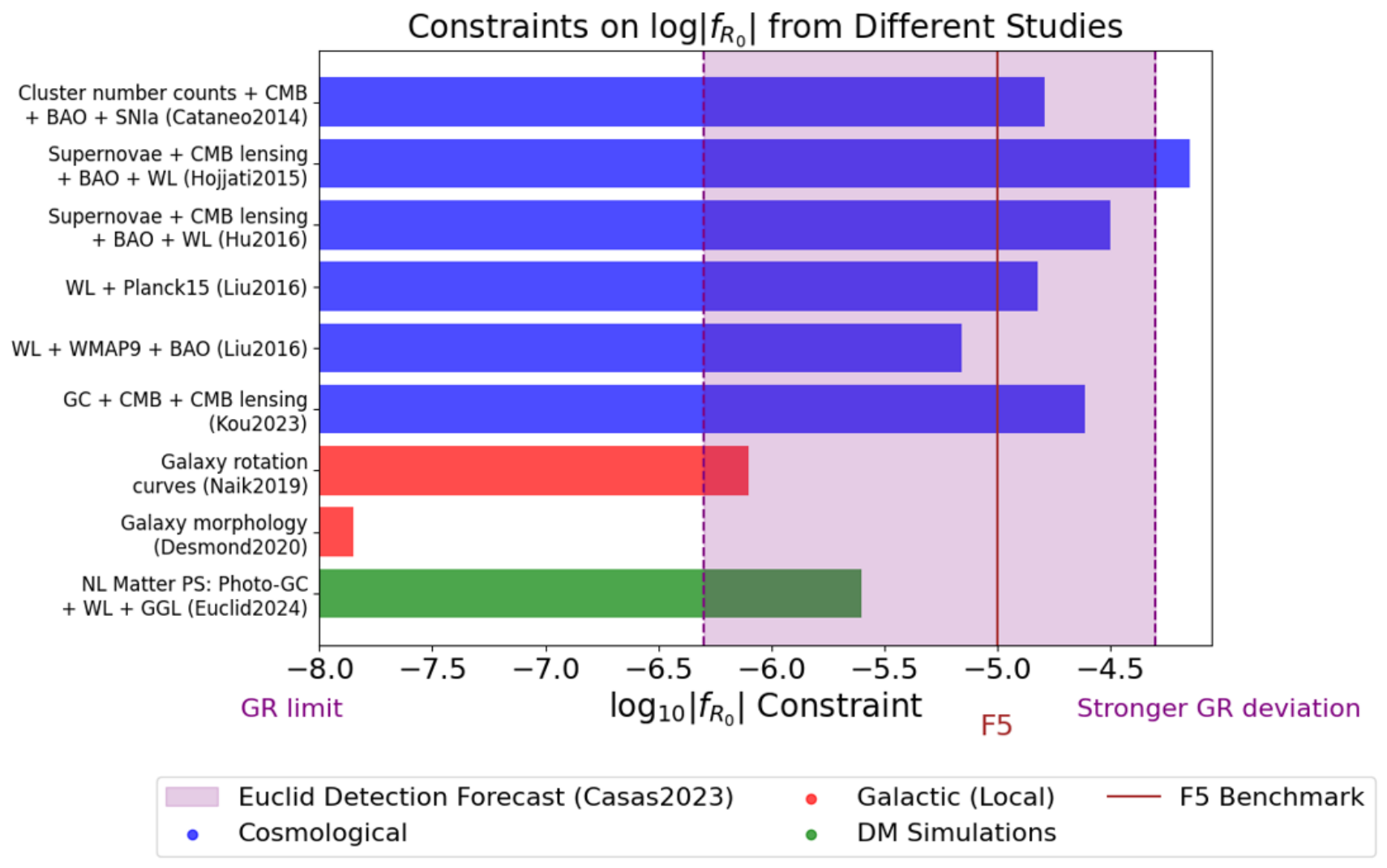}
    \vspace{-2mm}
    \caption{Current upper-bound constraints on the value of $\log |f_{R_{0}}|$ from various cosmological probes. The shaded region indicates the range that Euclid is expected to be able to distinguish. The black vertical line represents the value for F5.}
    \label{fig:constraints}
\end{figure*}

Ref.~\cite{Casas2023Euclid} performs a forecast for precise constraints on $f(R)$ for the Euclid mission. For a fiducial value of \(\log | f_{R_{0}} | = -5.30\), Euclid can constrain \(\log_{10} | \bar{f}_{R_0} |\) to 1$\%$ accuracy by combining both spectroscopic and photometric observations. Additionally, Euclid is expected to distinguish between larger values such as \(\log_{10} | \bar{f}_{R_0} | = -4.30 \), smaller values like \( \log_{10}| \bar{f}_{R_0} | = -6.30\), and \(\Lambda\)CDM with a confidence level exceeding \(3\sigma\). For the DESI mission we can refer to \cite{Alam_2021}. We summarize all these main constraints in Figure \ref{fig:constraints}

\subsection{Dark matter power spectrum boost}

The matter power spectrum, despite not being a direct observable, is one of the basic theoretical quantities that can be modeled to characterize the growth of cosmic structures. The latter allows the construction of predictions for actual observables, such as the galaxy power spectrum, that is a powerful compressed version of the data from which one can estimate cosmological parameters. Similarly an accurate estimation of the dark matter power spectrum for $f(R)$ is a key ingredient to model the $E_G$ estimator. Perturbation theory (PT) can be used to predict the matter power spectrum on quasi-linear scales \cite{Large-scalequasi} with great precision. In the non-linear regime, PT breaks down and one needs to resort to measurements from N-body simulations to derive accurate predictions. This makes the use of cosmic emulators crucial, as they allow for analytical predictions of non-linear scales by interpolating results from a vast number of N-body simulations covering a wide parameter space. In particular, these emulators enable a complex yet accurate modeling of matter and galaxy clustering on smaller scales.

Many emulators exists for $\Lambda$CDM, but in the past years some emulators have appeared for extended Dark Energy models, including modified gravity theories like $f(R)$. For linear matter power spectrum calculations, Boltzmann codes such as \textit{mgcamb} \cite{Zhao2008SearchingSurveys, Hojjati2011TestingCosmoMC, Zucca2019MGCAMBEnergy, Wang2023NewCobaya}, \textit{MGHalofit} \cite{Zhao2014ModelingGravity} and \textit{MGCLASS} \cite{Sakr2021CosmologicalII} are commonly used for different gravity theories including $f(R)$. Additionally, there are simulation-based emulators that extend into the mildly non-linear regime, like \textit{ELEPHANT} \cite{Winther2019EmulatorsCDM}, based on COLA 
(COmoving Lagrangian Acceleration) 
\cite{Ramachandra2020MatterCosmologies};  FORGE \cite{Arnold2021FORGEEmulator}, \textsc{e-mantis} \cite{Saez-Casares2023TheCosmology}, \textit{Sesame} \cite{Mauland2024Sesame:Models} and \textsc{FREmu} \cite{bai2024testingfrgravitycosmic}. Another prominent tool is \textit{ReACT} \cite{Bose2020LambdaCDM, Bose_2023}, which applies a halo model reaction framework validated using N-body simulations.

Cosmological simulations for \( f(R) \) models require significantly more computational time given that they have to compute the intrinsically non-linear evolution of the scalar field that mediates the modified gravity force. These emulators typically work by comparing the power spectrum results of modified gravity models to those of \(\Lambda\)CDM emulators. In this paper, we use the emulator \textsc{e-mantis} \citep{Saez-Casares2023TheCosmology} (Emulator for Multiple observable ANalysis in extended cosmological TheorIeS), which is specifically designed for the Hu \& Sawicki \(f(R)\) gravity model. The \textsc{e-mantis} emulator provides a boost for the \(f(R)\) gravity matter power spectrum, defined as:

\begin{equation}
    B(k) = \frac{P_{f(R)}(k)}{P_{\Lambda\mathrm{CDM}}(k)},
\label{eq:mg boost}
\end{equation}

where \(P_{f(R)}(k)\) and \(P_{\Lambda\mathrm{CDM}}(k)\) are the matter power spectra for \(f(R)\) gravity and \(\Lambda\)CDM, respectively. This boost is less sensitive to statistical and systematic errors and exhibits a smoother dependence on cosmological parameters than the power spectrum itself. Since both \(f(R)\) and \(\Lambda\)CDM simulations start from the same initial conditions, the minimal impact of \(f(R)\) gravity on large scales preserves the strong cancellation of cosmic variance and large-scale errors. Similarly, small-scale systematic errors due to limited mass resolution also cancel out. By focusing solely on this boost, the emulator significantly reduces computational demands, as less precise simulations are required to achieve the desired accuracy for the boost compared to the raw power spectrum.

The power spectrum boost is mainly influenced by three cosmological parameters: \(\bar{f}_{R_{0}}\), \(\Omega_m\), and \(\sigma_8\). Variations in other parameters, such as \(h\), \(n_s\), and \(\Omega_b\), have a negligible impact, with less than \(1\%\) variation up to scales of \(k = 10 h \, \mathrm{Mpc}^{-1}\) \cite{Saez-Casares2023TheCosmology}. The emulator does not account for the effect of baryonic physics on the matter distribution, which affects the matter power spectrum boost in \(f(R)\) gravity for scales \(k \gtrsim 2 \, h \, \mathrm{Mpc}^{-1}\) \cite{Arnold2019RealisticGravity}. However, Ref.~\cite{Saez-Casares2023TheCosmology} anticipates that using a \(\Lambda\)CDM emulator that incorporates the baryonic impact on the matter power spectrum can approximately correct for this effect for \(f(R)\) models. In this study we will use the Halofit matter power spectrum from Ref.~\cite{Takahashi2012RevisingSpectrum} which despite no including baryonic impact accounts for accurate nonlinear corrections.

Using simulations with an effective volume of \((560 \, h^{-1} \mathrm{Mpc})^3\) and a particle mass resolution of \(m_{\mathrm{part}} \sim 2 \times 10^{10} \, h^{-1} M_{\odot}\), the power spectrum boost can be determined with better than \(3\%\) accuracy for the range \(0.03 \, h \, \mathrm{Mpc}^{-1} < k < 7 \, h \, \mathrm{Mpc}^{-1}\) and redshifts \(0 < z < 2\). Although the systematic error on the boost varies with \(\bar{f}_R\), redshift, and scale, the \(3\%\) estimate is conservative, as most cases achieve better than \(1\%\) accuracy \cite{Saez-Casares2023TheCosmology}.

\subsection{Scale dependence in the growth rate}

At the linear perturbation level in the comoving gauge, the modified Einstein equations for \( f(R) \) gravity lead to the following equations in Fourier space for the evolution of matter overdensities. These describe how perturbations in the matter density evolve over time \cite{Tsujikawa2009TheGravity, Nojiri_2011, Nojiri_2017, Mirzatuny2019AnGravity}:

\begin{equation}
\begin{split}
\ddot{\delta}_m + \left( 2H + \frac{\dot{f_R}}{2f_R} \right) \dot{\delta}_m 
- \frac{\rho_m}{2f_R} \delta_m  \\
= \frac{1}{2f_R} \left[ \left( \frac{k^2}{a^2} - 6H^2 \right) \delta f_R \right.
+ 3H \dot{\delta f_R} + 3 \delta \ddot{f_R} \left. \right]
\end{split}
\end{equation}

\begin{equation}
\ddot{\delta f_R} + 3H \dot{\delta f_R} + \left( \frac{k^2}{a^2} + \frac{f_R}{3f_{RR}} - \frac{R}{3} \right) \delta f_R = \frac{1}{3} \rho_m \delta_m + \dot{f_R} \dot{\delta}_m.
\end{equation}

In these equations, \( k \) is the comoving wavenumber, \( a = (1+z)^{-1} \) is the scale factor (normalized to unity today), \( \rho_m \) is the matter density, and \( \delta_m(a) = \delta \rho_m / \rho_m \) is the matter density contrast. The Hubble parameter \( H \) is given by \( H = \dot{a} / a \), and dots represent derivatives with respect to cosmic time. Lastly, $f_{RR}$ is the derivative of $f_R$ with $R$

\begin{equation}
    f_{RR} \equiv \frac{df_R}{dR} \approx \frac{n(n+1)}{m^2} \frac{c_1}{c_2^2} \left( \frac{m^2}{R} \right)^{n+2},
\end{equation}

where we take the approximation from Eq.~\eqref{f_R approx}

For cosmologically viable \( f(R) \) models, \( f_R \) changes slowly, meaning \( |\dot{f_R}| \ll Hf_R \). Under this approximation, for \( f_R \), the time derivatives can be neglected and the oscillatory modes are insignificant compared to those driven by matter perturbations. Additionally, for modes well inside the Hubble radius, \( k^2/a^2 \gg H^2 \), further simplifying the equations. These approximations lead to the following equation for the evolution of the matter density contrast:

\begin{equation}
    \ddot{\delta}_m + 2H \dot{\delta}_m - 4 \pi G_{\text{eff}} \rho_m \delta_m \simeq 0,
\label{eq:diff eq}
\end{equation}

where \( G_{\text{eff}} \) replaces the standard gravitational constant \( G \) from \( \Lambda \)CDM cosmology. \( G_{\text{eff}} \) is the effective gravitational constant, defined as:

\begin{equation}
G_{\text{eff}}(k, z) = \frac{G}{f_R} \left[ 1 + \frac{1}{3} \left( \frac{k^2}{a^2 M^2 / f_R + k^2} \right) \right],
\end{equation}

The dependence of \( G_{\text{eff}} \) on scale introduces scale-dependent effects in the formation of cosmic structures, distinguishing \( f(R) \) models from standard cosmology. From Eq.~\eqref{eq:diff eq} the growth factor can we estimated, which is related to the growth rate of structures \textit{f}. The growth rate $f$ is typically defined as:

\begin{equation}
    f(a)=\frac{d \ln D}{d \ln a}=\frac{a}{D} \frac{d D}{d a}
\end{equation}

where $D(t)$ is the growth factor corresponding to the time factorization of the linear growth of matter perturbations. In particular, we assume that the growth of perturbations can be factored into time-dependent and spatially dependent parts:

\begin{equation}
\delta(\mathbf{x}, t)=\delta_0(\mathbf{x}) D(t)
\end{equation}

The growth factor can be calculated as solution to the  linear perturbation theory differential equation:

\begin{equation}
\ddot{D}(t)+2 H(t) \dot{D}(t)-4 \pi G_{\text{eff}} \rho_m(t) D(t)=0
\label{D diff}
\end{equation}

which happens to be the same than in $\Lambda$CDM but using $G_{\text{eff}}$ instead of G. This means that, in $f(R)$,  since $G_{\text{eff}}$ now depends on the scale $k$, the growth factor, and therefore the growth rate, will also depend on $k$. In Ref.~\cite{Garc_a_Farieta_2024} the value of the growth factor in $f(R)$ is calculated and studied in more detail. Instead, in this work we use the public code MGrowth\footnote{\url{https://github.com/MariaTsedrik/MGrowth/blob/main/docs/MGrowth.rst}} which allows to obtain the value of the growth rate for several models of gravity. One of the gravity models is HS f(R) for any value of redshift, wavenumber and $\bar{f}_{R_{0}}$ between $10^{-9}$ and $10^{-2}$. This code basically works by solving Eq.~\eqref{D diff} numerically using the $G_{\text{eff}}$ defined in this section. In Fig.~\ref{fig:MG_boost} we plot the values of the growth rate for the three redshifts bins that we will use for the analysis of GR and MG mock data. Within the Limber and small angle approximation, we can obtain a simple approximate relation between Fourier wavemodes and projected scales ($s$):

\begin{figure*}[t!]
\centering
\includegraphics[width=0.49\textwidth]{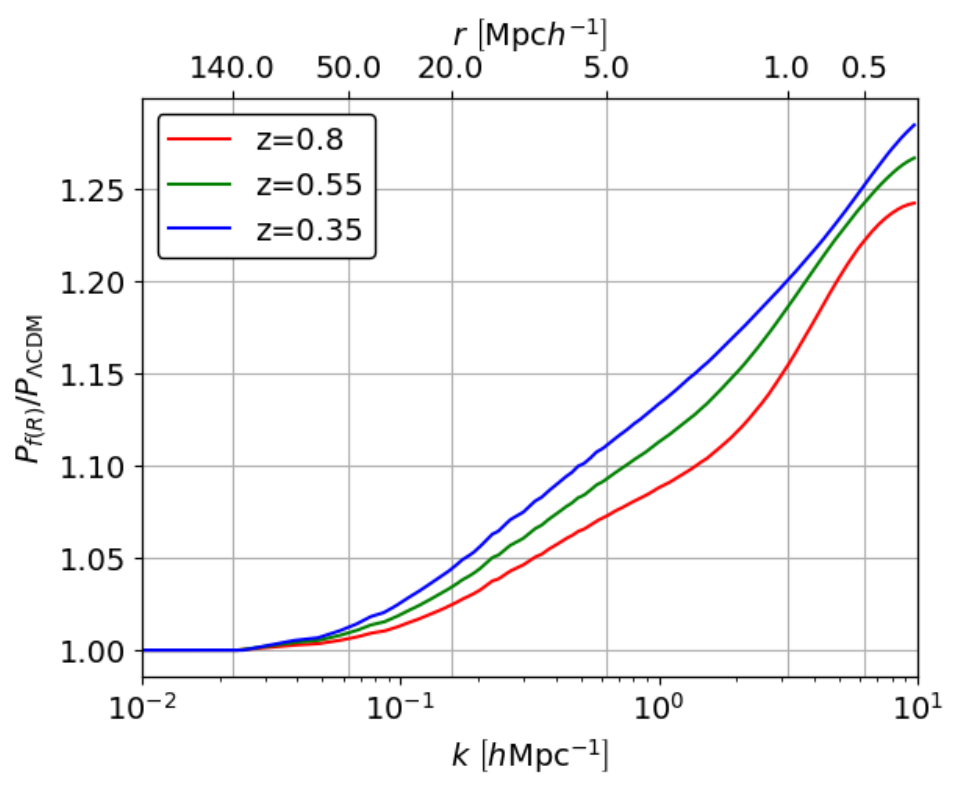}
\includegraphics[width=0.49\textwidth]{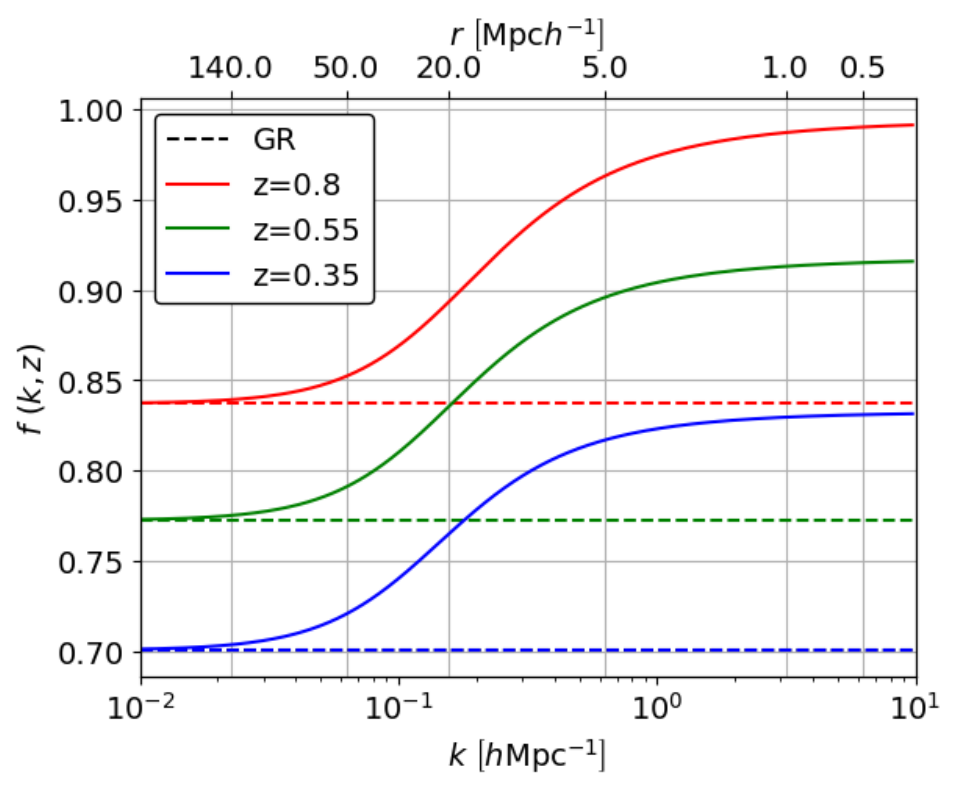}\vspace{-2mm}
\includegraphics[width=0.49\textwidth]{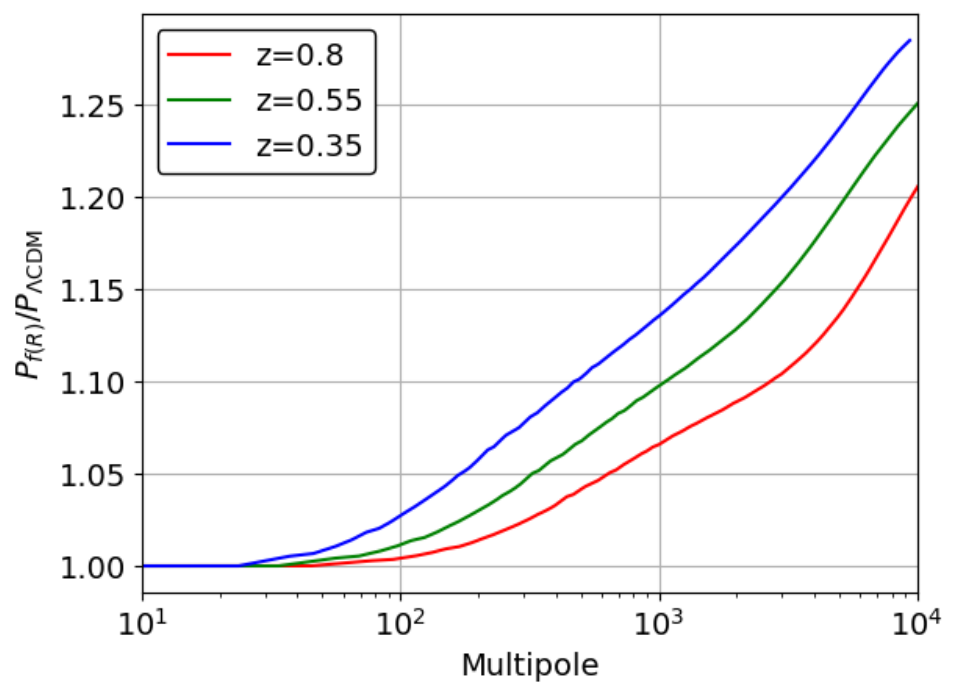}
\includegraphics[width=0.49\textwidth]{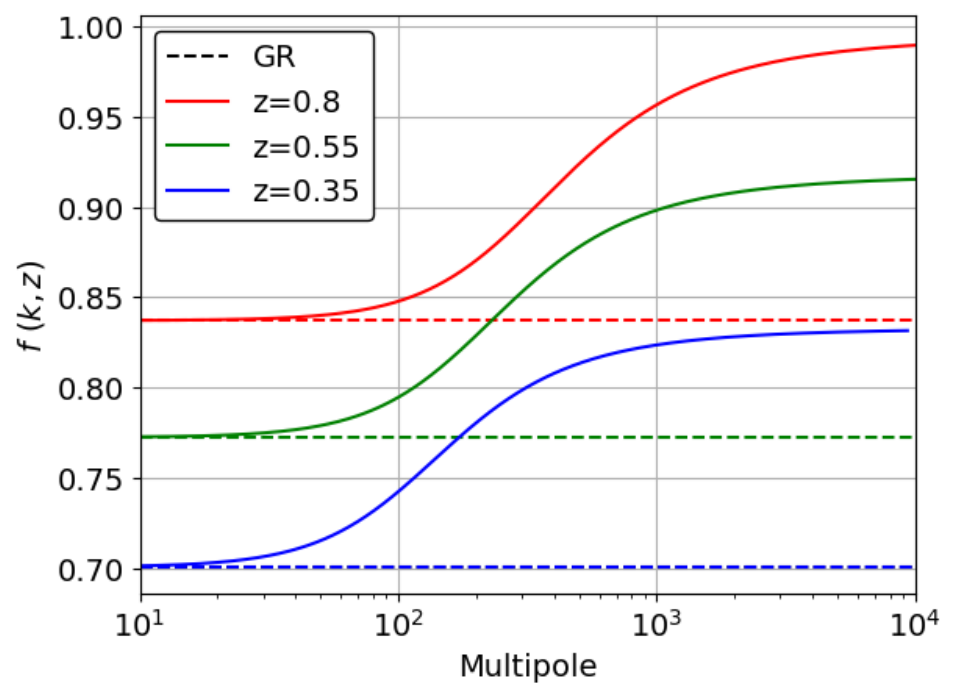}
\vspace{-2mm}
  \caption{Comparison of the effects of $f(R)$ gravity on the dark matter power spectrum (Left plots) and the growth rate (Right plots) for the three redshifts used in this analysis. We show the results on $k$-wavenumbers and the corresponding comoving distance separation (Top plots). On a separate plot (Bottom plots), due to the dependence on a different projected commoving transversal distance for each z, we show the equivalent multipoles. Note that for the DM spectrum, we show the ratio (boost), while for the growth rate we show the values for GR and MG separately since the ratio is quite similar for each z.}
  \label{fig:MG_boost}
\end{figure*}

\begin{equation}
    \ell \approx \frac{\pi}{\theta} 
\label{eq:ell trans}
\end{equation}

\begin{equation}
\theta \approx \frac{s}{\chi (z)}
\label{eq:theta trans}
\end{equation}

\begin{equation}
    k \approx \frac{\ell}{\chi (z)} \sim \frac{\pi}{s}
\label{eq:k trans}
\end{equation}

where $\chi (z)$ is the commoving distance and $\theta$ the angular scale. The final relation  between $k$ and $s$ (Eq.~\eqref{eq:k trans}) should be taken as a reference only since it comes from combining three approximations between variables. 

\section{The $E_G$ estimator}
\label{sec:EG estimator}

The original definition of the $E_G$ estimator in Fourier space as introduced in Ref.~\cite{Zhang2007AScales} is the following:

\begin{equation}
    E_G (k, z) \equiv \frac{c^{2} k^{2} ( \Phi-\Psi)} {3 H_{0}^{2} ( 1+z ) \nu( k )} \,, 
    \label{eq:EG_original}
\end{equation}

where $\nu$ is the divergence of the peculiar velocity field. The potentials $\Psi$ and $\Phi$ are, respectively, the time and spatial component of the perturbation fields of the metric. For a flat universe governed by the Friedmann–Lemaître–Robertson–Walker (FLRW) metric and under the assumption of negligible anisotropic stress and non-relativistic matter species, the Einstein field equations for time-time and momentum components in GR can be expressed in Fourier space as follows \cite{Hojjati2011TestingCosmoMC}:

\begin{equation}
\begin{aligned}
k^{2} \Psi &= -4 \pi G a^{2} \rho_m(a) \delta \\
\Phi &= -\Psi,
\label{eq:Poisson GR}
\end{aligned}
\end{equation}

\( \rho_m \) represents the background matter density, \( a \) is the scale factor and \( \delta \) denotes the matter density perturbation. In modified gravity models, these equations are usually generalized to:

\begin{equation}
\begin{aligned}
k^{2} \Psi &= -4 \pi G a^{2} \mu(k, a) \rho_m(a) \delta \\
\Phi &= -\gamma(k, a) \Psi,
\label{eq: Poisson MG}
\end{aligned}
\end{equation}

where \( \mu(k, a) \) and \( \gamma(k, a) \) are arbitrary functions of \( k \) and \( a \). The $\mu$ function parametrizes the effective strength of gravity, and $\gamma$ is the gravitational slip that quantifies the difference in the gravitational perturbation fields.  These functions reduce to \( \mu = \gamma = 1 \) in the GR case in order to recover Eq.~\eqref{eq:Poisson GR}.

Combining the equations in \eqref{eq: Poisson MG} we can then rewrite the numerator of \( E_G \) in Eq.~\eqref{eq:EG_original} as \cite{Pullen2014ProbingLensing}:

\begin{equation}
\begin{aligned}
k^{2} ( \Phi-\Psi)=\frac{3} {2} H_{0}^{2} \Omega_{m, 0} ( 1+z ) \mu( k, a ) [ \gamma( k, a )+1 ] \delta 
\label{eq: EG numerator}
\end{aligned}
\end{equation}

where \( \Omega_{m, 0} = \frac{8 \pi G \rho_{m0}}{3 H_0^2} \) with \( \rho_m(a) = \rho_{m0} a^{-3} \). The velocity perturbation \( \nu \) is given by \( \nu = f \delta \) at linear scales. Combining this relation with Eq.~\eqref{eq: EG numerator}, the expression for \( E_G \) becomes:

\begin{equation}
E_G(k, z) = \frac{\Omega_{m, 0} \mu(k, a) [\gamma(k, a) + 1]}{2 f(k,z)} =  \frac{\Omega_{m, 0} \Sigma(k, a)}{f(k,z)},
\label{eq: EG general}
\end{equation}

where we have re-parameterized as $\Sigma \equiv \frac{1}{2}\mu(1+\gamma)$ \cite{Wenzl2024ConstrainingBOSS}, which represents the lensing parameter. As mentioned earlier, since \( \mu = \gamma = 1 \), we have that $\Sigma=1$ in GR. Then we can clearly see that for GR, the value of $E_G$ is given by:

\begin{equation}
    E_G^{GR} = \frac{\Omega_{m, 0}}{f(z)},
\end{equation}

which can be readily computed, since $f$ in GR can be approximated as $f(z) \approx \Omega_m(z)^{0.55}$ \cite{Wang_1998, Linder_2005}. This means that for GR the value of $E_G$ is predicted from the background expansion only using a constraint on $\Omega_{m,0}$ and it is thus independent of sample-specific parameters like the galaxy bias and its potential systematic effects. This value is predicted to be scale independent for GR at linear scales. In contrast, for $f(R)$ and other MG theories this estimator depends on scale, what is a potential smoking gun for detecting deviations from GR.

Returning to the $f(R)$ model the functions $\mu$ and $\gamma$ can be parameterized as \cite{Pullen2014ProbingLensing}: 

\begin{equation}
{{{\mu^{{f(R)}} ( k, a, ) \!=\! \frac{1} {1-B_{0} a^{s-1} / 6} \left[ \frac{1+( 2 / 3 ) B_{0} \bar{k}^{2} a^{s}} {1+( 1 / 2 ) B_{0} \bar{k}^{2} a^{s}} \right]}}}
\end{equation}

\begin{equation}
{{{\gamma^{{f(R)}} ( k, a ) \!=\! \frac{1+( 1 / 3 ) B_{0} \bar{k}^{2} a^{s}} {1+( 2 / 3 ) B_{0} \bar{k}^{2} a^{s}} \,,}}}
\end{equation}

Where \( \bar{k} = k \cdot \left( \frac{c}{H_0}\right) = k \cdot \left[ 2997.9 \, \mathrm{Mpc}/h \right] \), and \( h = H_0 / \left[ 100 \, \mathrm{km/s/Mpc} \right] \), with \( s = 4 \) for models that follow the $\Lambda$CDM expansion history. The parameter \( B_0 \) is a free variable associated with the Compton wavelength of an additional scalar degree of freedom, and it is also proportional to the curvature of \( f(R) \) at present times. Current observational constraints place a limit of \( B_0 < 5.6 \times 10^{-5} \) at a \( 1 \sigma \) confidence level \cite{Pullen2014ProbingLensing}. Implementing this parametrization into the general expression from Eq.~\eqref{eq: EG general} one gets: 

\begin{equation}
    E_G^{f(R)}(k, z) = \frac{1}{1 - B_0 a^{s-1} / 6} \frac{\Omega_{m, 0}}{f^{f(R)}(k, z)}\,.
\end{equation}

Since the constraints establish that $B_0 a^{3} << 1$ then we can simply the expression to:

\begin{equation}
    E_G^{f(R)}(k, z) = \frac{\Omega_{m, 0}}{f^{f(R)}(k, z)}\,.
\label{eq: EG_fR teo}
\end{equation}

Therefore, in the large scale limit, the $E_G$ estimator in $f(R)$ will only differ from the one in GR through the scale and redshift dependence of the growth rate. Since as we have seen in the previous section, the growth rate depends on scale for $f(R)$, so will the value of $E_G^{f(R)}$. Going back to the $\Sigma$ parameter defined in Eq.~\eqref{eq: EG general}, we have that, similar to GR, $\Sigma \simeq1$ for the HS \(f(R)\) model. The reason behind this is that, unlike other modified gravity theories, HS \(f(R)\) has negligible impact on the propagation of light in the weak-field limit \cite{Hojjati2015SearchingData}, since $f (R)$ models have a conformal coupling. The actual value of $\Sigma$ in $f(R)$ is given by \cite{euclidcollaboration2024euclidpreparationsimulationsnonlinearities}: 

\begin{equation}
    \Sigma (z) = \frac{1}{1 + f_R(z)}
\label{eq:sigma fR}
\end{equation}  

Since the maximum value of |$f_R(z)$| is given by |$f_{R0}$| we can ignore this effect for the F5 model used.

\begin{figure*}[t!]
    \centering
    \includegraphics[width=\linewidth]{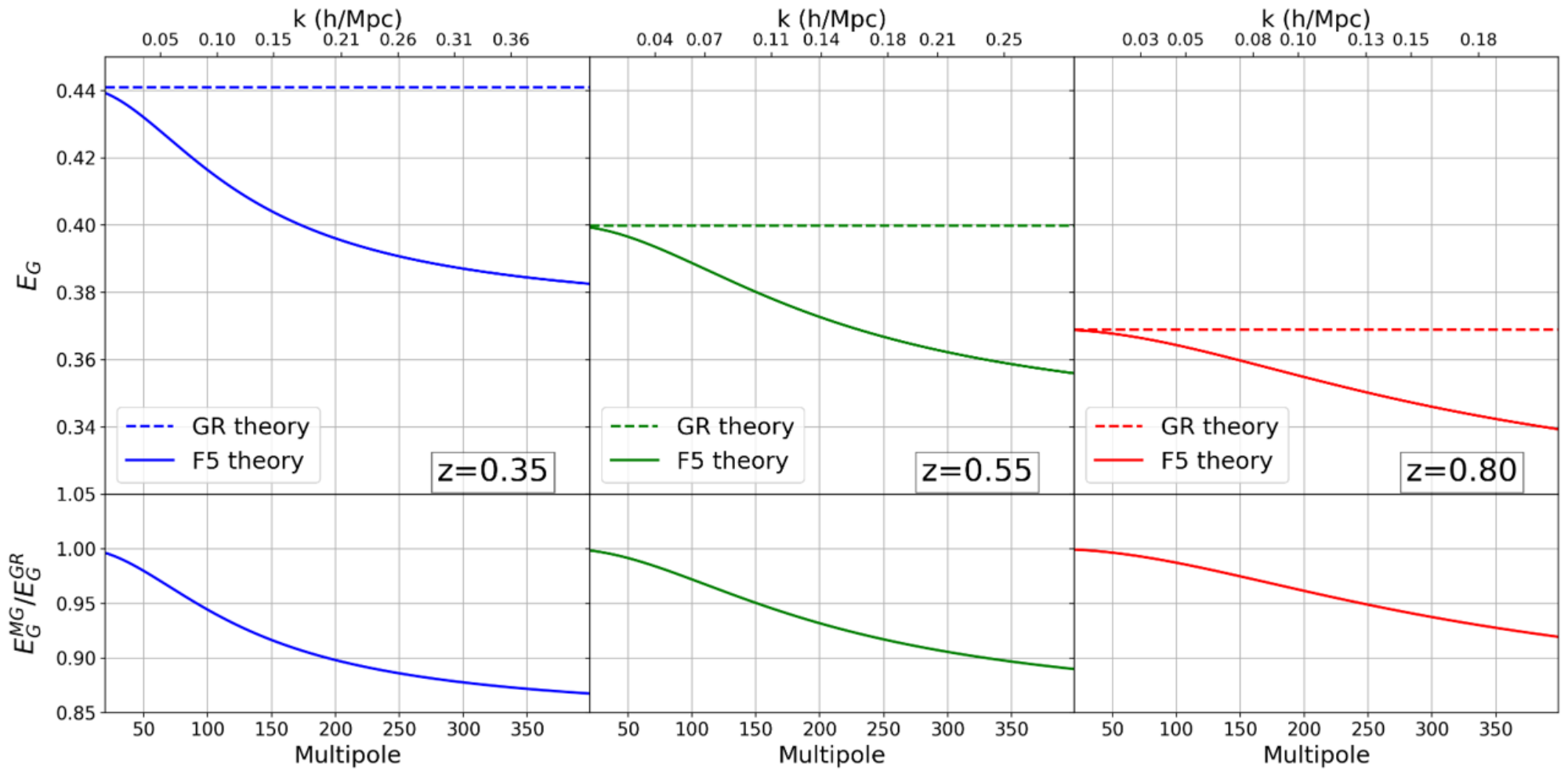}
    \vspace{-2mm}
    \caption{Plots showing the difference between the prediction of $E_G$ for GR (dashed line) and F5 (solid line) at the three redshift bins. The lower plots show the respective ratio of the prediction for F5 over GR.}
    \label{fig:ratio teo}
\end{figure*}

In summary, in order to differentiate HS $f(R)$ from GR we shall need to accurately estimate the linear growth rate of perturbations. In Fig.~\ref{fig:ratio teo} we plot the predictions of $E_G$ for F5 and GR and their ratio. We can see that contrary to the growth rate prediction, since $E_G$ is inverse to this parameter, now the prediction for F5 decreases with the scale and it is always lower than the one for GR which is scale independent. We also note that the ratio between both predictions increases as we move to lower redshifts. This suggests that, in principle, the optimal strategy is to to use the low redshift clustering measurements to constrain MG models. However, in practice, nonlinear effects at smaller scales significantly complicate their modeling. Besides, the amount of independent modes decrease due to gravitational coupling. These two factors increase the statistical errors on clustering measurements on small scales.

\section{Modeling RSD: Non-linear effects}
\label{sec:Comet}

The impact of (peculiar) velocities of galaxies away from the Hubble flow introduce a perturbation in the estimation of distances to galaxies as expected from the Hubble law. This systematic effect distorts the pattern of galaxy clustering in a way that depends on the growth rate. Therefore one can exploit the so-called redshift space distortions (RSD) as a powerful probe of dark energy and gravity. By analyzing these effects, we can directly measure the rate at which structures in the universe grow, since it is related to the growth rate \textit{f}.

In the linear regime, the galaxy clustering pattern suffers from a characteristic squashing distortion along the line of sight, known as the Kaiser effect \cite{1984ApJ...284L...9K}. In this limit, one can obtain a rather simple expression that relates the linear power spectrum in real and redshift space through the growth rate. Previous works like Ref.~\cite{Hern_ndez_Aguayo_2019} remarked the limitations of only considering linear theory to determine the value of \textit{f} at small scales (s < 40 h$^{-1}$Mpc). But modeling also the nonlinear contribution, known as the Finger of God effect, is significantly more challenging due to the complicated nature of non-linear gravitational growth. There are numerous models to describe these interactions, each with its own parameter space. These models address nonlinearities in different ways and may perform better or worse depending on the galaxy sample and the scales being analyzed. Regarding the galaxy power spectrum in redshift space, effective field theory (EFT) \cite{Ivanov2019CosmologicalSpectrum, DAmico2020LimitsCode} has gained popularity recently since it is very versatile. 

The generality of EFT allows to incorporate several kind of extra parameters to describe models of modified gravity. It also introduces counterterms parameters in order to correct that the energy-momentum tensor is no longer homogeneous and isotropic at small scales, \emph{i.e.} off-diagonal elements no longer vanish. Although many models try to incorporate as many parameters to have more freedom to describe nonlinear processes this comes at the prize of having to evaluate a much larger parameter space. On top of that, there may be prior volume effects in the language of Bayesian statistics, also called "projection effects", where different parameters contribute to the model, or the power spectrum in this case, in a similar manner, \emph{i.e.} the parameters are degenerated with each other. This can significantly bias the estimation of the cosmological parameters. 

In this section we will give an introduction to another perturbative model called the velocity difference generating function (VDG) model \cite{Sanchez2016TheWedges, Eggemeier2023COMETTheory}. We will use this model in this paper to estimate the $\beta$ parameter for the calculation of the $E_G$ estimator. We choose this model over others due to the reported level of accuracy at small scales when comparing results with simulated catalogs (see Ref.~\cite{eggemeier2025boostinggalaxyclusteringanalyses}). The reason behind this is related to the better description of the PDF of "pairwise velocities" (mean value of the peculiar velocity difference of a galaxy pair at a given separation) \cite{Ferreira_1999, Cabr2} compared to the EFT model, which primarily captures the Fingers of God (FoG) effect \cite{Eggemeier2023COMETTheory}.  The COMET emulator \cite{Eggemeier2023COMETTheory} implements the VDG model, alongside an EFT implementation with a similar parameter space. In what follows, we shall use the VDG emulator as our reference non-linear RSD model.

\subsection{COMET-EMU}

The emulator contains a reduced parameter space thanks to the \textit{evolution mapping} approach from Ref.~\cite{Sanchez2022EvolutionRegime} which separates the parameter space in: shape parameters, which determine the shape of the linear power spectrum in terms of the physical densities $\omega_i$ and the spectral index $n_s$, and evolution parameters, which determine the amplitude and evolution with redshift, which in turn depend on the scalar amplitude of the primordial power spectrum, $A_s$, and the parameters defining the curvature and the dark energy model.

The emulator also works with the parameter $\sigma_{12}$ as the root-mean-square (RMS) of matter fluctuations in spheres of radius R=12 Mpc. This parameter is determined from the evolution and shape parameters when emulating a $\Lambda$CDM model, but it becomes a free parameter (in which case $A_s$ is no longer needed) when selecting a non-$\Lambda$CDM cosmology.  The emulator uses $\sigma_{12}$ instead of the most conventional $\sigma_8$ (RMS of matter fluctuations in spheres of radius R=8 Mpc/$h$) since, as shown in Ref.~\cite{Sanchez2020ArgumentsCosmology}, it is better to express the power spectrum in Mpc instead of Mpc/$h$ to have the correct scaling in the evolution mapping approach. The predictions of COMET are limited to the range of scales $k \in [6.95 \cdot 10^{-4}, 0.35028]$ Mpc$^{-1}$, although it can make power-law extrapolations beyond this range of scales. 

Unfortunately, COMET does not allow the input of a custom matter power spectrum outside of those already incorporated into the code, which limits our ability to modify this aspect of the analysis. Since \textsc{e-mantis} only calculates the boost for the dark matter power spectrum, we are uncertain whether it can be directly applied to the final multipoles, which account for galaxies and RSD effects.  Looking at appendix \ref{annex:VDG} and the final expression for the multipole wedges, Eq.~\eqref{eq: wedges}, it appears that the relationship remains linear with the matter power spectrum, as the galaxy power spectrum terms, Eq.~\eqref{eq:galaxy power spectrum}, are also linear with the matter power spectrum. The final expression for the multipoles, Eq.~\eqref{eq: multipoles final comet}, integrates the wedges over the angle $\mu$, but since the boost is independent of $\mu$, we can simply factor it out of the integral. In order to have a bound on the error induced by this modeling choice we also perform the analysis on the F5 data using the GR template. We will further discuss the impacts of this approximation on section \ref{sec:Discussion4}. 

\subsection{The VDG model}
\label{sec:VDG}

This model differs from the EFT approach only on its treatment of RSD. While EFT performs a full expansion of the real-to-redshift space mapping, VDG partly retains the non-perturbative nature of this mapping including the exponential-type PDF (damping factor) for the pairwise velocities. The model was originally proposed in Ref.~\cite{Sanchez2016TheWedges} to describe the matter power spectrum in RSD. In Ref.~\cite{Eggemeier2023COMETTheory} it is referred as VDG due to its relation to the velocity difference generating function (see sec.~\ref{sec:Velocity generating function}) to account for the virialized velocity (dynamical equilibrium velocity in a gravitationally bound system) of galaxies  impact on the power spectrum. For this, the model uses an effective damping function (Eq.~\eqref{eq:velocity generating}) to describe non-perturbatively the  the impact of the distribution of velocities at small scales due to non-linear effects corresponding mostly to the fingers-of-God (FOG) effect due to virial motions.

The VDG model implemented in COMET is rather complex, so in appendix \ref{annex:VDG} we summarize the key points of the implementation from Ref.~\cite{Eggemeier2023COMETTheory}, and we refer to this paper (or Ref.~\cite{eggemeier2025boostinggalaxyclusteringanalyses}) for further  details . In section \ref{sec:MCMC params} we give a brief description of the space of COMET's parameters that we shall use to perform the model fits to the simulation data vectors using a MCMC approach. 

\section{Simulations}
\label{sec:Data}

\subsection{Nbody simulations}

We take our data from the General Relativity and $f(R)$ Modified Gravity mock presented in Ref.~\cite{Arnold2018TheDistributions}. The $f(R)$ simulation was obtained using the cosmological simulation code \textsc{mg-gadget3}, which is a modification of the code \textsc{p-gadget3} that allows to run collisionless simulations in the Hu-Sawicki $f(R)$-gravity model. Four collisionless cosmological simulations were conducted. Each simulation was run twice: once using the $f(R)$ model and once with a $\Lambda$CDM cosmology, both utilizing identical initial conditions. In this paper we use the pair of simulations with the higher resolution placing $2048^3$ particles in a 768 Mpc/$h$ side-length box, which give a mass resolution of $M_{part}=3.6 \ x \ 10^9 M_\odot/h$. The mocks have a fiducial Planck-like cosmology following Ref.~\cite{Ade2016Planck2015Parameters}, with $\Omega_m = 0.3089$, $\Omega_\Lambda = 0.6911$, $\Omega_b = 0.0486$, $h = 0.6774$, $\sigma_8 = 0.8159$ and $n_s = 0.9667$.

To solve the equation for the scalar degree of freedom in modified gravity (Eq.~\eqref{Eequn}),  \textsc{mg-gadget} applies an iterative Newton-Raphson method combined with multi-grid acceleration on an adaptive mesh refinement (AMR) grid. Instead of solving directly for \(f_R\), the code solves for \(u = \log(f_R / f_{R_0})\) to avoid nonphysical positive values of \(f_R\) in the simulation, a technique first introduced by Ref.~\cite{Oyaizu2008NonlinearMethodology}. Once \(f_R\) is determined, it is used to calculate an effective mass density, incorporating all $f(R)$ effects, including the chameleon mechanism \cite{Arnold2018TheDistributions}:

\begin{equation}
\delta \rho_{\rm eff} = \frac{1}{3} \delta \rho - \frac{1}{24\pi G} \delta R.
\end{equation}

The total gravitational acceleration can then be computed by adding this effective density to the real mass density and using the standard Tree-PM Poisson solver implemented in \textsc{p-gadget3}.

Each simulation includes a 2D lightcone output, consisting of 400 \textsc{HEALPix} maps \cite{GorskiHEALPix:SPHERE} between redshifts \(z = 80\) and \(z = 0\). These maps, evenly spaced in lookback time, have a resolution of $nside=8192$ (\emph{i.e.}, a pixel angular extent of 0.43 arcmin). They are constructed using the 'Onion Universe' method \cite{Fosalba2008TheShells}, where the simulation box is repeated in all directions around the observer located at one of the corners of the original box to cover the volume up to a given redshift \(z_i\), and particles within a thin spherical shell at \(z_i\) are binned onto a \textsc{HEALPix} map. The shell thickness is chosen to ensure a space-filling lightcone output. From this lightcone decomposition in concentric shells projected onto a set of HEALPix maps, the convergence maps can be generated for each redshift bin. Other lensing properties map, such as the deflection and the cosmic shear, can be computed using simple algebraic operations, valid on the full-sky limit, on the harmonic decomposition of the convergence maps \cite{Hu_2000}.

From the simulations, a 3D halo catalog is generated on the fly. Halos are identified using a shrinking sphere method applied to objects found by the Friends-of-Friends (FOF) halo finder in \textsc{p-gadget3}. The catalog stores properties such as the halo's mass, position, velocity, center of mass, and tensor of inertia. The simulations also produce time-slice outputs and halo catalogs generated using the \textsc{subfind} algorithm \cite{SpringelPopulating0}.

\subsection{Galaxy mocks}

The galaxy assignment method used for the simulations of this paper are described in detail in Tutusaus et al. (in preparation). Below we summarize the main steps in the galaxy mock production and calibration. Following Ref.~\cite{Carretero}, the galaxy population is added to these halos based on a combination of models: the Halo Occupation Distribution (HOD) and Sub-Halo Abundance Matching (SHAM). In this setup, the model includes two key parameters for populating galaxies within each halo. The first parameter, \( M_1 \), establishes a mass threshold that controls the presence of central galaxies in the halos. For each halo that surpasses this mass threshold, a central galaxy is assigned. The second parameter, \( \Delta L_M \), introduces scatter in the pseudo-luminosity assigned to galaxies, which helps match the observed galaxy distribution and luminosity scatter seen in the survey data. The number of satellite galaxies are then assigned to each halo following a Poisson distribution, and they are positioned according to a Navarro-Frenk-White (NFW) profile within each halo to approximate the realistic clustering patterns. Then luminosity, galaxy properties such as positions, velocities, and colors are assigned to each galaxy. Galaxy colors are determined by dividing populations into red, green, and blue groups, with fractions (for centrals and satellites) calibrated against observational data from the Sloan Digital Sky Survey (SDSS). These assignments result in detailed mocks that mimic real galaxy distributions across properties at low redshifts, such as luminosity, clustering as a function of color and magnitude, and the color-magnitude diagram.

After generating this initial galaxy catalog, calibration is applied to ensure that the mock galaxy catalog closely mirrors observed galaxy clustering statistics (Tutusaus et al. (in preparation)). Calibration of the mocks involves adjusting 23 free parameters to minimize discrepancies with observational data, including luminosity functions, clustering measurements, and color distributions. Observational data from sources like the SDSS serve as the benchmark for calibration. To manage the high-dimensional parameter space, the Differential Evolution Algorithm is used. This stochastic optimization method iteratively refines a population of candidate parameter sets by combining them to minimize a $\chi^2$ discrepancy metric. The algorithm ensures convergence to optimal parameters by leveraging mutation, recombination, and selection processes, even in the presence of stochastic variations inherent in the mock generation process. This calibration yields a galaxy catalog that faithfully represents the spatial distribution and clustering properties observed in the universe.

As shown in Ref.~\cite{Arnold2019RealisticGravity}, most of the differences in dark-matter clustering between $f(R)$ and GR models appear at low redshift. Therefore in our analysis we study the $E_G$ estimator in 3 redshift bins at $z<1$. In particular we choose bins centered at $z=0.35, 0.55$ and $0.8$, each one with bin-width of $\Delta_z = 0.1$. The lensing source sample is selected at $z=1.0$ with a bin-width of $\Delta_z=0.2$. The redshifts selected and bin cuts performed are always spectroscopic observed redshifts. We always select all the galaxies over the full sky within each case, imposing a relative magnitude cut on the SDSS r-band of $r<24$.

In order to test the robustness of our results to sample selection, we select six different galaxy samples: 1) the full sample of galaxies, which is our baseline or reference case, 2) only the central galaxies of the halos, 3) a red galaxy sample, 4) a blue galaxy sample, 5) a faint sample, selected by imposing a magnitude cut, $23<r<24$, 6) a bright sample, obtained with a relative magnitude cut of $r<22.5$. 
The color classification is defined using a $g-r$ cut \cite{euclidcollaboration2024euclidvflagshipgalaxy}. In table \ref{tab:densitty} we present the number of galaxies and number densities for both catalogs for each galaxy sample and redshift bin. 

\begin{table*}
\begin{tabular}{cc|ccc|ccc|}
\cline{3-8}
 &
   &
  \multicolumn{3}{c|}{\textbf{GR}} &
  \multicolumn{3}{c|}{\textbf{MG}} \\ \hline
\multicolumn{1}{|c|}{\textbf{Sample}} &
  \textbf{\begin{tabular}[c]{@{}c@{}}Observed\\ Redshift\end{tabular}} &
  \multicolumn{1}{c|}{\textbf{Nº galaxies}} &
  \multicolumn{1}{c|}{\begin{tabular}[c]{@{}c@{}}\textbf{Nº density}\\ (h/Mpc)$^3$\end{tabular}} &
  \textbf{\begin{tabular}[c]{@{}c@{}}Angular\\ density ($sr^{-1}$)\end{tabular}} &
  \multicolumn{1}{c|}{\textbf{Nº galaxies}} &
  \multicolumn{1}{c|}{\begin{tabular}[c]{@{}c@{}}\textbf{Nº density}\\ (h/Mpc)$^3$\end{tabular}} &
  \textbf{\begin{tabular}[c]{@{}c@{}}Angular\\ density ($sr^{-1}$)\end{tabular}} \\ \hline
\rowcolor[HTML]{E4F8E4} 
\multicolumn{1}{|c|}{\cellcolor[HTML]{E4F8E4}} &
  \textbf{$\mathbf{0.35 \pm 0.05}$} &
  \multicolumn{1}{c|}{\cellcolor[HTML]{E4F8E4}125,758,470} &
  \multicolumn{1}{c|}{\cellcolor[HTML]{E4F8E4}$4.35\times 10^{-2}$} &
  $1.00\times 10^{7}$ &
  \multicolumn{1}{c|}{\cellcolor[HTML]{E4F8E4}119,326,008} &
  \multicolumn{1}{c|}{\cellcolor[HTML]{E4F8E4}$4.13\times 10^{-2}$} &
  $9.50\times 10^{6}$ \\ \cline{2-8} 
\rowcolor[HTML]{E4F8E4} 
\multicolumn{1}{|c|}{\cellcolor[HTML]{E4F8E4}} &
  \textbf{$\mathbf{0.55 \pm 0.05}$} &
  \multicolumn{1}{c|}{\cellcolor[HTML]{E4F8E4}171,038,806} &
  \multicolumn{1}{c|}{\cellcolor[HTML]{E4F8E4}$3.01\times 10^{-2}$} &
  $1.36\times 10^{7}$ &
  \multicolumn{1}{c|}{\cellcolor[HTML]{E4F8E4}182,214,423} &
  \multicolumn{1}{c|}{\cellcolor[HTML]{E4F8E4}$3.21\times 10^{-2}$} &
  $1.45\times 10^{7}$ \\ \cline{2-8} 
\rowcolor[HTML]{E4F8E4} 
\multicolumn{1}{|c|}{\multirow{-3}{*}{\cellcolor[HTML]{E4F8E4}\textbf{All galaxies}}} &
  \textbf{$\mathbf{0.80 \pm 0.05}$} &
  \multicolumn{1}{c|}{\cellcolor[HTML]{E4F8E4}71,500,296} &
  \multicolumn{1}{c|}{\cellcolor[HTML]{E4F8E4}$7.94\times 10^{-3}$} &
  $5.69\times 10^{6}$ &
  \multicolumn{1}{c|}{\cellcolor[HTML]{E4F8E4}75,675,455} &
  \multicolumn{1}{c|}{\cellcolor[HTML]{E4F8E4}$8.41\times 10^{-3}$} &
  $6.02\times 10^{6}$ \\ \hline
\rowcolor[HTML]{E0E2FF} 
\multicolumn{1}{|c|}{\cellcolor[HTML]{E0E2FF}} &
  \textbf{$\mathbf{0.35 \pm 0.05}$} &
  \multicolumn{1}{c|}{\cellcolor[HTML]{E0E2FF}73,832,083} &
  \multicolumn{1}{c|}{\cellcolor[HTML]{E0E2FF}$2.55\times 10^{-2}$} &
  $5.88\times 10^{6}$ &
  \multicolumn{1}{c|}{\cellcolor[HTML]{E0E2FF}76,231,335} &
  \multicolumn{1}{c|}{\cellcolor[HTML]{E0E2FF}$2.64\times 10^{-2}$} &
  $6.07\times 10^{6}$ \\ \cline{2-8} 
\rowcolor[HTML]{E0E2FF} 
\multicolumn{1}{|c|}{\cellcolor[HTML]{E0E2FF}} &
  \textbf{$\mathbf{0.55 \pm 0.05}$} &
  \multicolumn{1}{c|}{\cellcolor[HTML]{E0E2FF}117,723,717} &
  \multicolumn{1}{c|}{\cellcolor[HTML]{E0E2FF}$2.07\times 10^{-2}$} &
  $9.37\times 10^{6}$ &
  \multicolumn{1}{c|}{\cellcolor[HTML]{E0E2FF}124,580,224} &
  \multicolumn{1}{c|}{\cellcolor[HTML]{E0E2FF}$2.19\times 10^{-2}$} &
  $9.91\times 10^{6}$ \\ \cline{2-8} 
\rowcolor[HTML]{E0E2FF} 
\multicolumn{1}{|c|}{\multirow{-3}{*}{\cellcolor[HTML]{E0E2FF}\textbf{Central galaxies}}} &
  \textbf{$\mathbf{0.80 \pm 0.05}$} &
  \multicolumn{1}{c|}{\cellcolor[HTML]{E0E2FF}55,051,493} &
  \multicolumn{1}{c|}{\cellcolor[HTML]{E0E2FF}$6.12\times 10^{-3}$} &
  $4.38\times 10^{6}$ &
  \multicolumn{1}{c|}{\cellcolor[HTML]{E0E2FF}61,317,982} &
  \multicolumn{1}{c|}{\cellcolor[HTML]{E0E2FF}$6.81\times 10^{-3}$} &
  $4.88\times 10^{6}$ \\ \hline
\rowcolor[HTML]{FEE5E3} 
\multicolumn{1}{|c|}{\cellcolor[HTML]{FEE5E3}} &
  \textbf{$\mathbf{0.35 \pm 0.05}$} &
  \multicolumn{1}{c|}{\cellcolor[HTML]{FEE5E3}42,963,500} &
  \multicolumn{1}{c|}{\cellcolor[HTML]{FEE5E3}$1.49\times 10^{-2}$} &
  $3.42\times 10^{6}$ &
  \multicolumn{1}{c|}{\cellcolor[HTML]{FEE5E3}31,651,738} &
  \multicolumn{1}{c|}{\cellcolor[HTML]{FEE5E3}$1.09\times 10^{-2}$} &
  $2.52\times 10^{6}$ \\ \cline{2-8} 
\rowcolor[HTML]{FEE5E3} 
\multicolumn{1}{|c|}{\cellcolor[HTML]{FEE5E3}} &
  \textbf{$\mathbf{0.55 \pm 0.05}$} &
  \multicolumn{1}{c|}{\cellcolor[HTML]{FEE5E3}47,592,520} &
  \multicolumn{1}{c|}{\cellcolor[HTML]{FEE5E3}$8.38\times 10^{-3}$} &
  $3.79\times 10^{6}$ &
  \multicolumn{1}{c|}{\cellcolor[HTML]{FEE5E3}46,889,007} &
  \multicolumn{1}{c|}{\cellcolor[HTML]{FEE5E3}$8.26\times 10^{-3}$} &
  $3.73\times 10^{6}$ \\ \cline{2-8} 
\rowcolor[HTML]{FEE5E3} 
\multicolumn{1}{|c|}{\multirow{-3}{*}{\cellcolor[HTML]{FEE5E3}\textbf{Red galaxies}}} &
  \textbf{$\mathbf{0.80 \pm 0.05}$} &
  \multicolumn{1}{c|}{\cellcolor[HTML]{FEE5E3}15,828,156} &
  \multicolumn{1}{c|}{\cellcolor[HTML]{FEE5E3}$1.76\times 10^{-3}$} &
  $1.26\times 10^{6}$ &
  \multicolumn{1}{c|}{\cellcolor[HTML]{FEE5E3}16,745,570} &
  \multicolumn{1}{c|}{\cellcolor[HTML]{FEE5E3}$1.86\times 10^{-3}$} &
  $1.33\times 10^{6}$ \\ \hline
\rowcolor[HTML]{F0F6FE} 
\multicolumn{1}{|c|}{\cellcolor[HTML]{F0F6FE}} &
  \textbf{$\mathbf{0.35 \pm 0.05}$} &
  \multicolumn{1}{c|}{\cellcolor[HTML]{F0F6FE}67,289,748} &
  \multicolumn{1}{c|}{\cellcolor[HTML]{F0F6FE}$2.33\times 10^{-2}$} &
  $5.35\times 10^{6}$ &
  \multicolumn{1}{c|}{\cellcolor[HTML]{F0F6FE}70,419,462} &
  \multicolumn{1}{c|}{\cellcolor[HTML]{F0F6FE}$2.44\times 10^{-2}$} &
  $5.60\times 10^{6}$ \\ \cline{2-8} 
\rowcolor[HTML]{F0F6FE} 
\multicolumn{1}{|c|}{\cellcolor[HTML]{F0F6FE}} &
  \textbf{$\mathbf{0.55 \pm 0.05}$} &
  \multicolumn{1}{c|}{\cellcolor[HTML]{F0F6FE}105,853,959} &
  \multicolumn{1}{c|}{\cellcolor[HTML]{F0F6FE}$1.86\times 10^{-2}$} &
  $8.42\times 10^{6}$ &
  \multicolumn{1}{c|}{\cellcolor[HTML]{F0F6FE}115,424,023} &
  \multicolumn{1}{c|}{\cellcolor[HTML]{F0F6FE}$2.03\times 10^{-2}$} &
  $9.19\times 10^{6}$ \\ \cline{2-8} 
\rowcolor[HTML]{F0F6FE} 
\multicolumn{1}{|c|}{\multirow{-3}{*}{\cellcolor[HTML]{F0F6FE}\textbf{Blue galaxies}}} &
  \textbf{$\mathbf{0.80 \pm 0.05}$} &
  \multicolumn{1}{c|}{\cellcolor[HTML]{F0F6FE}52,725,554} &
  \multicolumn{1}{c|}{\cellcolor[HTML]{F0F6FE}$5.86\times 10^{-3}$} &
  $4.20\times 10^{6}$ &
  \multicolumn{1}{c|}{\cellcolor[HTML]{F0F6FE}56,066,342} &
  \multicolumn{1}{c|}{\cellcolor[HTML]{F0F6FE}$6.23\times 10^{-3}$} &
  $4.46\times 10^{6}$ \\ \hline
\rowcolor[HTML]{FFE7B7} 
\multicolumn{1}{|c|}{\cellcolor[HTML]{FFE7B7}} &
  \multicolumn{1}{c|}{\cellcolor[HTML]{FFE7B7}\textbf{$\mathbf{0.35 \pm 0.05}$}} &
  \multicolumn{1}{c|}{\cellcolor[HTML]{FFE7B7}84,652,259} &
  \multicolumn{1}{c|}{\cellcolor[HTML]{FFE7B7}$2.93\times 10^{-2}$} &
  \multicolumn{1}{c|}{\cellcolor[HTML]{FFE7B7}$6.74\times 10^{6}$} &
  \multicolumn{1}{c|}{\cellcolor[HTML]{FFE7B7}87,599,150} &
  \multicolumn{1}{c|}{\cellcolor[HTML]{FFE7B7}$3.03\times 10^{-2}$} &
  \multicolumn{1}{c|}{\cellcolor[HTML]{FFE7B7}$6.97\times 10^{6}$} \\ \cline{2-8} 
\rowcolor[HTML]{FFE7B7} 
\multicolumn{1}{|c|}{\cellcolor[HTML]{FFE7B7}} &
  \multicolumn{1}{c|}{\cellcolor[HTML]{FFE7B7}\textbf{$\mathbf{0.55 \pm 0.05}$}} &
  \multicolumn{1}{c|}{\cellcolor[HTML]{FFE7B7}51,431,683} &
  \multicolumn{1}{c|}{\cellcolor[HTML]{FFE7B7}$9.06\times 10^{-3}$} &
  \multicolumn{1}{c|}{\cellcolor[HTML]{FFE7B7}$4.09\times 10^{6}$} &
  \multicolumn{1}{c|}{\cellcolor[HTML]{FFE7B7}53,406,408} &
  \multicolumn{1}{c|}{\cellcolor[HTML]{FFE7B7}$9.41\times 10^{-3}$} &
  \multicolumn{1}{c|}{\cellcolor[HTML]{FFE7B7}$4.25\times 10^{6}$} \\ \cline{2-8} 
\rowcolor[HTML]{FFE7B7} 
\multicolumn{1}{|c|}{\multirow{-3}{*}{\cellcolor[HTML]{FFE7B7}\textbf{Bright galaxies}}} &
  \multicolumn{1}{c|}{\cellcolor[HTML]{FFE7B7}\textbf{$\mathbf{0.80 \pm 0.05}$}} &
  \multicolumn{1}{c|}{\cellcolor[HTML]{FFE7B7}3,302,610} &
  \multicolumn{1}{c|}{\cellcolor[HTML]{FFE7B7}$3.67\times 10^{-4}$} &
  \multicolumn{1}{c|}{\cellcolor[HTML]{FFE7B7}$2.63\times 10^{5}$} &
  \multicolumn{1}{c|}{\cellcolor[HTML]{FFE7B7}3,581,404} &
  \multicolumn{1}{c|}{\cellcolor[HTML]{FFE7B7}$3.98\times 10^{-4}$} &
  \multicolumn{1}{c|}{\cellcolor[HTML]{FFE7B7}$2.85\times 10^{5}$} \\ \hline
\rowcolor[HTML]{EFEFEF} 
\multicolumn{1}{|c|}{\cellcolor[HTML]{EFEFEF}} &
  \multicolumn{1}{c|}{\cellcolor[HTML]{EFEFEF}\textbf{$\mathbf{0.35 \pm 0.05}$}} &
  \multicolumn{1}{c|}{\cellcolor[HTML]{EFEFEF}19,848,022} &
  \multicolumn{1}{c|}{\cellcolor[HTML]{EFEFEF}$6.86\times 10^{-3}$} &
  \multicolumn{1}{c|}{\cellcolor[HTML]{EFEFEF}$1.58\times 10^{6}$} &
  \multicolumn{1}{c|}{\cellcolor[HTML]{EFEFEF}10,869,529} &
  \multicolumn{1}{c|}{\cellcolor[HTML]{EFEFEF}$3.76\times 10^{-3}$} &
  \multicolumn{1}{c|}{\cellcolor[HTML]{EFEFEF}$8.65\times 10^{5}$} \\ \cline{2-8} 
\rowcolor[HTML]{EFEFEF} 
\multicolumn{1}{|c|}{\cellcolor[HTML]{EFEFEF}} &
  \multicolumn{1}{c|}{\cellcolor[HTML]{EFEFEF}\textbf{$\mathbf{0.55 \pm 0.05}$}} &
  \multicolumn{1}{c|}{\cellcolor[HTML]{EFEFEF}84,989,583} &
  \multicolumn{1}{c|}{\cellcolor[HTML]{EFEFEF}$1.50\times 10^{-2}$} &
  \multicolumn{1}{c|}{\cellcolor[HTML]{EFEFEF}$6.76\times 10^{6}$} &
  \multicolumn{1}{c|}{\cellcolor[HTML]{EFEFEF}93,461,750} &
  \multicolumn{1}{c|}{\cellcolor[HTML]{EFEFEF}$1.65\times 10^{-2}$} &
  \multicolumn{1}{c|}{\cellcolor[HTML]{EFEFEF}$7.44\times 10^{6}$} \\ \cline{2-8} 
\rowcolor[HTML]{EFEFEF} 
\multicolumn{1}{|c|}{\multirow{-3}{*}{\cellcolor[HTML]{EFEFEF}\textbf{Faint galaxies}}} &
  \multicolumn{1}{c|}{\cellcolor[HTML]{EFEFEF}\textbf{$\mathbf{0.80 \pm 0.05}$}} &
  \multicolumn{1}{c|}{\cellcolor[HTML]{EFEFEF}58,350,416} &
  \multicolumn{1}{c|}{\cellcolor[HTML]{EFEFEF}$6.48\times 10^{-3}$} &
  \multicolumn{1}{c|}{\cellcolor[HTML]{EFEFEF}$4.64\times 10^{6}$} &
  \multicolumn{1}{c|}{\cellcolor[HTML]{EFEFEF}61,392,659} &
  \multicolumn{1}{c|}{\cellcolor[HTML]{EFEFEF}$6.82\times 10^{-3}$} &
  \multicolumn{1}{c|}{\cellcolor[HTML]{EFEFEF}$4.89\times 10^{6}$} \\ \hline
\rowcolor[HTML]{FEFEDC} 
\multicolumn{1}{|c|}{\cellcolor[HTML]{FEFEDC}\textbf{Source galaxies}} &
  \textbf{$\mathbf{1.0 \pm 0.1}$} &
  \multicolumn{1}{c|}{\cellcolor[HTML]{FEFEDC}58,903,889} &
  \multicolumn{1}{c|}{\cellcolor[HTML]{FEFEDC}$3.28\times 10^{-3}$} &
  $4.69\times 10^{6}$ &
  \multicolumn{1}{c|}{\cellcolor[HTML]{FEFEDC}64,311,850} &
  \multicolumn{1}{c|}{\cellcolor[HTML]{FEFEDC}$3.58\times 10^{-3}$} &
  $5.12\times 10^{6}$ \\ \hline
\end{tabular}
\caption{Table with all the galaxy samples: count number, number density and angular density; for each sample and gravity simulation.}
\label{tab:densitty}
\end{table*}

\section{Methodology}
\label{sec:methodology}

In this section we specify how we calculated the $E_G$ estimator for both catalogs (F5 and GR). The expression in Eq.~\eqref{eq:EG_original} cannot be directly calculated since the quantities involved correspond to fields. A good alternative is to use the 3D power spectra to estimate $E_G$ with observables \cite{Pullen2014ProbingLensing, Wenzl2024ConstrainingBOSS}:

\begin{equation}
    \hat{E}_G ( k, z ) = \frac{c^2 \hat{P}_{\nabla^2 (\Psi - \Phi) g} ( k, z )}{3 H_0^2 ( 1 + z ) \hat{P}_{\nu g} ( k, z )}, 
\end{equation}

where $P_{\nabla^{2} ( \Psi-\Phi) g}$ is the galaxy  and gravitational potential perturbations cross-power spectrum, $\nabla^{2} ( \Psi-\Phi)$, and $P_{\nu g}$ is the galaxy-peculiar velocity cross-power spectrum, while the hats denote estimates based on observable quantities. One advantage of this the estimator is that it does not depend on the (linear) galaxy bias since \( \hat{P}_{\nu g} \propto b_g \) and \( \hat{P}_{\nabla^2(\Psi - \Phi)g} \propto b_g \) as well. From Eq.~\eqref{eq: EG numerator} we get that:

\begin{equation}
    \nabla^2 (\Psi - \Phi) = \frac{3} {2} H_{0}^{2} \Omega_{m, 0} ( 1+z ) \mu( k, a ) [ \gamma( k, a )+1 ] \delta 
\label{eq: potential pertubation}
\end{equation}

which is valid for both GR and $f(R)$ models, so we can change the galaxy-potential cross-spectrum by:  $\langle \nabla^{2} ( \Psi-\Phi) g \rangle \propto  \langle \delta g \rangle \propto \langle\kappa g \rangle$, where $\kappa$ is the convergence, which is an actual observable, unlike $\delta$. Later we will perform the connection between these two properties when building the final expression for the $E_G$. Projecting 3D power spectra into (2D) angular quantities, we can estimate $E_{G}$ as:

\begin{equation}
\hat{E}_G^\ell ( z ) = \frac{2 c^2 \hat{C}_\ell^{\nabla^{2} ( \Psi-\Phi) g}}{3 H_0^2 \beta \hat{C}_\ell^{gg}},
\label{eq:EGCl}
\end{equation}

where \( \hat{C}_\ell^{gg} \) is the angular auto-power spectrum of the galaxy sample, while \( \hat{C}_\ell^{\nabla^{2} ( \Psi-\Phi) g} \) is the gravitational potential perturbations-galaxy cross-spectrum. The correlation in the denominator comes from the approximation \(\hat{P}_{\nu g}(k, z) = \beta \hat{P}_{gg}(k, z) \), where \( \beta = f / b_1 \) is derived from a RSD analysis at the same effective redshift as the auto-correlation. To derive the above expression we also assumed the linear continuity equation ( \( \nu = - \beta \delta_g \)). Under the Limber approximation, the angular power spectra can be expressed as:

\begin{equation}
\hat{C}_{\ell}^{\nabla^{2} ( \Psi-\Phi) g}=\frac{1} {2} \int\mathrm{d} z \frac{H ( z )} {c ( 1+z )} \frac{W_{g}^{2} ( z )} {\chi^{2} ( z )} \hat{P}_{\nabla^{2} ( \Psi-\Phi) g} ( k, z ), 
\label{eq: Cltg}
\end{equation}

\begin{equation}
    C_{\ell}^{g g}=\int\mathrm{d} z \, \frac{H ( z )} {c} \frac{W_{g}^{2} ( z )} {\chi^{2} ( z )} P_{g g} \biggl( k=\frac{\ell+1 / 2} {\chi( z )}, z \biggr),
\label{eq: Clgg}
\end{equation}

\begin{equation}
    C_{\ell}^{\kappa g}=\int\mathrm{d} z \, \frac{W_{\kappa} ( z ) W_{g} ( z )} {\chi^{2} ( z )} P_{\delta g} \biggl( k=\frac{\ell+1 / 2} {\chi( z )}, z \biggr),
\label{eq: Clkg}
\end{equation}

the last term is the convergence-galaxy angular power spectrum. The terms \( W_g(z) \) and \( W_\kappa(z) \) are the window function for the galaxy sample and lensing, respectively, and \( \chi(z) \) is the comoving distance at a given redshift. These kernels are given by:

\begin{equation}
    W_{g}(z) = \frac{\mathrm{d}N}{\mathrm{d}z},
\end{equation}

\begin{equation}
    W_{\kappa}(z) = \frac{3 H_0^2 \Omega_{\mathrm{m}, 0}}{2 c^2} \hat{W}_{\kappa}(z),
\end{equation}

\begin{equation}
    \hat{W}_{\kappa}(z) \equiv (1 + z) \chi(z) \int d\chi' \frac{n_s(\chi')}{\chi'} \left( 1 - \frac{\chi(z)}{\chi'} \right),
\label{eq:lensing efficiency}
\end{equation}

where $n_s$ is the redshift distribution of sources.

From Eqs. \eqref{eq: Clkg} and \eqref{eq: Clgg} is easy to see that the dependence of $E_G$ estimator on the linear galaxy bias still cancels out when expressed in terms of angular power spectra (\emph{i.e.}, in terms of projected quantities).  in particular, we have that \( \hat{P}_{gg} \propto b_g^2 \) and \( \hat{P}_{\delta g} \propto b_g \), while the $\beta$ term adds an extra galaxy bias on the numerator. This $\beta$ parameter needs to be estimated in such a way that is consistent with the $C_\ell$'s. The effective redshifts of auto- and cross-correlations for the same galaxy sample generally differ, which may introduce a bias in the estimation of \( E_{G} \). These effective redshifts are given by the following expressions \cite{Chen2022CosmologicalTheory, Wenzl2024ConstrainingBOSS}:

\begin{equation}
z_{\mathrm{eff}}^{\mathrm{cross}} = \frac{\int \mathrm{d} z\, \chi^{-2} \hat{W}_{\kappa}(z) W_{g}(z) z}{\int \mathrm{d} z\, \chi^{-2} \hat{W}_{\kappa}(z) W_{g}(z)},
\label{eq: z_cross_eff}
\end{equation}
\begin{equation}
z_{\mathrm{eff}}^{\mathrm{auto}} = \frac{\int \mathrm{d} z\, \chi^{-2}(z) H(z) c^{-1} W_{g}^{2}(z) z}{\int \mathrm{d} z\, \chi^{-2}(z) H(z) c^{-1} W_{g}^{2}(z)}.
\label{eq: z_eff}
\end{equation}

We follow the same methodology than in Ref.~\cite{Wenzl2024ConstrainingBOSS} and we equate the effective redshift by weighting the galaxy sample of the cross-correlation as follows:

\begin{equation}
    W_g^* \equiv \frac{\mathrm{d}N^*}{\mathrm{d}z} =  \frac{\mathrm{d}N}{\mathrm{d}z} w_{\times}(z)
\end{equation}
with:
\begin{equation}
w_{\times}(z) = W_g \frac{1}{\hat{W}_{\kappa}(z) I}, \tag{25}
\label{eq:wx}
\end{equation}
where:
\begin{equation}
I = \int \mathrm{d} z \frac{W_{g}^{2}(z)}{\hat{W}_{\kappa}(z)},
\end{equation}

where $I$ is introduced to normalize the weighted galaxy distribution $\int W_g^* dz = 1$. Following the steps in Ref.~\cite{Wenzl2024ConstrainingBOSS}, we use this re-weighting and express Eq.~\eqref{eq:EGCl} as,

\begin{equation}
\begin{aligned}
    \hat{C}_{\ell}^{\nabla^2(\Psi - \Phi) g} &\approx \frac{H(z_{\mathrm{eff}}) I}{c} \left[ \int \mathrm{d}z \frac{\hat{W}_{\kappa}(z) W_{g}^{*}(z)}{\chi^2(z)} \hat{P}_{\kappa g}(k, z) \right] \\
    & = \frac{H(z_{\mathrm{eff}}) I}{c} C_{\ell}^{\kappa g *},
\end{aligned}
\label{eq: kg*}
\end{equation}

where we have used that \( W_g^*(z) = W_g(z) w_{\times}(z) \) and \( \hat{P}_{\nabla^2(\Psi - \Phi) g} \approx \hat{P}_{\kappa g} \), which holds for both GR and \( f(R) \) (Eq.~\eqref{eq: potential pertubation}). Here, the term \( C_{\ell}^{\kappa g *} \), equivalent to the expression in brackets in Eq.~\eqref{eq: kg*}, represents the cross-correlation measurement using the reweighted galaxy sample. The term \( H(z) \) is moved outside the integral as it varies slowly over the redshift range of the sample.

The final estimator for \( E_G \) is given by Ref.~\cite{Wenzl2024ConstrainingBOSS}:

\begin{equation}
\hat{E}_{G}^{\ell}(z_{\mathrm{eff}}) \approx \Gamma(z_{\mathrm{eff}}) \frac{C_{\ell}^{\kappa g *}}{\beta C_{\ell}^{g g}},
\label{eq:EG final}
\end{equation}

where:

\begin{equation}
\Gamma(z_{\mathrm{eff}}) \equiv \frac{2cH(z_{\mathrm{eff}})}{3H_0^2} \int \mathrm{d} z \frac{W_{g}^2(z)}{\hat{W}_{\kappa}(z)},
 \label{eq:Gamma factor}
\end{equation}

and \( z_{\mathrm{eff}} \) is the effective redshift of the observables as defined by Eq.~\eqref{eq: z_eff}.

\begin{figure}
\centering
\includegraphics[width=\columnwidth]{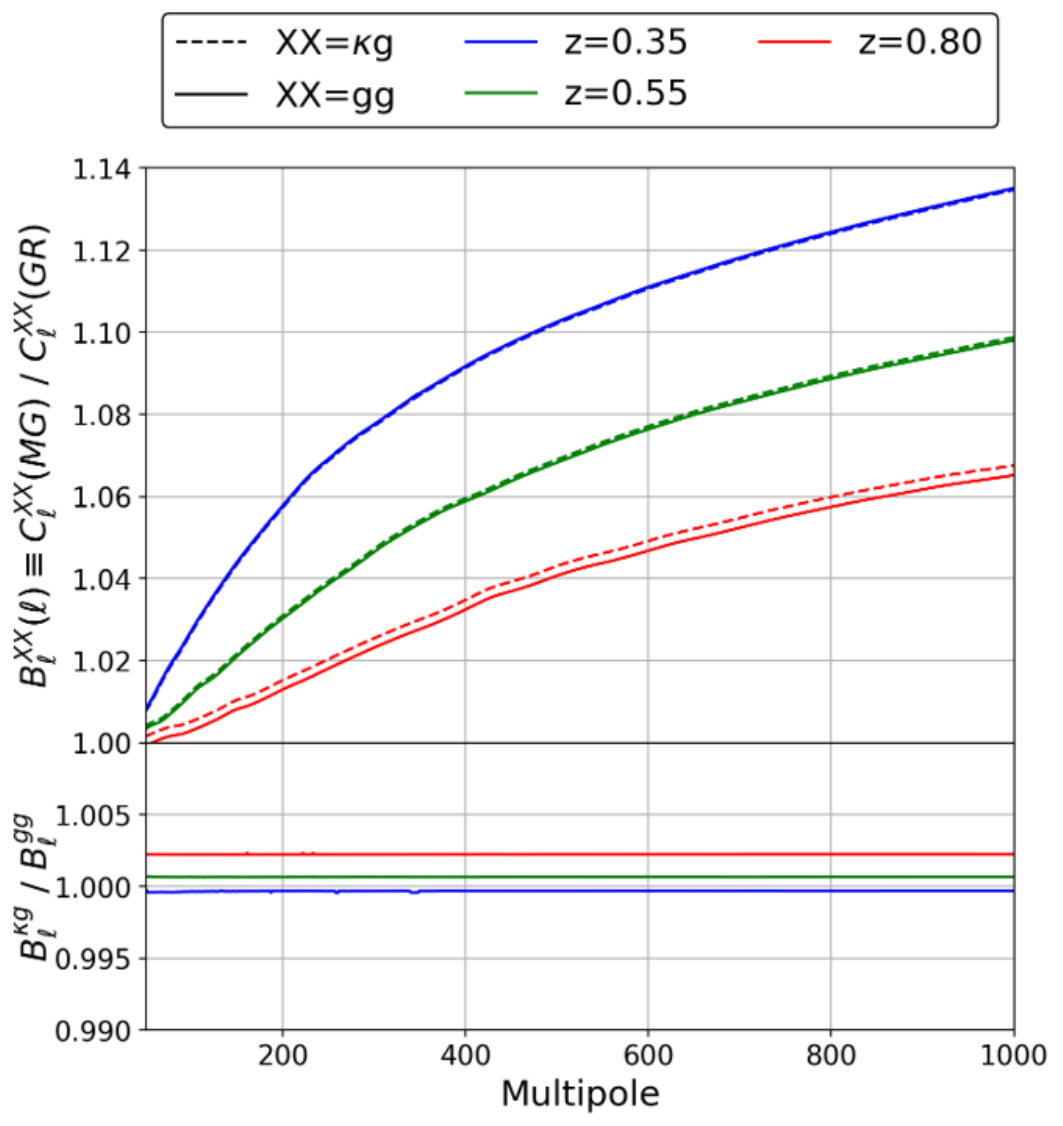}
\caption{F5 boost ($B_\ell$ ratio, analogous to Eq.~\eqref{eq:mg boost} impact on the $C_\ell$s for auto-galaxy correlation (continuous line) and convergence-galaxy cross-correlations (dashed line) for the 3 redshifts bins. The lower panel shows the ratio between the auto and cross- spectra for each z. One can appreciate how the boost ratio is roughly unity at all redshifts. 
The theoretical $C_\ell$s are obtained with \textit{pyCCL}, whereas the F5 boost, applied to the 3D matter power spectrum, is calculated with \textsc{e-mantis}.}
\label{Fig:Cl_boost}
\end{figure}

In Fig.~\ref{Fig:Cl_boost} we can appreciate how the $f(R)$ boost on $C_\ell^{gg}$ is the same than in $C_\ell^{\kappa g}$ for any redshift. This confirms the theoretical prediction from Eq.~\eqref{eq: EG_fR teo} since the MG boost cancels out according to Eq.~\eqref{eq:EG final} 
Therefore, re-weighting the angular power spectra does not change the fact that the $E_G$ estimator is only sensitive to the underlying gravity model through the growth rate (\emph{i.e.}, it is not sensitive to the ratio between the galaxy-lensing cross-correlation over the galaxy auto-correlation).

This definition of the $E_G$ estimator is arguably the most accurate yet, as claimed by Ref.~\cite{Wenzl2024ConstrainingBOSS}, when modeling observable quantities. Previous to this work, most analysis were using the expressions found by Ref.~\cite{Pullen2014ProbingLensing} and Ref.~\cite{Pullen2016ConstrainingVelocities}, where they average the actual redshift of each probe involved in the gravity estimator, instead of using the effective redshift. Ref.~\cite{Pullen2016ConstrainingVelocities} proposes a correction term but it can only be obtained numerically using N-body simulations. Without this correction the estimated value is biased at the 5$\%$ level from the theoretically predicted value for $E_G$.

Ref.~\cite{Wenzl2024ConstrainingBOSS} already compared the accuracy of Pullen and their own estimator and showed that their estimator has an accuracy better than $3\%$ on all scales. We also calculate the relative difference between the analytic value calculated with Eq.~\eqref{eq: EG general} and the value obtained with theoretical predictions from \textit{pyCCL} using Eq.~\eqref{eq:EG final}. To calculate the $C_\ell$s and $\Gamma$, Eq.~\eqref{eq:Gamma factor}, we use the fiducial values of the simulations, the $n(z)$ from our full sample for GR and F5, respectively, and, since the linear galaxy bias is irrelevant for $E_G$, we leave the default value of 1. For the F5 calculation we use \textsc{e-mantis} \cite{Saez-Casares2023TheCosmology} to estimate the $C_\ell$s boost, although as shown in Fig.~\ref{Fig:Cl_boost} this is also irrelevant when doing the ratio, and \textit{MGrowth} to estimate the scale dependent growth rate. The Pullen estimator is calculated following Eq.~(16) from Ref.~\cite{Pullen2016ConstrainingVelocities} where they use $C_\ell^{\kappa g}$ instead of $C_\ell^{\kappa g*}$ and a different $\Gamma$ which does not take into account the effective redshift. In Fig.~\ref{Fig:acuracy} we show the accuracy results and we find that, as expected, the Wenzl estimator is more accurate than the one used by Pullen. Although it depends on the redshift, in the two extreme cases the relative difference for the Pullen estimator is doubled over the relative difference of the Wenzl estimator. One thing to note is that the accuracy gets worse as we go to higher redshifts, although for the Wenzl estimator the difference does not exceed 2$\%$.

In the following sections we detail the calculation of all the ingredients needed to estimate $E_G$ from the expression \eqref{eq:EG final}.

\subsection{Linear galaxy bias}
\label{sec:galaxy bias}
The linear galaxy bias $b_g \equiv b_1$, can be obtained from the RSD analysis using clustering multipoles.  In practice, since we want to avoid as much as possible the degeneracy among RSD-related parameters we shall estimate the linear galaxy bias independently, that is then used as a prior to better constrain the other parameters from the RSD analysis.

In order to estimate the galaxy bias, we calculate the galaxy angular auto-correlation and compare with theory predictions from the public code \textit{pyCCL}\footnote{\url{https://github.com/LSSTDESC/CCL/blob/master/readthedocs/index.rst}} \cite{Chisari_2019}.

The measurement of the galaxy angular auto-correlation is performed with the routine \textit{anafast} from the \textsc{healpy}\footnote{\url{https://healpy.readthedocs.io/}} package. For each sample (\emph{i.e.}, for each case, redshift, galaxy sample) we generate a HEALPix map with nside=1024 where each pixel value is given by,

\begin{equation}
    \delta = \frac{n_{pix} - \bar{n}}{\bar{n}}
 \label{eq:overdensity HEALPix}
\end{equation}

where $n_{pix}$ is the number of galaxies that fall inside that pixel, while $\bar{n}$ is the mean galaxy value of all the pixels. Although we are working with full sky, which would indicate that we do not need any mask, we still generate 100 masks to define Jackknife regions. Each Jackknife mask extracts a different 1/100 same size region of the sky for which we use a k-means algorithm called \textit{kmeans-radec}\footnote{\url{https://github.com/esheldon/kmeans_radec}} to determine the regions. The $C_\ell$'s are calculated with the software \textsc{Polspice}\footnote{\url{https://www2.iap.fr/users/hivon/software/PolSpice/}} for each Jackknife region.  We then use linear multipole bins of $\Delta \ell=20$ between $\ell_{min}=30$ and $\ell_{max}=1024$ to get a smoother estimate of the $C_\ell$'s on linear scales, and subtract the shot noise,

\begin{equation}
n_{shot} = \frac{4 \pi}{n_{gal}} \cdot \frac{1}{ \text{pixelwindow($\ell$)}^2},
\end{equation}

which is the standard definition of area/$n_{gal}$, with $n_{gal}$ being the total number of galaxies (see Table \ref{tab:densitty}), and the area is given by the full sky (4$\pi$ radians) with a correction by the pixel window to be consistent with the correction that is applied to the (signal plus noise) $C_\ell$'s.

The binned values are then averaged over each Jackknife region to obtain the final $C_\ell^{gg}$'s, while the Jackknife covariance matrix is given by:

\begin{equation}
\sigma^2_{ij} (C_\ell^{gg}) = \frac{N_{JK} -1}{N_{JK}} \sum\limits^{N_{JK}}_{i,j=1} [C_\ell(\ell)_i - \Bar{C_\ell}(\ell)][C_\ell(\ell)_j - \Bar{C_\ell}(\ell)]
\end{equation}

where the top bar indicates the mean value over all the Jackknife regions while the suffix $i, j$ indicates $i,j$-region calculated $C_\ell$.

Using the code \textit{pyCCL} we obtain the prediction using the fiducial cosmology of the mocks since we wish to get an independent estimate the galaxy bias prior to the RSD analysis. For the theory prediction of the matter power spectrum, we use the CAMB code to compute the transfer function and the non-linear power spectrum given by Halofit \cite{Takahashi2012RevisingSpectrum}. The $n(z)$ for the prediction are obtained from the same sample and it is binned in 15000 equidistant z-bins in order to have a precise definition of the n(z). We integrate the $C_\ell$'s exactly, \emph{i.e.}, without the Limber approximation in order to have a more accurate estimation on linear scales.

\begin{figure*}
  \centering
\includegraphics[width=\columnwidth]{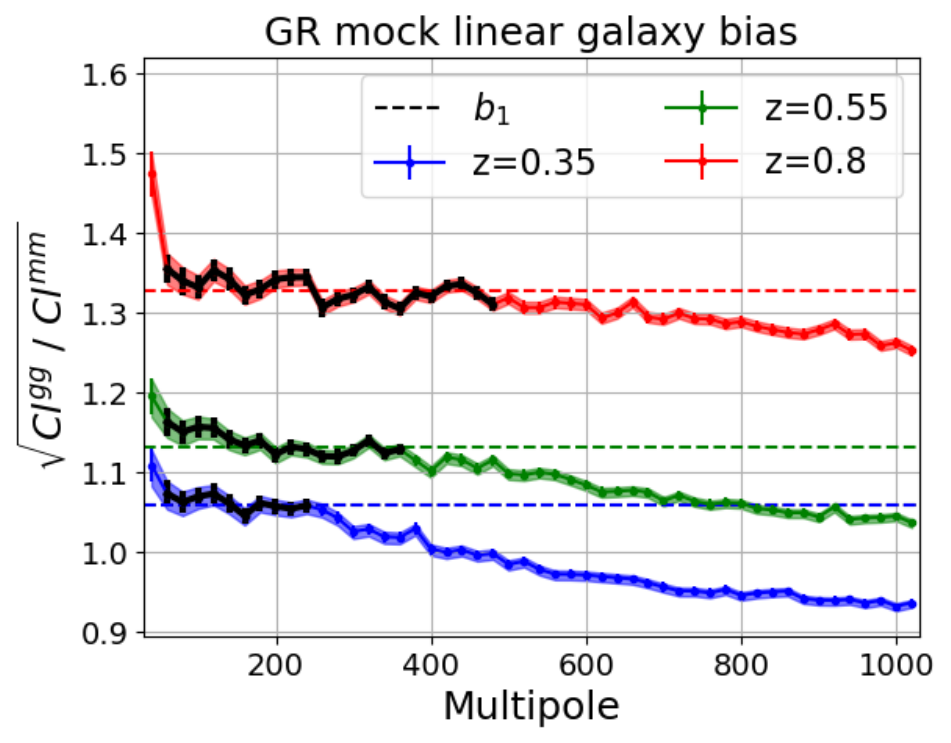}
\includegraphics[width=\columnwidth]{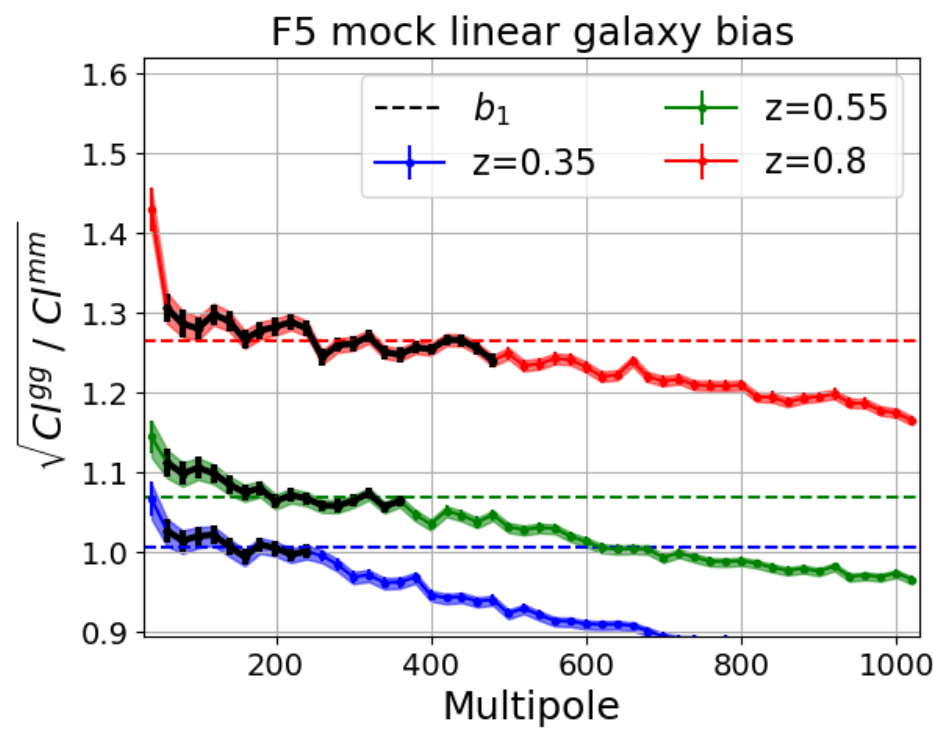}
  \caption{Linear galaxy bias estimated as root squared ratio of of the galaxy-galaxy and matter-matter $C_\ell$s, for the three redshifts bins using the full sample. The black lines show the linear regions we have considered to estimate the galaxy bias. The dashed lines show the value of the estimated bias for each redshift using $\chi^2$ minimization. The left plot shows the results for the GR mock while the right plot for the F5 mock. The errorbars are given by 100 Jackknife regions. The shaded areas correspond to the theoretical Gaussian error.}
  \label{Fig:b1}
\end{figure*}

We estimate the linear galaxy bias as,

\begin{equation}
    b_1 = \sqrt{\frac{C_\ell^{gg}}{C_\ell^{\delta \delta}}}
\label{eq: b1}
\end{equation}

where $C_\ell^{\delta \delta}$ is the matter angular power spectrum. Given that we assign no error on the estimate of $C_\ell^{\delta \delta}$, we can obtain the covariance matrix of $b_1$ directly from that of $C_\ell^{gg}$. With the covariance matrix we can estimate the bias as the  value, $E$, that minimizes the following $\chi^2$, assuming that the Jackknife errors are Gaussian distributed:

\begin{equation}
    \chi^2 = (\Bar{b_1} - E)^T (\sigma^2_{ij} (b_1) [\ell_{min}:\ell_{max}])^{-1} (\Bar{b_1} - E)
    \label{eq:chi2}
\end{equation}

where $\Bar{b_1}$ is the mean value of $b_1$ given by Eq.~\eqref{eq: b1} in the region $[\ell_{min}:\ell_{max}]$ considered. We take $\ell_{min}=50$ to avoid the large sample variance from the lowest multipoles, while $\ell_{max}$ sets the maximum scale where the galaxy bias remains linear, which depends on redshift. For $z$=0.8 we estimate the ratio between $C_\ell^{gg}$ and $C_\ell^{\delta \delta}$ deviates significantly from the linear model at $\ell\approx500$. Using the small-angle approximation, $k\approx \ell/\chi_d(z)$, where $\chi_d$, we see that this multiple corresponds to a wavenumber of, $k_{lin} < 0.119 h^{-1}$Mpc. Similarly, at $z$=0.35 and $z$=0.55 we obtain a $l_{max}$ of $250$ and $372$, respectively. The values of $b_1$ obtained for each sample are shown in Table \ref{tab:b1}, while the individual ratios for the Full Sample are shown in Fig.~\ref{Fig:b1}. The errors are given by the 1-$\sigma$ error, \emph{i.e.} $\mathcal{O}(b_1) = |b_1( \chi^2_{\text{min}}) - b_1 (\chi^2_{\text{min}} \pm 1)|$.

For the F5 mock data, we use the emulator \textsc{e-mantis} to apply the $f(R)$ boost, with $|\Bar{f}_{R0}| = 10^{-5}$, to the matter power spectrum at each $n(z)$ bin, from which the $C_\ell$s are calculated. We also perform an extra case where we assume that the F5 mock follows GR and we estimate $b_1$ without using the boost for this mock. As we can see in Table \ref{tab:b1} this does not impact the estimated linear galaxy bias. This is to be expected since the boost is only noticeable at smaller non-linear scales (see Fig.~\ref{fig:MG_boost} and Fig.~\ref{Fig:Cl_boost}). We also notice that the galaxy bias in GR is usually significantly larger than F5. The reason behind this is that the matter clustering is stronger in F5 so the estimated galaxy bias is lower in order to match the same calibration of the observed galaxy clustering amplitude at low redshift implemented in both mocks.

\begin{table*}
\centering
\begin{tabular}{|l|c|c|c|}
\hline
\textbf{Case} & \textbf{z=0.35} & \textbf{z=0.55} & \textbf{z=0.8} \\
\hline
\textbf{All F5} & 1.047 $\pm$ 0.004 & 1.104 $\pm$ 0.003 & 1.305 $\pm$ 0.002 \\
\textbf{All GR} & 1.077 $\pm$ 0.004 & 1.147 $\pm$ 0.003 & 1.353 $\pm$ 0.002 \\
\hline
\textbf{Central F5} & 0.778 $\pm$ 0.003 & 0.856 $\pm$ 0.002 & 1.185 $\pm$ 0.002 \\
\textbf{Central GR} & 0.815 $\pm$ 0.004 & 0.895 $\pm$ 0.002 & 1.213 $\pm$ 0.002 \\
\hline
\textbf{Red F5} & 1.246 $\pm$ 0.006 & 1.342 $\pm$ 0.004 & 1.393 $\pm$ 0.003 \\
\textbf{Red GR} & 1.282 $\pm$ 0.006 & 1.441 $\pm$ 0.004 & 1.639 $\pm$ 0.003 \\
\hline
\textbf{Blue F5} & 0.830 $\pm$ 0.004 & 0.888 $\pm$ 0.002 & 1.140 $\pm$ 0.002 \\
\textbf{Blue GR} & 0.825 $\pm$ 0.004 & 0.900 $\pm$ 0.002 & 1.158 $\pm$ 0.002 \\
\hline
\textbf{Bright F5} & 1.057 $\pm$ 0.004 & 1.196 $\pm$ 0.003 & 1.705 $\pm$ 0.004 \\
\textbf{Bright GR} & 1.090 $\pm$ 0.004 & 1.233 $\pm$ 0.003 & 1.780 $\pm$ 0.004 \\
\hline
\textbf{Faint F5} & 1.047 $\pm$ 0.004 & 1.041 $\pm$ 0.003 & 1.232 $\pm$ 0.002 \\
\textbf{Faint GR} & 1.067 $\pm$ 0.004 & 1.086 $\pm$ 0.003 & 1.278 $\pm$ 0.002 \\
\hline
\textbf{All (F5 assumed as GR)} & 1.047 $\pm$ 0.004 & 1.104 $\pm$ 0.003 & 1.305 $\pm$ 0.002 \\
\hline
\end{tabular}
\caption{Linear galaxy bias values and standard deviation for all the different cases studied.}
\label{tab:b1}
\end{table*}

\subsection{Multipoles of the correlation function}
\label{sec:cfm}

Since the growth rate is completely degenerated with $\sigma_8$ at the the angular power spectrum level \cite{Wenzl2024ConstrainingBOSS}, in this work we concentrate on the multipoles of the correlation function, which allow to break this degeneracy. In particular, combining the monopole ($\ell=0$) and the quadrupole ($\ell=2$) moments once can  break the degeneracy with the linear galaxy bias. Although the hexadecapole ($\ell=4$) further breaks this degeneracy with additional information, we have decided to leave this contribution out of the analysis as its measurement in our mocks turns out to be very noisy in practice.

We calculate the multipoles using the recent code Fast-Correlation-Function-Calculator \cite{Zhao_2023} due to its superior performance with large data sets. Even though the code allows for the direct calculation of the even multipoles, we decide to calculate the 2D anisotropic correlation function $\xi(r, \mu)$  (where $\mu$ represents the cosine of the angle between the line of sight direction and the direction between the galaxy pairs) since this will be useful to estimate the Jackknife errors.

We use 20 radial projected bins with projected separations from $s_{min}=0.5 \ Mpc/h$ to $s_{max}=200 \ Mpc/h$. We also use 200 bins on the $\mu$ parameter in the positive range [0, 1], since the code only calculates even multipoles which are symmetric, which means that Eq.~\eqref{eq: multipoles final comet} should be multiplied by 2.

Similarly to the linear galaxy bias, the errors are calculated using $N_{jk}=100$ Jackknife resampling of the data. We use once again the code \textit{kmeans-radec} to generate 100 equidistant regions, \emph{i.e.} with the same area, on the sky, and we assign each galaxy to a region. In order to speed up the calculation, instead of performing the calculation of each individual Jackknife region, we take a different approach: we calculate the $\xi(r, \mu)$ correlation of the full dataset with each individual region, \emph{i.e.} each region that is subtracted from each Jackknife region, we then subtract the pair count of this cross-correlation from the full dataset auto-correlation. In our correlation estimator (see below), we must also apply the same process to the random dataset, which we choose to be twice as dense as that of the mock data for an accurate account of the mask. We verified that increasing the number of random points does not affect the final results. This is expected, as we are not using a complex mask and there are no systematic effects. Since we are using the full Landy-Szalay estimator \cite{Landy1993BiasFunctions}, we also need to subtract the cases for the data-random (DR) terms, where D corresponds to the data and R to the randoms. Our estimator for each Jackknife region is thus given by:

\begin{equation} \label{LS jack}
    \xi(r, \mu)_{N_{i}} = \frac{ (DD - DD_i) - (2DR - DR_i - D_iR) + (RR - RR_i) }{(RR - RR_i)}
\end{equation}

where the suffix on $D_i$ (and $R_i$) correspond to the data (random) of the region $i$ not included on Jackknife region $N_{i}$. Each set of pairs in expression Eq.~\eqref{LS jack} are normalized by dividing individually the total number of pairs by the total number of data points, \emph{i.e.}  galaxies, of each data pair. This method has allowed us to speed-up the error computation as we avoid repeating common pairs in each Jackknife region.

Since calculations are done in configuration space, the definition of the multipoles is given by:

\begin{equation}
\xi_{\ell} (r) = \frac{(2\ell + 1)}{2} \int^1_{-1} \xi(r, \mu) P_\ell(\mu) d\mu
\label{eq:cfmultipoles}
\end{equation}

where $L_\ell$ is the Legendre polynomial of degree $\ell$. The Jackknife covariance matrix is obtained in a similar fashion than in the previous section.

\begin{figure*}
    \centering
    \includegraphics[width=0.8\linewidth]{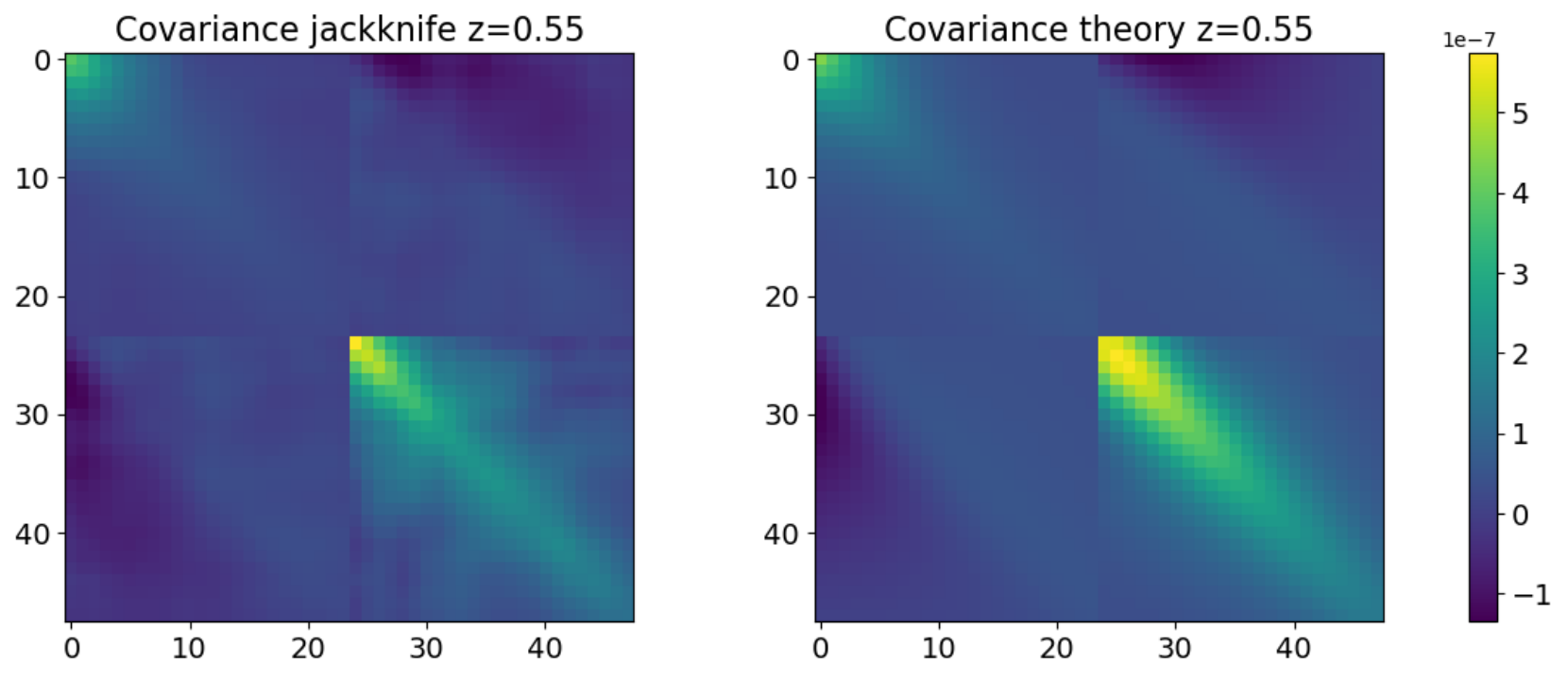}
    \caption{Monopole-quadrupole joint covariance matrix of the multipoles of the correlation function for the GR mock at z=0.55 for the "all galaxies" case. The values are in configuration space for the 24 linear projected separation bins chosen for the full range of scales used, $[20, 140] \rm{Mpc/h}$. Left figure shows the Jackknife covariance obtained from the data while the right plot shows the theoretical Gaussian covariance obtained with BeXiCov (see text for details).}
    \label{fig:covariance matrix}
\end{figure*}

In Fig.~\ref{fig:covariance matrix} we show the Jackknife covariance matrix for the GR mock at z=0.55 for the joint monopole and quadrupole analysis. We will also fit the results using a Gaussian theoretical covariance matrix, also shown in the same figure, obtained with the code BeXiCov\footnote{\url{https://gitlab.com/esarpa1/BeXiCov/-/tree/main}} with the same simulation volume and particle density. We can check that both matrices are similar, specially when considering the global amplitude, which is a good indicator that the Jackknife resampling was correctly generated. Both matrices are very close to be singular which may complicate the inversion for likelihood determination. We find that despite this, both matrices can be properly inverted using the python module \textit{mpmath}\footnote{\url{https://mpmath.org/}}. Although we find that the theoretical covariance, which is closer to singular than the Jackknife one, has difficulty in correctly estimating the likelihood of the chains, so we needed to apply a negligible perturbation of a $0.1\%$ increase to the diagonal elements. We have checked that using the SVD and Cholesky decompositions does not improve this, and we need to add this same small perturbation to the diagonal elements to be able to invert the covariance matrix.

As stated in Ref.~\cite{li2025testinggeneralrelativityusing}, the inverse of an estimated multivariate Gaussian covariance with a finite sample size, \( \hat{C}^{-1} \), follows an inverse Wishart distribution and provides a biased estimate of the true inverse covariance matrix \( C^{-1} \). The unbiased estimate of the inverse matrix is given by:

\begin{equation}
    \hat{C}^{-1}_{\text{unbiased}} = M \left( 1 - \frac{N_d + 1}{N_{\text{JK}} - 1} \right) \hat{C}^{-1},
\label{eq:covfix}
\end{equation}

where \( N_d \) denotes the number of band-powers used, with values of 6, 18, 24 for the small, large and full scales, respectively. Besides, M accounts for the errors on the model parameters and is defined as, 

\begin{equation}
    M = \frac{1 + B (N_d - N_p)}{1 + A + B (N_p + 1)},
\end{equation}

where \( N_p \) denotes the number of parameters. When estimating the single parameter \( E_G \), we take \( N_p = 1 \). The constants \( A \) and \( B \) are given by,

\begin{equation}
    A = \frac{2}{(N_{\text{JK}} - N_d - 1)(N_{\text{JK}} - N_d - 4)},
\end{equation}
\begin{equation}
    B = \frac{N_{\text{JK}} - 2}{(N_{\text{JK}} - N_d - 1)(N_{\text{JK}} - N_d - 4)}.
\end{equation}

\subsection{Estimating the growth rate using MCMC}
\label{sec:MCMC params}

With the multipoles calculated, we need a method to compare them to theoretical predictions in order to constrain the growth rate and the galaxy bias. As previously mentioned, we use the VDG model implemented in \textsc{comet-emu} Ref.~\cite{Eggemeier2023COMETTheory} to obtain a reliable estimate of nonlinear effects. Our objective is to calculate the growth rate across different scales to investigate potential scale dependence. To this end, we define two sets of scales, categorized as large and small, to test for scale dependence in $f(R)$ gravity. It is important to note that, despite the chosen terminology, both scale ranges remain within the linear (or quasi-linear) regime for the redshifts considered. Otherwise, the linear galaxy bias estimation would be biased by nonlinear effects, and the definition of $E_G$, Eq.~\eqref{eq:EG final}, would no longer hold. We also perform the calculation on both scales on GR in order to check that there is no scale dependence in this case, as expected, to validate our methodology. For F5, as indicated in section \ref{sec:Comet}, we simply multiply the final multipoles by the boost provided by \textsc{e-mantis} at the respective redshift $z$.

Estimating the growth rate at small scales can be quite challenging since we do not have many (largely independent) data points to work with and there are more parameter degeneracies due to the higher amount of non-linear parameters, on top of linear ones, that affect these scales. For this reason, in general we try to constraint as many parameter as possible. Fortunately \textsc{comet-emu} let us work with a reduced parameter space from which we select the following: 

\begin{itemize}
    \item $\bm{b_1}$ : We will use a Gaussian prior centered on the value of the linear galaxy bias calculated with the $C_\ell$s and with a deviation given by the obtained standard deviation of the measurements. Although we also tried to set it as a free parameter in the range [0.5,2], it did not produce any significant changes on the posteriors.  
    \item $\bm{f}$ : Since the growth rate is the parameter of interest, we leave this parameter as completely free between the broad limits of the emulator ([0.5, 2]).
    \item $\bm{\sigma_{12}}$ : Since we are not assuming a cosmological model with \textsc{comet-emu} for any of the mocks we can leave this parameter as a free parameter. Although in the final setup we ended up fixing this value to the fiducial value of the mocks, which is similar to what Ref.~\cite{Cabr1} or Ref.~\cite{Wenzl2024ConstrainingBOSS} did in order to avoid degeneracies between $f$, $b_1$ and $\sigma_8$. For the $f(R)$ estimation this should not be a problem since, as mentioned in Ref.~\cite{Saez-Casares2023TheCosmology}, the value of $\sigma_8$ (and consequently $\sigma_{12}$) are taken to be the same for both $\Lambda$CDM and $f(R)$. 
    \item $\bm{b_2}$ : The second order bias can be estimated from the $C_\ell$s using non-linear models like Eulerian or Lagrangian Perturbation theory. In some other analysis, is set with a Gaussian prior centered at 0 with a variance of 1. We tried some of these approaches, but we decided to leave it as a free parameter (between [-2, 2]), since we have checked our analysis is robust to different choices for the prior range used.
    \item $\bm{\gamma_2, \gamma_{21}}$ : The tidal biases are highly non-linear parameters, that in most cases are completely degenerated with $b_2$ and between themselves. In GR it exists an empirical relation with the linear galaxy bias \cite{Sanchez2016TheWedges}:  $\gamma_{2}=0.524 - 0.547b_1 + 0.046b_1^2$ and $\gamma_{21}=(2/21)(b_1-1)+ (6/7)\gamma_2$, which helps break the degeneracy. Originally, we tested that fixing the values of the tidal biases with the galaxy bias helped when using only the monopole and quadrupole, since otherwise the degrees of freedom (number data points minus the number of parameters), when not using the hexadecapole, seemed to be very small for the amount of parameters, but we found significant projection effects for the bias parameters. However when we fixed the value of $\sigma_{12}$, the degeneracies, with or without fixing the values of the tidal bias, these projection effects disappeared so we opted to leave them as free parameters since it is unclear whether these relations are valid for our MG model.
    \item $\bm{c_0, c_2, c_4}$ : The counter-terms are implemented to correct the assumption of a zero stress-energy tensor. So these parameters are defined as non-linear parameters but they actually affect significantly all scales depending on their value.  Each $c_i$ parameter affects mostly the corresponding $\ell=i$ correlation function multipole, while its effect on the other multipoles is negligible unless its value is extremely high. Since we opt to leave the hexadecapole out of the analysis we do not consider $c_4$, otherwise we add degeneracy to the other parameters. 
    \item $\bm{a_{\text{vir}}}$ : This parameter is exclusive to the VDG model and it is related to the virialized velocities. Since it only appears on Eq.~\eqref{eq:velocity generating} as a squared quantity, we restrict it to positive values. We set the limits given by the emulator to the range [0, 8]. 
    \item $\bm{q_{\text{lo}}, q_{\text{tr}}}$ : The Alcock-Paczynski parameters quantify how well we recover the fiducial or input cosmological parameters used in our simulations.
\end{itemize}

In summary, for the full scales standard case we use the 9-dimensional parameter space $\{b_1, f, b_2, \gamma_2, \gamma_{21}, c_0, c_2, a_{\text{vir}}, q_{\text{lo}}, q_{\text{tr}}\}$ with a Gaussian prior on $b_1$. A summary of the parameters and the priors used ca be found at Table \ref{tab:RSD_priors}.

\begin{table*}
\caption{Summary of parameters and priors adopted in the RSD modeling.}
\label{tab:RSD_priors}
\centering
\begin{tabular}{lccp{6cm}}
\hline\hline
Parameter & Description & Prior type & Prior range / specification \\
\hline
$b_1$ & Linear galaxy bias & Gaussian & Informative prior from $b_1$ estimated with $C_\ell$s to reduce degeneracy with $f$ and $\sigma_{12}$ \\
$f$ & Growth rate & Top-hat & [0.5, 2.0] \\
$\sigma_{12}$ & Pairwise velocity dispersion & Fixed & Fiducial mock value to avoid degeneracy with $f$ and $b_1$ \\
$b_2$ & Second-order bias & Top-hat & [-2, 2] \\
$\gamma_2$ & Tidal bias & Top-hat & [-4, 4]\\
$\gamma_{21}$ & Tidal bias combination & Top-hat & [-4, 4] \\
$c_0, c_2$ & EFT counterterms & Top-hat & [-500, 500] \\
$a_{\rm vir}$ & Virial velocity parameter (VDG only) & Top-hat & [0, 8] \\
$q_{\rm lo}, q_{\rm tr}$ & Alcock--Paczynski parameters & Top-hat & [0.8, 1.2] \\
\hline
\end{tabular}
\end{table*}

We use the code \texttt{MultiNest} \cite{Feroz_2009} to perform the MCMC fitting with 1200 live points, a sampling efficiency of 0.8 and an evidence tolerance of 0.01, where these settings are guided from previous cosmological analyses (see \emph{e.g.}, Ref.~\cite{DESY3}). As mentioned above, we do not use the hexadecapole since we considered it to be very noisy, so it would not bring much more additional information to our analysis. One last consideration is that \textsc{comet-emu} obtains predictions in Fourier space and our data is in configuration space. Then we use the algorithm \textit{hankl}\footnote{\url{https://github.com/minaskar/hankl/blob/master/docs/source/index.rst}} to re-express the comet multipoles in configuration space by using a Fast Fourier Transform for each evaluation:

\begin{equation}
\xi_{l} ( r )=i^{l} \int_{0}^{\infty} k^{2} d k / ( 2 \pi^{2} ) P_{l} ( k ) j_{l} ( k r ) 
\end{equation}

where $j_{l}$ are the Spherical Bessel function of order $l$. To avoid numerical instabilities in the above integral, the COMET multipoles are first smoothed by multiplying by the factor $\exp\left(-\left(k \cdot r_{\text{smooth}}\right)^2\right)$, where $r_{\text{smooth}} = 0.25$ Mpc/$h$.

The reason behind using configuration space instead of Fourier space to compare directly with COMET are the stochastic terms (see sec.~\ref{sec: stochastic}). In configuration space these terms disappear since they average out. This helps reducing the parameter space eliminating three terms: $N_{0}^P$, $N_{2,2}^P$, $N_{2,0}^P$. The first term $N_{0}^P$ is completely degenerated with parameters such as $b_1$ that control the full amplitude of the multipoles while the other two are partially degenerated with parameters that depend on the scale.

For the small scale we select the data points within the range [20, 50] Mpc/$h$, while for the large scales we use scales within [50, 140] Mpc/$h$ for all redshifts. We choose 50 Mpc/$h$ as the splitting scale to ensure a sufficient number of data points in the small-scale regime, while also isolating the region where we expect the largest differences between the two gravity models (see Figure \ref{fig:MG_boost}). We limit the minimum scale to 20 Mpc/$h$ to avoid too large nonlinear effects, although we also perform a calculation of the $E_G$ with a minimum scale of 10 Mpc/$h$. This leaves us with 6 data points for small scales and 18 data points for large scales. We performed some validation tests by fitting the COMET parameters with a synthetic data vector generated with COMET. We deduced that for small scales 6 data points were not enough, even in an idealized case like this, to correctly estimate the 9 parameters of the model. For this reason we came up with the following methodology: 1) We first perform a fit on the full range of scales [20, 140] Mpc/$h$ leaving all parameters free except for $b_1$, which contains a Gaussian prior from the $C_\ell$s as detailed earlier, and $\sigma_{12}$ which is fixed to the fiducial value, 2) then we use the estimated mean values and their associated standard deviation for $b_1$, $b_2$, $\gamma_2$, $\gamma_{21}$, $c_0$ and $c_2$ as Gaussian priors for the subsequent fits of the model parameters when we split the full dynamic range on large and small scales. Using this approach we recovered the cosmological parameters in an unbiased way, so we apply this methodology as the baseline for this work.

\subsection{Calculating the lensing power spectrum}

Having $C_\ell^{gg}$ already calculated when estimating the linear galaxy bias, the last ingredient to calculate the $E_G$ estimator is the $C_\ell^{\kappa g*}$ corresponding to the weighted convergence-galaxy angular power spectrum defined in Eq.~\eqref{eq: kg*}.

The procedure is the same than when calculating the $C_\ell^{gg}$, with similar HEALPix maps and Jackknife regions. In this case we are dealing with a cross-correlation of the source sample at z=1.0 (see Table \ref{tab:densitty}) with a $\Delta z =0.2$. We have chosen this particular source z-bin as a working example, but we do not expect our main results to change significantly by selecting another source sample. The convergence $\kappa$ is directly obtained from the catalog for each galaxy and it is averaged on each pixel as the average convergence value of all the galaxies inside that pixel.

Since we are using the weighted estimator $C_\ell^{\kappa g*}$ from Ref.~\cite{Wenzl2024ConstrainingBOSS} Eq.~\eqref{eq: kg*} we have to account for re-weighting of the galaxy sample. For that we just calculate the weight $w_{\times}(z)$ using Eq.~\eqref{eq:wx} and we assign it to each source galaxy at its corresponding z. Then we proceed as in sec.~\ref{sec:galaxy bias} where $n_{pix}$ Eq.~\eqref{eq:overdensity HEALPix} is given by the average value of the weights of all galaxies falling in that pixel, and similarly for $\bar{n}$. We note that no shot-noise correction is needed since this is a cross-correlation estimator.

\subsection{Calculating the $E_G$ estimator}

In order to combine the previous observables that comprise the $E_G$ estimator we use a similar approach to that in Ref.~\cite{Wenzl2024ConstrainingBOSS} with the ratio distribution. As mentioned in that work, the ratio between Gaussian distributed quantities (Jackknife resampling, which mostly contains sample variance directly from the data) it is not guaranteed to be also Gaussian distributed, so it is necessary to explicitly compute the ratio distribution. They performed a double ratio distribution by first combing the angular power spectra as:

\begin{equation}
    R \equiv \Gamma C_\ell^{\kappa g*}/C_\ell^{gg},
\label{eq:ratio}
\end{equation}

which includes the $\Gamma$ parameter, Eq.~\eqref{eq:Gamma factor}, which we treat as factor without attributed error, just like Ref.~\cite{Wenzl2024ConstrainingBOSS}. Then the second ratio distribution is perform by combining $R$ with the RSD parameter as $R/\beta \equiv {\hat{E_G}}$. We instead perform only one ratio distribution since we directly combine the observables $C_\ell^{\kappa g*}/C_\ell^{gg}$ given that we use the same collection of Jackknife masks for each kind of $C_\ell$. In this regard, we are actually doing the ratio in each Jackknife region and then obtaining directly the Jackknife variance for this ratio. This a luxury that Ref.~\cite{Wenzl2024ConstrainingBOSS} could not afford since they did not have enough area to calculate the Jackknife properly so they had to resort to the theoretical Gaussian covariance estimate. Then the probability distribution of the $E_G$ estimator is given by the ratio distribution of $\beta \equiv f / b_1$ and $R \equiv \Gamma C_\ell^{\kappa g*}/C_\ell^{gg}$. The ratio distribution is then,

\begin{equation}
p_{\hat{E}_{G}} ( \hat{E}_{G} | \hat{C}_{\ell}^{\kappa g}, \hat{C}_{\ell}^{g g}, \beta)=\int\mathrm{d} \beta^{\prime} | \beta^{\prime} | p_{R} ( \hat{E}_{G} \cdot\beta^{\prime} ) p_{\beta} ( \beta^{\prime} ), 
\label{eq:ratio distribution}
\end{equation}

where $p_R$ is considered a multivariate Gaussian distribution with mean given by $\widebar{\left( \frac{C_{\ell, i}^{\kappa g}}{C_{\ell, i}^{gg}} \right)}$ calculated with the $N_{\text{JK}}=100$ regions for each of the $i$ $\ell$-bins. The variance of this distribution is given by,

\begin{equation}
\sigma^2_{ij} = \frac{N_{\text{JK}} -1}{N_\text{JK}} \sum\limits^{N_\text{JK}}_{i,j=1} \left[ \frac{C_{\ell, i}^{\kappa g}}{C_{\ell, i}^{gg}} - \widebar{\left( \frac{C_{\ell, i}^{\kappa g}}{C_{\ell, i}^{gg}} \right)} \right] \left[ \frac{C_{\ell, j}^{\kappa g}}{C_{\ell, j}^{gg}} - \widebar{\left( \frac{C_{\ell, j}^{\kappa g}}{C_{\ell, j}^{gg}} \right)}\right].
\end{equation}

\section{Results}
\label{sec:Results}

In this section, we provide the final results for the \(E_G\) estimator and the observables that comprise it for each sample analyzed in this study.

\subsection{Angular power spectrum ratio}
\label{sec:angular_power_spectra_measurements}

Below we discuss how we compute the parameter $R$, defined in Eq.~\eqref{eq:ratio}. The $\Gamma$ parameter, Eq.~\eqref{eq:Gamma factor}, uses the n(z) computed from the mock catalogs, that enters in the calculation of $W_g$. For the rest of the parameters we use the fiducial values of the simulation to set $H_0$, $H(z_{\text{eff}})$ and to calculate the comoving distances needed for $W_\kappa$. The mean effective redshifts for the samples at $z = [0.35, 0.55, 0.8]$, are, respectively, \( z_{\text{eff}} = [0.353, 0.546, 0.791] \). We obtain the same results using equations \eqref{eq: z_cross_eff} and \eqref{eq: z_eff}. In view of this, we do not assign any additional theoretical error to the value of $\Gamma$ in relation to the total error budget of $E_G$.

\begin{figure}
    \centering
\includegraphics[width=0.9\linewidth]{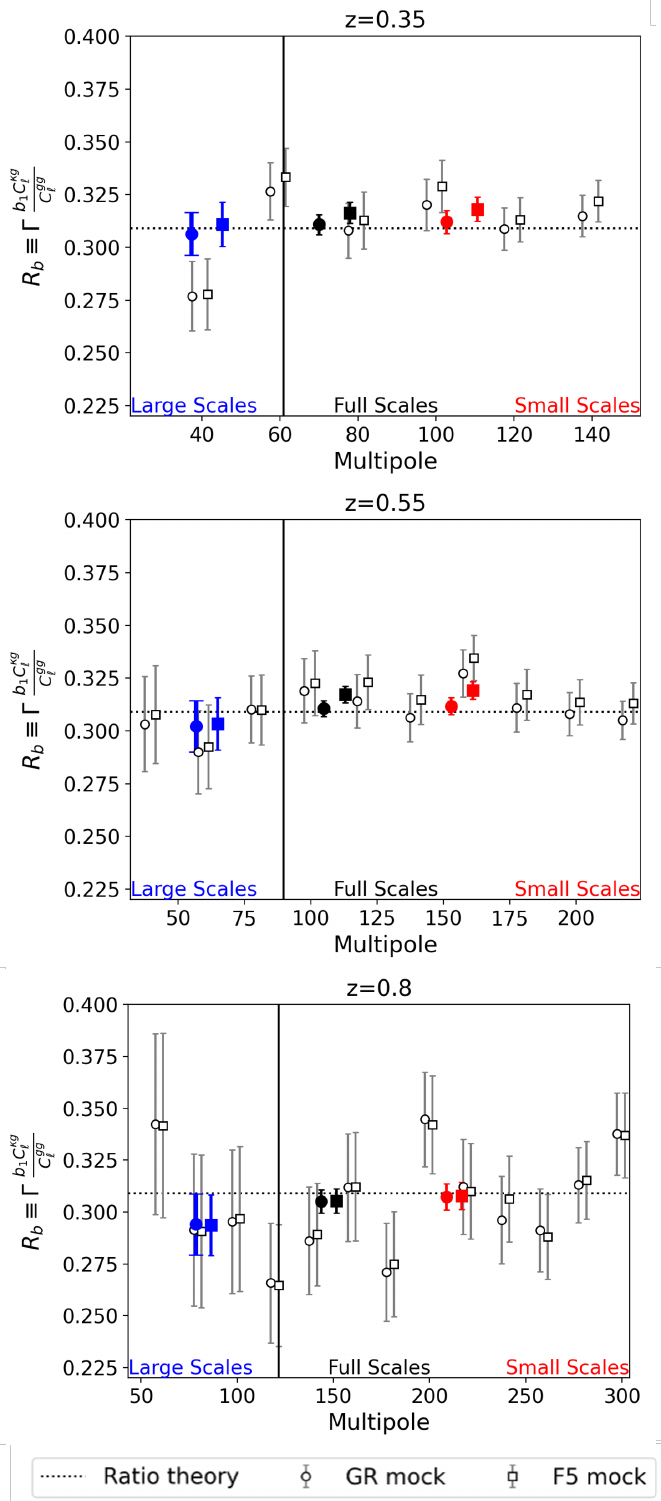}
    \caption{Results for the growth rate independent part of the $E_G$ estimator for the all galaxies case. The white filled data points show the values at which the $C_\ell$s are binned. The black, blue, and red data points represent the full, large and small scales mean result, respectively. The solid vertical black line represents the scale at which we separate the small and large scales.}
    \label{fig:ratio list}
\end{figure}

\begin{figure*}
    \centering
    \includegraphics[width=0.8\linewidth]{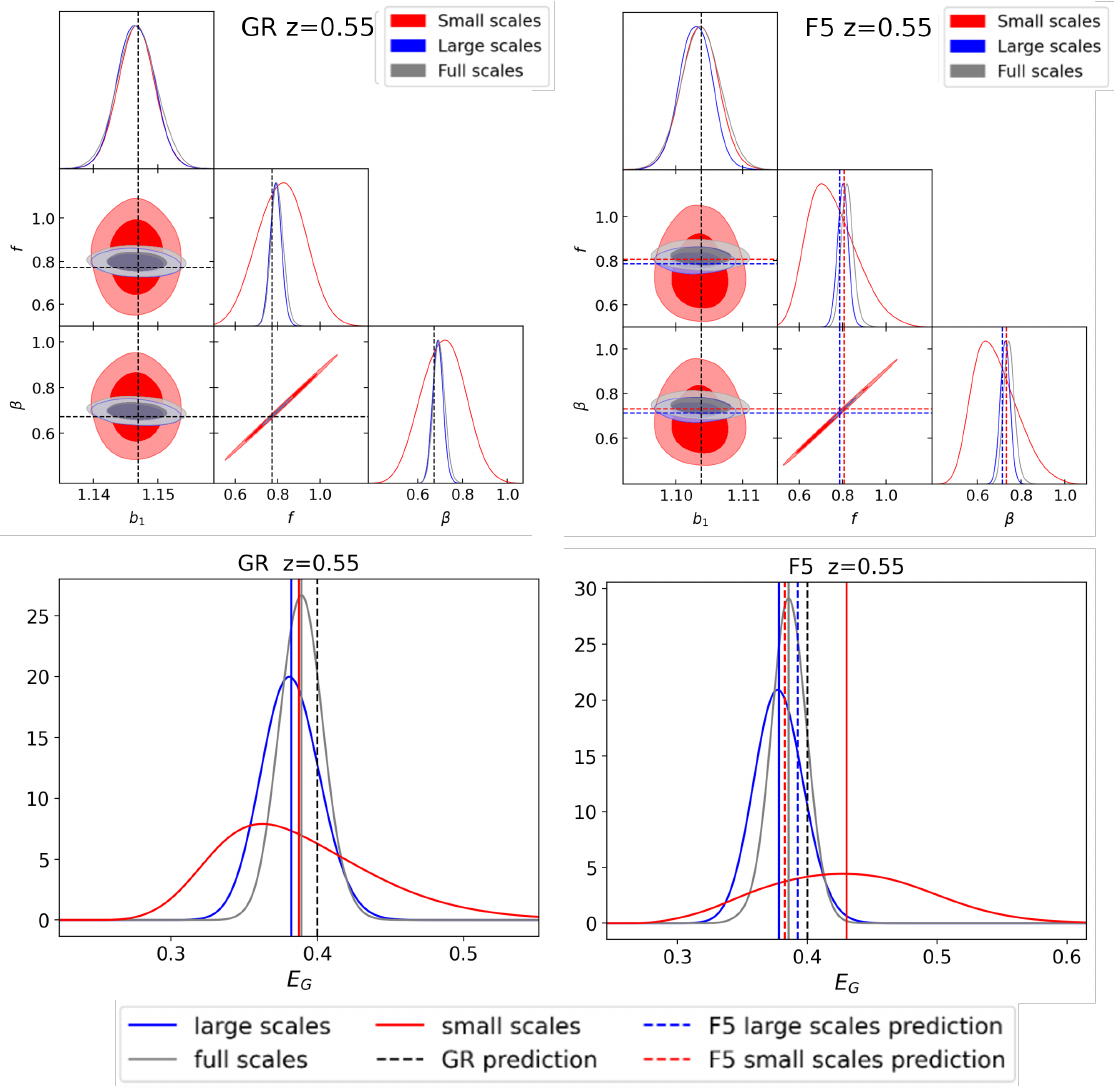}
    \caption{Top plots show the contours results for $b_1$ and $f$ and the corresponding derived $\beta$ parameter for small, large and full scales for GR (Left) and F5 (Right). The dashed lines show the predictions for $f$ and $b_1$ (estimated from the $C_\ell$s) for GR (black), and F5 small (red) and large (blue) scales. The bottom plots show the PDF for the EG estimator for the respective scales, the solid lines show the mean of the respective same color PDF while the dashed lines represent the corresponding theory predictions.}
    \label{fig:EG0.55}
\end{figure*}

\begin{figure*}
    \centering
    \includegraphics[width=0.88\linewidth]{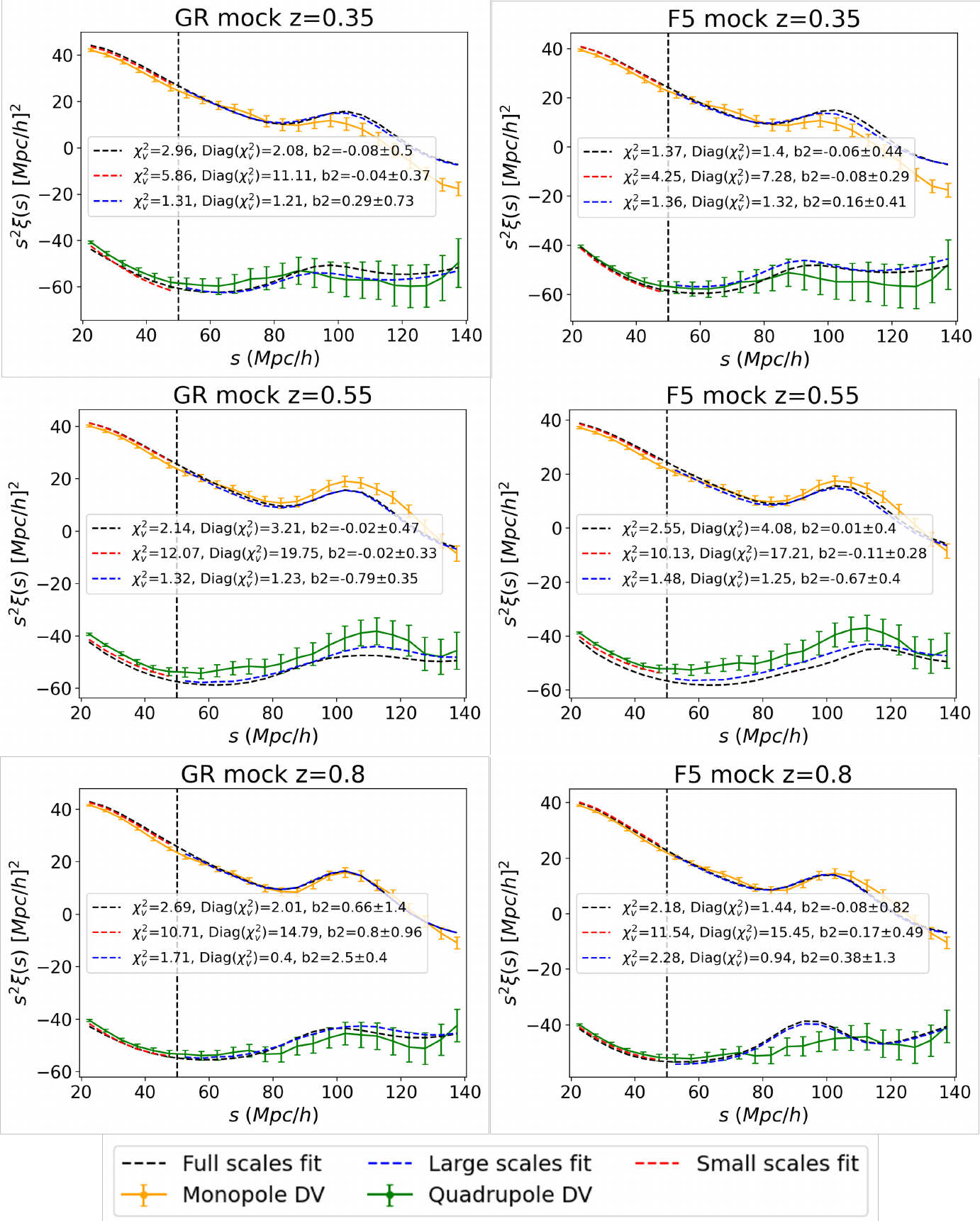}
    \caption{Multipole data vectors values for the F5 and GR mock using the all galaxy sample. The monopole is the yellow line and the quadrupole is the green line. The best fit to the VDG model using the Jackknife covariance matrix is shown as dashed lines for the full (black), small (red) and large (blue) scales. The corresponding $\chi^2_\nu$ of the best fit is displayed in the legend, while Diag($\chi^2_\nu$) is calculated with only the diagonal terms of the covariance matrix.}
    \label{fig:fits}
\end{figure*}

In order to build the $E_G$ estimator, we decompose it in two different contributions. The first involves the ratio of (real space) angular power spectra, as given by Eq.~\eqref{eq:ratio}, and a second term given by the linear growth rate. Focusing on the first of these contributions, we have already shown above the results for the galaxy auto-correlation, $C_\ell^{gg}$, when calculating the linear galaxy bias in Fig.~\ref{Fig:b1}. We then include the cross-correlation between source shears and lens galaxy positions, $C_\ell^{\kappa g}$, to estimate the ratio estimator, $R$. More interestingly, we can define a (linear) galaxy-bias independent  estimator, $R_b \equiv b_1 R$, where the linear galaxy bias, $b_1$, is estimated using the full range of scales (within the scale-cuts used, see Fig \ref{Fig:b1}). Within the approximation that the galaxy bias is linear (for the range of scales considered here) $R_b$ does not depend on scale for both gravity models used (F5 and GR). This is shown in Fig.~\ref{fig:ratio list} where most measurements of $R_b$ are within errors for all redshift bins explored, and thus consistent with the theory prediction. At larger scales the values fluctuate more due to the sample variance which in turn increases the error-bars.

On the other hand, using our small-scale cut set in configuration space, $s_{min}=20$  Mpc/$h$, we can define the corresponding largest multipole at each redshift bin. Using Eqs. \eqref{eq:ell trans} and \eqref{eq:theta trans},  we get $\ell \approx [150, 225, 300]$ at z=[0.35, 0.55, 0.8]), respectively. We assume that within the resulting multipoles ranges (see Fig.~\ref{Fig:b1}), the galaxy bias is approximately linear, what is in agreement with the fact that we observe no deviation from a scale-independent behavior for the ratio estimator, $R_b$ (see Fig.~\ref{fig:ratio list}). We note that this is consistent with our assumption that our  $E_G$ gravity estimator is also defined on linear scales. The observed multipole bin-to-bin fluctuations largely cancel out when taking the average over the range of scales of interest (\emph{i.e.}, for the used split in the so-called" large" or "small" scales in the plot). In fact, the statistical average for $R_b$ follows a ratio distribution (see Eq.~\eqref{eq:ratio distribution}) with $\beta = 1/b_1$. Although we see that for F5, in the low redshift bins, some small-scale values tend to be above the theory prediction beyond the 1-$\sigma$ errors, overall the measurements agree with theory.

The errors attributed to $R_b$ come mostly from the $C_\ell^{kg}$, since $b_1$ and $C_\ell^{gg}$ have negligibly small uncertainties. In particular, the statistical errors scale with the lensing efficiency (see Eq.~\eqref{eq:lensing efficiency}), \emph{i.e.}, the larger the lensing signal, the smaller the errors in $C_\ell^{kg}$, and vice versa. Given that we choose our source sample at $z=1$, i.e., at a distance $\approx 2300 \ \rm{Mpc/h}$, the lensing efficiency peaks at half this distance, $\approx 1150 \ \rm{Mpc/h}$. Therefore, given that  the lens planes for the three redshift bins ($z=[0.35, 0.55, 0.8]$) are located at $\approx [960, 1430, 1940] \rm{Mpc/h}$ respectively, the highest z-bin has about a factor of 2 lower signal and correspondingly larger statistical error in its measurement of the $C_\ell^{kg}$. In turn this reflects in a larger error in $R_b$, as shown in Fig \ref{fig:ratio list}.

\subsection{RSD $\beta$ parameter fitting results}

\begin{figure}
    \centering
\includegraphics[width=0.9\linewidth]{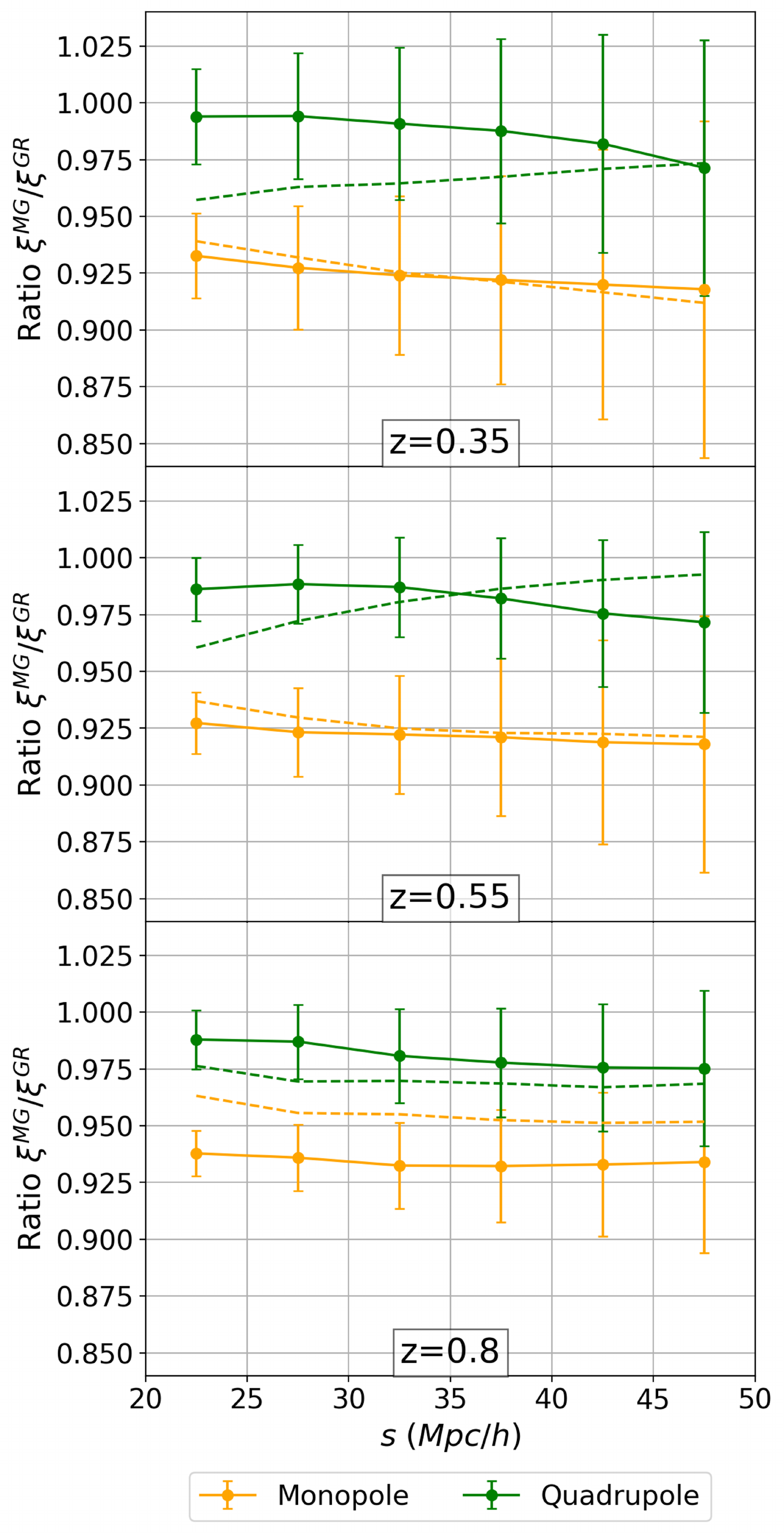}
    \caption{Comparison of the multipoles: monopole (orange) and quadrupole (green) at small scales for GR and F5. The solid lines represent the ratio of the data vectors while the dashed line is the ratio of the best fit of each case (for both the ratio is defined as F5 over GR). The errors represent the propagated Jackknife errors of each data vector.}
    \label{fig:ratio small}
\end{figure}

The redshift space contribution to the $E_G$ estimator is given by the linear growth rate-related quantity, $\beta$, see Eq.~\eqref{eq:EG final}, which is a derived parameter. It is calculated from the fitted posterior distributions for the linear galaxy bias, $b_1$, and the linear growth rate, $f$, using an Markov chain Monte Carlo (MCMC) approach with the \texttt{MultiNest} sampler. In the upper plots of Fig.~\ref{fig:EG0.55} we show the marginalized posterior profiles (over the full parameter space) for $b_1$ and $f$, for both catalogs at $z=0.55$ (for the equivalent plots for the other two redshifts check the appendix for Fig.~\ref{fig:EG0.35} and Fig.~\ref{fig:EG0.8}). We also show the density profile for $\beta$ which is derived from the posterior distributions of the other two parameters. We can see in these plots that the growth rate at small scales has a variance around 3 times bigger than the other two cases (\emph{i.e.}, the large and full scales), as expected, since there are much fewer (uncorrelated) modes available for measurements on these small (non-linear) scales.

For completeness, in appendix \ref{annex:triangle} we show the posterior distribution of the full 10-dimensional parameter space of the VDG model for the F5 z=0.8 case. As shown in  Fig.~\ref{fig:F5triangle08}, most of the model parameters seem to converge well although we observe a double peak on the local quadratic bias parameter, $b_2$, for GR when constrained to large scales only. We note that this is not the case for the F5 model, where even this parameter exhibits a well defined peak, and away from zero, at variance with the large-scale fit. 
The fact that this non-linear parameter shows this double peak at large scales could be the result that these particular scales do not have enough non-linear information to properly fit this parameter. For the rest of parameters, they seem to follow the expected behavior. The counterterms ($c_0, c_2, \gamma2, \gamma{21}$) are consistent with zero and the Alcock-Paczynski parameters are consistent with 1 (although at the 2-sigma level only for the full scales case) so we largely recover the fiducial cosmology of the mock.

As mentioned before, for small scales we can appreciate how the contours are much bigger for parameters such as the growth rate or the Alcock-Paczynski parameter. This is arguably caused by the few data points included in the "small scales" range, and the large correlation between them induced by the non-linear clustering, which results in very few effectively independent modes to constrain the model parameters. This is usually called "projection effects" from the prior volume of the multi-dimensional parameter space that is intrinsically largely degenerate and, thus, it needs many independent modes to break such degeneracies. In order to alleviate this, we first derive Gaussian priors for the counterterms of the VDG model by first running a MCMC chain using the full scales with fix cosmology (as mentioned before). Then we use the derived constraints on these counterterms to run the full MCMC chains letting both the cosmology and counterterms parameter change for the small and large scales cases. In particular, we have observed that this prior step is crucial to avoid a large amount of projection effects in our constraints on the growth rate.

In Fig.~\ref{fig:fits} we show the results for the multipoles of the correlation function alongside the COMET theoretical data vectors using the corresponding best-fit parameters. The best-fit shown reproduces overall the measurements, although there are discrepancies specially on large-scales. We note however that the large covariance between scales makes the fit appear worse than it is in practice. We can see that the quadrupole is quite noisy in comparison with the monopole, specially at large scales. This is induced by the larger error-bars at these scales due to sample variance. It seems that COMET can hardly fit both the monopole and the quadrupole simultaneously. Since the monopole have smaller error-bars the MCMC chains seem to favor fitting the monopole over the quadrupole. This is illustrated by the fact that the monopole exhibits a good fit overall, with only a poorer model fitting around the baryon acoustic oscillation feature at $100 \ \rm{Mpc/h}$ scales. Similarly, the quadrupole is well described by the non-linear theory model overall, except for the redshift bin at $z=0.55$ where the theory deviates from the simulation on large scales.

In order to quantify the model fitting we define a reduced $\chi_\nu^2$ test as,

\begin{equation}
    \chi_\nu^2 = \frac{\chi^2}{n_d - n_p}
\label{eq:chi-reduced}
\end{equation}

where $n_d$ is the number of data points combined between monopole and quadrupole (48 for the standard case) and $n_p$ the number of free parameters ($\approx$10 for most cases), while $\chi^2$ has the same definition as in Eq.~\eqref{eq:chi2}.

Overall the results for all redshift bins and scales yield large $\chi_\nu^2$ values (see legends of the different panels in Fig.~\ref{fig:fits}). This, at face value, suggest a poor fit to the theory model. There might be several reasons for this. On one hand, the particularly small error-bars, specially at small scales, exacerbate this excess in the resulting $\chi_\nu^2$ value. But this is also true, although with somewhat lower values, for the full and large scales. This is specially surprising for the case at $z=0.8$, where the mock data vectors seem to be in very good agreement with the best-fit, at least visually (\emph{i.e.}, in chi-by-eye). In fact, one would expect to obtain better fits as we go to higher redshift where clustering has a lower degree of non-linearity. However, one thing that seems to artificially increase the value of $\chi_\nu^2$ is that the Jackknife covariance is close to singular. Moreover if we only take the diagonal elements of the covariance matrix the value of $\chi_\nu^2$ drops by a factor of 2 or more at large scales, yielding a value consistent with unity in all cases, except for the GR model in the lowest redshift bin (see Fig.~\ref{fig:fits}).

Since it is on the smallest scales that we expect the specific gravity theory has larger impact on the clustering (related to the non-linear power boost in the matter power spectrum in MG with respect to GR), we further investigate in Fig.~\ref{fig:ratio small}, where we plot the ratio of the data vectors and the corresponding ratio of the best fits of F5 over GR. This provides a direct way to see to what extent it is possible to distinguish between the two models. Note that the errors, obtained with the Jackknife method applied to each catalog, are big enough as compared to any small numerical uncertainty in the ratio of the data vectors. Looking at the correlation function multipoles ratios we find that the theory prediction for the monopole ratios closely follow the ratio of data vectors. For the quadrupole there is some degree of discrepancy between theory and simulations, although not significant, given the errors. This puts into perspective that, despite the small errors, they are still large enough to shadow possible deviations with respect to standard gravity even for (ideal, \emph{i.e.}, systematics free) all sky surveys as we model in this paper.  In section \ref{sec:nulltest} we further quantify our results from the small-scale clustering in redshift space in terms of a null test of gravity.

\subsection{Results for the $E_G$ estimator}
\label{sec:egresults}

\begin{figure*}
\centering
\includegraphics[width=0.816\textwidth]{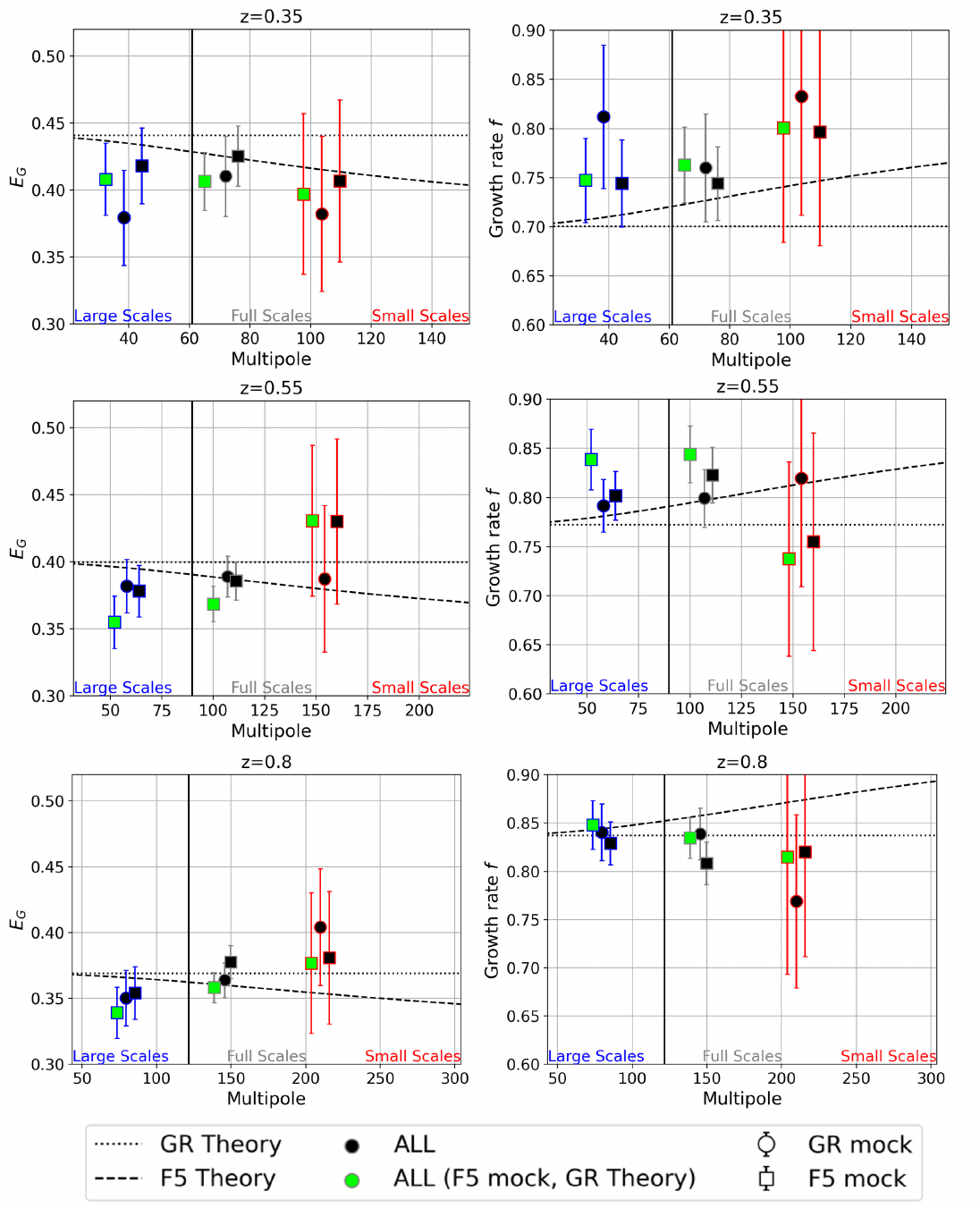}
\caption{
Final results for the $E_G$ parameter (left plots) and the growth rate f (right plots) for the GR (circles) and F5 (squares) mocks at the three redshift bins studied. The results are split according to the various scales considered: small (red symbols and errors), large (blue) and full range (gray). The vertical solid line separates the small and the large scales. The black filling of the data points represent the baseline case (All) for both GR and F5 (using the corresponding gravity for the theory model), while the green one represent the case that uses the F5 mock data but assumes the GR theory to perform the fits (GR theo). The dot (dashed) line represent the theoretical prediction for GR (F5). The estimated values are centered around the mean $\ell$ of the respective range of scales (large, small or full scales) although, for a given range of scales, the different cases (\emph{i.e.}, depending on gravity model and theory assumption) are slightly shifted  to the left and right to avoid to clutter.}
\label{Fig:EG list}
\end{figure*}

\begin{figure*}
    \centering
    \includegraphics[width=0.88\linewidth]{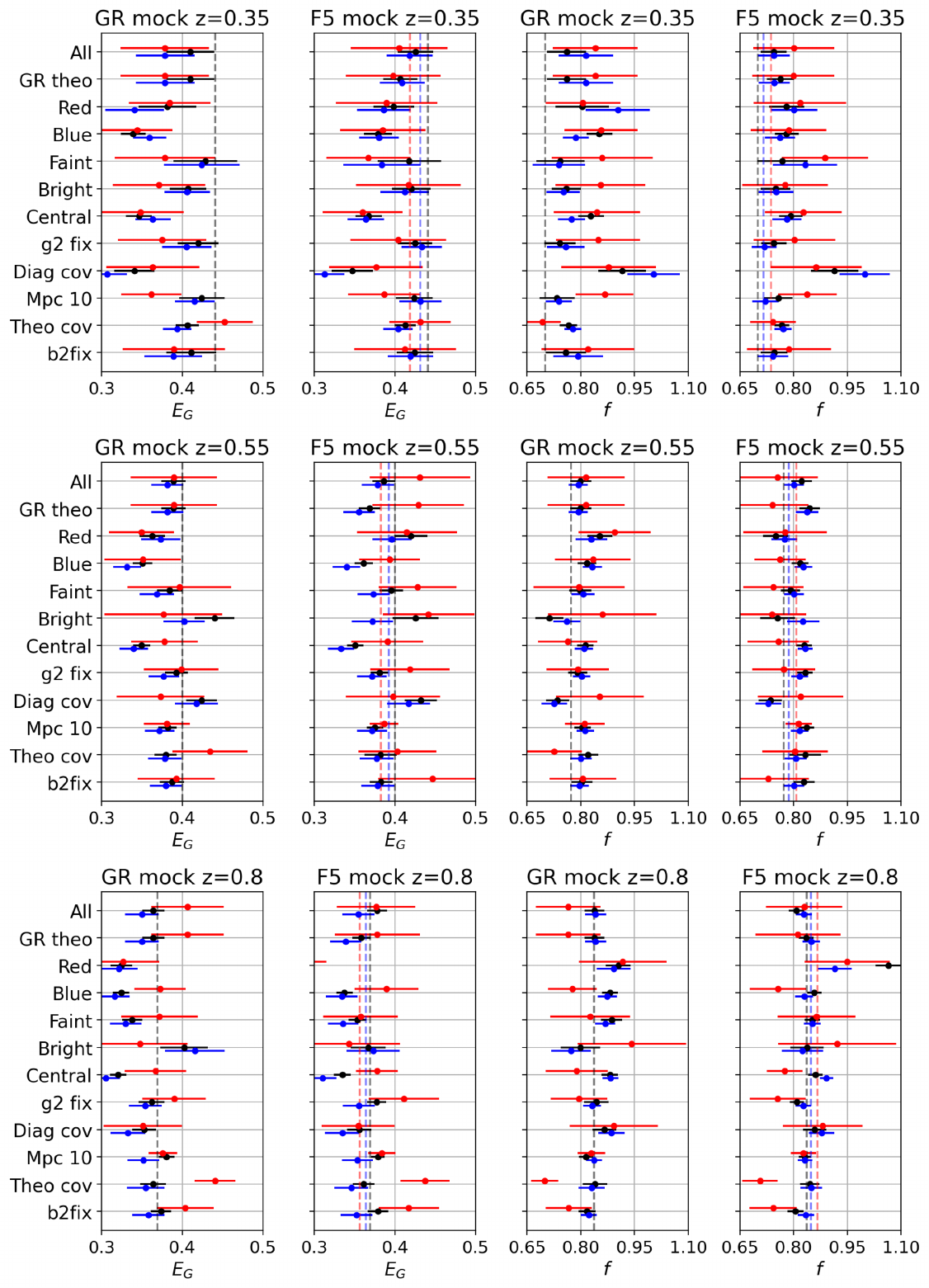}
    \caption{Summary of the main results for the various cases analyzed in this paper. The black, blue, red points and error-bars represent the full, large and small scales result, respectively. Results displayed are for both gravity theories, with the left plots showing the EG estimator and the right plots showing the growth rate results. The dashed lines show the fiducial values where the F5 have the respective values for small and large scales.}
    \label{fig:Cases}
\end{figure*}

Having estimated all the ingredients of the gravity estimator, \emph{i.e.}, the RSD $\beta$ parameter\footnote{As described in Ref.~\cite{Chen2022CosmologicalTheory} we can ensure that we can combine all our results since the correlations have the same effective redshift. The same goes for the  $\beta$ parameter which is estimated from a 3D distribution at the same effective redshift.}, as discussed in the  previous section, and the real-space clustering ratio, $R_b$ (see Eq.~\eqref{eq:ratio}), we are ready to compute the $E_G$ statistic. In order to derive the posterior for this estimator, we shall apply the ratio distribution, as given by Eq.~\eqref{eq:ratio distribution}.

In the lower panels of Fig.~\ref{fig:EG0.55} (and Fig.~\ref{fig:EG0.35} and Fig.~\ref{fig:EG0.8}), we show the posterior PDF obtained for the $E_G$ estimator using the ratio distribution given by Eq.~\eqref{eq:ratio distribution}. We simultaneously plot the results for the small and large scales alongside the full range of scales. The $E_G$ estimator seems to follow a Gaussian distribution as obtained in Ref.~\cite{Wenzl2024ConstrainingBOSS}. The corresponding best-fit values with error-bars are provided in Fig.~\ref{Fig:EG list} what provides a more direct comparison of the marginal differences between the gravity theories fitted to the simulations. Given that the $E_G$ estimator is inversely proportional to the growth rate its value decreases when going to smaller scales at a given redshift. Taking into account that the clustering ratio $R$, Eq.~\eqref{eq:ratio}, is not expected to change in F5 with respect to GR the (see \ref{sec:angular_power_spectra_measurements}), the $E_G$ estimator is only sensitive to the underlying gravity theory through the linear growth rate parameter. This is illustrated in detail in Fig.~\ref{Fig:EG list} which summarizes the main results of this paper and tests the robustness of them with respect to the analysis choices used, as we shall discuss in more detail below.

Overall, as shown in Fig.~\ref{fig:EG0.55}, Fig.~\ref{fig:EG0.35} and Fig.~\ref{fig:EG0.8}, the values of the gravity estimator for the full scales of GR and F5 seem to agree with the respective prediction within 1-$\sigma$ errors. However, for F5 at $z=0.8$, the estimated value of the growth rate is biased low at around 2-$\sigma$, but since the value of $R_b$ for F5 is slightly biased high, this tends to compensate for the estimated $E_G$ value. Moreover, for this case, the GR and F5 simulation results can not be distinguished given the statistical errors. For the other redshift bins, $z = 0.35$ and $z = 0.55$, the values for both simulations appear to align closely with the best-fit models. 
However, at $z=0.55$, the estimated value of the gravity parameter appears to be biased high on small scales (specially for F5), although in rough agreement with theory, given the errors. For the redshift bin at $z=0.35$, it follows the opposite trend, making the estimated F5 values agree better with the theory prediction. For completeness, we have also included the case where we try to fit an F5 simulated data vector with a GR model. This is a direct way of testing (sort of a "null test") how well the estimator is able to distinguish between close gravity theories such as F5 and GR. Our results yield best-fit values to GR and F5 theory that are statistically consistent, suggesting that assuming the wrong theory might not significantly bias our results. We have checked that even at higher redshifts (\emph{e.g.}, using a mock sample at $z=1$ with lensing sources at $z=1.2$), where the clustering is closer to the linear regime, we find that the F5 results tend to be biased high at all scales, while for GR results are unbiased except on large scales.

The results in all the previous plots were for the all galaxies case, where we consider the full sample of galaxies in a given redshift bin and the relative magnitude cut on the r-band of $r<24$. As stated in the sec.~\ref{sec:Data}, we also perform the same calculations on several different samples and analysis choices, as a set of robustness tests. In Fig.~\ref{fig:Cases}, we present the mean and standard deviation for the gravity estimator for all the cases considered. Below we summarize the cases explored,
\begin{itemize}
    \item 1) $\mathbf{All}$: reference case used in this. It uses the full-sky mock, all galaxies sample,  with a cut in relative for $r<24$ for the SDSS r-band. The rest of sample-specific cases \-3)-7)\- are derived from this one by applying additional selecting criteria. 
    \item 2) $\mathbf{GR \ theo}$: same as above but  using a best-fit GR theory to analyze the F5 simulated data. 
    \item 3) $\mathbf{Red}$: same as reference, but for red-color galaxies classified using a $g-r$ cut following Ref.~\cite{Carretero}.
    \item 4) $\mathbf{Blue}$: same as reference, but for a blue galaxy sample, following Ref.~\cite{Carretero}. 
    \item 5) $\mathbf{Faint}$: same as reference sample, but for faint galaxies, obtained by imposing a relative magnitude bin, $23 < r < 24$.
    \item 6) $\mathbf{Bright}$: same as reference, but for a bright galaxy sample defined by imposing a relative magnitude cut $r<22.5$.
    \item 7) $\mathbf{Central}$: same as reference case, but only selecting the central galaxy of each halo. 
    \item 8) $\mathbf{\gamma_2 \ fix}$: same as reference case, but setting the tidal bias parameters, $\gamma_2$ and $\gamma_{21}$ as derived parameters from the linear galaxy bias $b_1$, using the relations: $\gamma_2=0.524 - 0.547b_1 + 0.046b_1^2$ and $\gamma_{21}=(2/21)(b_1-1)+ (6/7)\gamma_{2}$ from Ref.~\cite{Sanchez2016TheWedges}.
    \item 9) $\mathbf{Diag \ cov}$: same as the reference case but using only the diagonal elements of the Jackknife covariance matrix.
    \item 10) $\mathbf{Mpc \ 10}$: same as the reference case but extending the minimum scale to 10 Mpc/$h$. The full range is now defined as $[10, 140]$ Mpc/$h$ and the small scale goes from $[10, 40]$ Mpc/$h$, while the large scale range remains the same. 
    \item 11) $\mathbf{Theo \ cov}$: same as the reference case but using the theoretical Gaussian covariance matrix to calculate the likelihood.
    \item 12) $\mathbf{b2fix}$: Same as the reference case, but setting the second order bias as a derived parameter from the linear galaxy bias using the Local Lagrangian relations from Ref.~\cite{Lazeyras_2016}: $b_2 = 0.412 - 2.143 \ b_1 + 0.929 \ b_1^2 + 0.008 \ b_1^3$.
\end{itemize}

In summary, our extended analysis shows that our results are robust to changes in the galaxy sample selection and analysis choices made. The general trend is usually the same as the reference "All" galaxies case, where at small scales the values for $E_G$ are over-predicted and for large scales are under-predicted, although largely within errors. Similar results hold for both gravity models, although a larger tension is observed for F5 than for GR. In particular, the recovered scale dependence goes in the opposite direction to that predicted by theory. In the cases where we put specific cuts on the sample such as for the red, blue, faint, bright and central samples, the values deviate more from the prediction than the "All" galaxies case. The rest of cases depict different analysis choices relative to the reference case. When we fix the tidal biases ($\gamma2$, $\gamma{21}$), it does not seem to have significant impact on the growth rate posteriors. The tidal biases obtained are slightly different, e.g for $z=0.55$ the estimated value for $\gamma{21}$ turn out positive instead of the negative value we find in Fig.~\ref{fig:GRtriangle} and Fig.~\ref{fig:F5triangle}. It seems that these parameters are highly nonlinear and are not largely degenerate with a linear parameter like the growth rate. In addition, we have also tested the impact of the choice of covariance matrix in our results. For this purpose we have compared the case when using the theoretical (Gaussian) and diagonal matrix and they seem to produce similar results to the Jackknife covariance, with the exception at $z=0.35$ where results worsen significantly. A possible explanation for this is that at lower redshifts the off-diagonal terms of the covariance become larger, and the accuracy in the computation of these may have a larger impact in our results. Lastly, we have tested how including even smaller scales in the analysis may impact our findings. This is specially relevant since we expect these small non-linear scales to be the ones most sensitive to the underlying gravity model. With this in mind we have extended the range of the "small scales" case from $20 \ \rm{Mpc/h}$ (in the reference case) to $10 \ \rm{Mpc/h}$. However, including these smaller scales does not appear to significantly reduce the estimator biases; in fact, it worsens the fit in certain cases. This suggests that the non-linear model used is already breaking down at the minimum scale used, for all the redshifts explored. Finally, fixing the value of $b_2$ does not seem to affect our results in any significant way.

\section{A null test of gravity from small scales clustering in z-space}
\label{sec:nulltest}

\begin{table}[]
\begin{tabular}{c|c|c|c|c}
\cline{2-4}
                                                              & \textbf{z=0.35} & \textbf{z=0.55} & \textbf{z=0.8} &                                            \\ \hline

\multicolumn{1}{|c|}{\multirow{2}{*}{\textbf{$\boldsymbol{\chi_{\nu,D}^2}$}}}    & \cellcolor{red!20}5.78           & \cellcolor{red!20}10.99          & \cellcolor{red!20}10.59          & \multicolumn{1}{c|}{\textbf{Small Scales}} \\ \cline{2-5} 
\multicolumn{1}{|c|}{}                                        & 1.94           & 4.32          & 3.54          & \multicolumn{1}{c|}{\textbf{Full Scales}}  \\ \hline
\multicolumn{1}{|c|}{\multirow{2}{*}{\textbf{$\boldsymbol{\chi_{\nu,T}^2}$}}}    & \cellcolor{red!20}6.09           & \cellcolor{red!20}10.30          & \cellcolor{red!20}8.20            & \multicolumn{1}{c|}{\textbf{Small Scales}} \\ \cline{2-5} 
\multicolumn{1}{|c|}{}                                         & 5.21          & 3.85          & 4.46           & \multicolumn{1}{c|}{\textbf{Full Scales}}  \\ \hline
\multicolumn{1}{|c|}{\multirow{2}{*}{\textbf{$\boldsymbol{\chi_{\nu,GR}^2}$}}} & \cellcolor{red!20}6.49           & \cellcolor{red!20}12.12          & \cellcolor{red!20}9.02             & \multicolumn{1}{c|}{\textbf{Small Scales}} \\ \cline{2-5} 
\multicolumn{1}{|c|}{}                                        & 4.46          & 6.13          & 4.37           & \multicolumn{1}{c|}{\textbf{Full Scales}}  \\ \hline
\end{tabular}
\caption{$\chi_\nu^2$ values for a comparison between data vectors, combining monopole and quadrupole, (MG vs GR, denoted as $\chi_{\nu,D}^2$), for the best-fit theory data vectors ($\chi_{\nu,T}^2$), and for best-fit theory data vectors assuming that F5 data follows the GR model  ($\chi_{\nu,GR}^2$). The values are calculated using the same range of scales defined for the small and full scales for the multipoles of the correlation function (see Figure \ref{fig:fits}) and the Jackknife covariance of the GR catalog.}
\label{tab:chi2}
\end{table}

From the comprehensive analysis presented in section \ref{sec:egresults} above, we conclude that the $E_G$ has clear limitations to constrain gravity, possibly due to a combination of cosmic variance \emph{i.e.}, noise from the fact that we only have one simulated universe, and projection effects across the multi-dimensional parameter space of the non-linear model used. Besides, most of the deviations in the clustering statistics are expected to show up on small (non-linear) scales, where the perturbative model we have used (VDG model) is expected to break-down. In Fourier space, the limiting scale of the model is at $k>0.35 \rm{h/Mpc}$, see Ref.~\cite{Eggemeier2023COMETTheory,eggemeier2025boostinggalaxyclusteringanalyses}, what should translate, according to Eq.~\eqref{eq:k trans}, into projected scales of about $10 \ \rm{Mpc/h}$.  In particular, we have also checked that this conclusion is robust to the specific perturbative model used (EFT or VDG), although we have found that the VDG model outperforms the EFT for the RSD modeling, what has been our criteria to select the former as our reference model for the analysis (see also Ref.~\cite{eggemeier2025boostinggalaxyclusteringanalyses}).

\begin{figure*}
    \centering
    \includegraphics[width=0.92\linewidth]{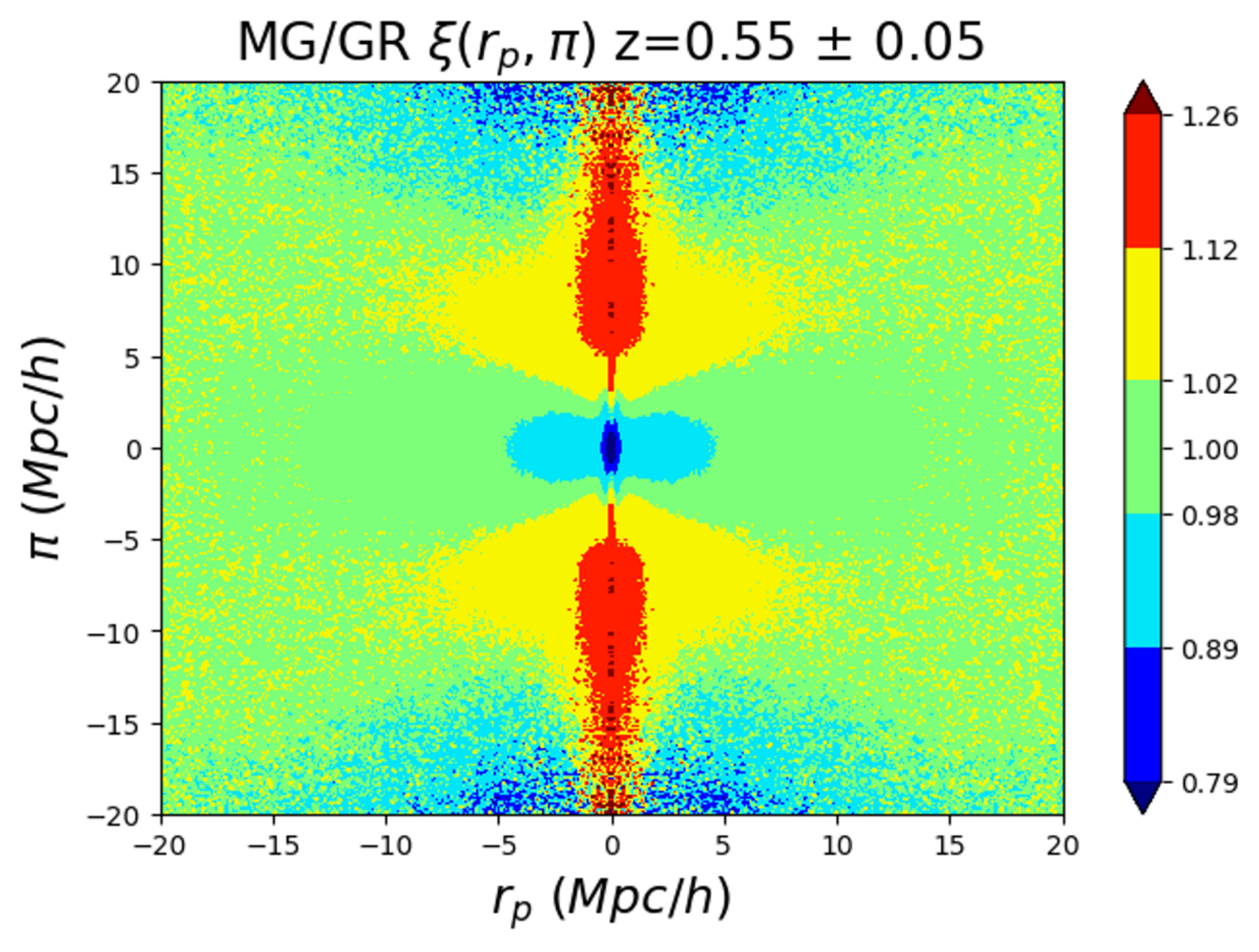}
    \caption{2D correlation function in $\pi$ (LOS distance), $r_p$ (projected distance). The color scale indicates the clustering amplitude ratio of the F5 over the GR mocks, where each correlation is normalized by its corresponding linear galaxy bias, $b_1$, at $z=0.55$.}
    \label{fig:2Dplot}
\end{figure*}

All things considered, we find that, with the current state-of-the-art modeling tools, the $E_G$ estimator is not well suited to differentiate between both gravity theories, even for an ideal survey set-up. Alternatively, in this section we explore whether one can set constraints on gravity using all the information contained in the basic 2-point clustering statistics in redshift space.  In order to quantify the expected differences in RSD clustering, we compute the ratio of the 2D correlation function $\xi(\pi, r_p)$ between the F5 and the GR mocks, where $\pi$ is the line of sight (LOS) distance and $r_p$ is the project distance on the plane of the sky. Fig.~\ref{fig:2Dplot} shows this projected clustering ratio, where we factor out the dependence on the linear galaxy bias of both mocks by multiplying by $(b_1^{GR}/b_1^{MG})^2$. Our results show the only significant ($>10\%$) differences show up at rather small projected scales $<10$ Mpc/$h$, where the Finger of God effect becomes prominent, what renders the widely-used perturbative models (VDG or EFT) largely inaccurate. This is a good indicator that properly modeling even smaller scales than those investigated in this work would be a critical improvement to detect potential deviations from standard gravity. Unfortunately, current models do not allow to predict these scales with enough accuracy.

In order to quantify the observed differences from the small scales in the projected clustering, we decompose the data vectors into their correlation function multipoles, according to Eq.~\eqref{eq:cfmultipoles}, 
to quantify the differences between both gravity simulations. Specifically, we compute the $\chi_\nu^2$ statistic for the difference between the data vectors of GR and F5 (monopole and quadrupole), as well as for the difference between the best-fit theory data vectors. The number of degrees of freedom, $ \nu$, is defined according to Eq.~(\ref{eq:chi2}). The total number of data points, $n_d$, (monopole + quadrupole) is 48 for the full scale range and 12 for the small-scale range. In our multipole analysis, we fitted 9 parameters, $n_p$. The $\chi_\nu^2$ estimator thus defined provides a simple "null test" to measure the ability to distinguish between the two theories. Since the simulations were generated using the same cosmology, except for the gravity theory. Once these mocks are calibrated against observations at very low redshifts (see Ref.~\cite{Carretero}), any observed differences in the clustering at higher redshift are expected to be caused by the different underlying gravity model used to produce each synthetic galaxy catalog.

Table \ref{tab:chi2} summarizes the results for this section, where we define three $\chi_\nu^2$ statistics: $\chi_{\nu,D}^2$, representing the difference between the F5 and GR data vectors; $\chi_{\nu,T}^2$, quoting the difference between F5 and GR best-fit theory data vectors; and $\chi_{\nu,GR}^2$, the same as $\chi_{\nu,T}^2$ but using the best-fit theory data vector assuming that F5 data follows the GR model. In all cases the $\chi_\nu^2$ values are computed using the Jackknife covariance derived from the GR mock data vectors, which closely resemble that of F5. The computation follows the method outlined in Eq.~\eqref{eq:chi-reduced}. The high $\chi_{\nu,D}^2$ values demonstrate that the theories can be clearly distinguished, in terms of such null test, despite the associated uncertainties. We get that the data vectors are drawn from distributions that differ at a significance level of $\sim$3$\sigma$, for each of the redshift bins studied. If we use the best-fit theory models instead, the null test ($\chi_{\nu,T}^2$, and even $\chi_{\nu, GR}^2$) yields similar significances, what suggests that the VDG model can distinguish between theories. Additionally, the table includes results for the same test applied to the full scales range. Here the $\chi_\nu^2$ values are approximately half those obtained for small scales, confirming that most of the discriminating power resides on small scales ($< 50 \ \rm{Mpc/h})$.

The results indicate that the differences between the models are significant enough to distinguish the underlying gravity model. Developing more precise theoretical models that can accurately predict behavior at even smaller scales would be a valuable step toward improving model selection. However, while these findings hold for the idealized case of simulations, it remains essential to test their applicability to real observations, where higher levels of uncertainty and sources of systematic error would degrade the ideal survey case presented in this work.

\section{Discussion}
\label{sec:Discussion} 
  
\subsection{\textbf{Challenges of growth rate estimation at different scales}}  

This work provides a systematic assessment of the growth rate and the 
$E_G$ gravity estimator across multiple scales in a modified gravity scenario, using a unified analysis pipeline applied to matched GR and 
$f(R)$ simulations. While some previous studies \cite{Pullen2014ProbingLensing, Pullen2016ConstrainingVelocities, Wenzl2024ConstrainingBOSS} assumed a scale-independent growth rate, calculated from the full range of scales, this approach limits the fundamental purpose of $E_G$ as a general test of gravity models. Probing the scale dependence is crucial since the gravity estimator and the linear growth rate are predicted to vary with scale in most non-standard theories and thus require a proper framework to accurately model from large (linear) to small (quasi-linear or fully non-linear) scales. From the observational standpoint, this is what past analyses have studied, e.g Ref.~\cite{Pullen2016ConstrainingVelocities} and Ref.~\cite{Wenzl2024ConstrainingBOSS} measured the value of $E_G$ at different harmonic multipoles, $\ell$, in a similar way to what we have done with the clustering ratio, $R_b$, as shown in Fig.~\ref{fig:ratio list}. However these analyses have overlooked the consistent estimation of the scale of the growth rate, specifically the $\beta$ parameter, and its possible scale dependence, to accurately derive the $E_G$ gravity estimator.

However, this is inherently challenging due to the limited availability of accurate non-linear models in redshift space. The present study develops a  well defined framework to address this issue, using state-of-the-art perturbative models, and emphasizing the importance of separating the analysis into large- and small-scale regimes. In practice, small scales are especially difficult to model due to strong non-linear growth related the so-called Fingers-of-God effect, that demand a proper account of the distribution of pairwise velocities and how those impact the 2-point clustering in redshift space. On top of that,  our simulation only has one realization which can add a noise term affecting cosmic variance and even stochastic components from the simulation, like galaxy placement, that may significantly impact small scales.

\subsection{\textbf{The Role of priors in the parameter estimation}}

This study also remarks the pivotal role of imposing (Gaussian) priors to the counterterms of the VDG perturbation theory model for improving the robustness of parameter fits, especially when working with small-scale data. Small scales present significant challenges due to the limited amount of uncorrelated information and complex parameter degeneracies within the multi-dimensional model parameter space, the so-called "projection effects". One example of such degeneracies arise between the growth rate parameter and the clustering amplitude ($\sigma_8$ or $\sigma_{12}$). The inclusion of Gaussian priors helps to stabilize the fitting procedure, enabling faster convergence and reducing variance in the results. For large scales, the posteriors can still be reasonably accurate without Gaussian priors, particularly if nonlinear biases are not included. However, the large error bars at these scales result in high variance in the posteriors.

Most of the considerations discussed here were analyzed separately, although some may be specific to our dataset. Certain aspects have already been addressed in previous studies, such as fixing the value of $\sigma_{12}$, as done in Ref.~\cite{Wenzl2024ConstrainingBOSS} for $\sigma_{8}$. This step was crucial to mitigate the strong degeneracy between $f$ and $\sigma_{12}$. The latter parameter can, however, be estimated through alternative analyses, such as a $3x2$pt approach, that combines 2-point statistics of clustering and weak-lensing. This reasoning also underpins our decision to use the fiducial cosmological parameters of the simulation for the background cosmology $\{\Omega_m$, $\Omega_b$, $H_0\}$ in COMET, and in the $C_\ell$ theory predictions to fix the amplitude of clustering parameter, $\sigma_{12}$. In any case, we incorporate Alcock-Paczynski parameters to account for deviations in cosmological parameters, and these have been observed to align reasonably well with 1 (\emph{i.e.}, unbiased cosmology) within statistical errors.

Regarding the linear galaxy bias, a tight Gaussian prior based on the estimates from the harmonic space galaxy clustering is not strictly necessary in most cases, as fits to the correlation function multipoles tend to yield consistent values. This is further validated in Appendix \ref{annex:bias}, where we apply a more sophisticated non-linear model to jointly fit the linear ($b_1$) and quadratic bias ($b_2$), improving the estimate of $b_1$ from the $C_\ell$. Using this method, we finding consistent values, and thus similar Gaussian priors, for $b_1$. While the Gaussian prior on $b_1$ is not critical, it significantly accelerates chain convergence and helps avoid degeneracies with non-linear biases in specific cases. For the rest of non-linear parameters, although these priors could theoretically introduce parameter biases, the analysis confirms they allow sufficient wide sampling of the parameter space. The triangular plots for the derived parameter constraints (Appendix \ref{annex:triangle}) demonstrate that the priors are not overly constraining (\emph{i.e.}, not too informative), as we can see that the parameters for the small and large scales are able to vary over the values (priors) for the full scales. This particular approach proves particularly important for small scales, where degeneracies and noisy posteriors are typically found.

\subsection{\textbf{Small Scales as the critical testing ground for gravity models}}

When analyzing the full range of scales, the results are generally robust. This agrees with previous work where $\beta$ is calculated over an extensive range of scales, such as in Ref.\cite{Pullen2016ConstrainingVelocities} and Ref.~\cite{Wenzl2024ConstrainingBOSS}. In our case, the use of all-sky lightcone simulations yields relatively small error bars, with a relative error of approximately 3–5$\%$ for $E_G$, which makes the agreement with theoretical predictions even more remarkable. For comparison, Ref.~\cite{Wenzl2024ConstrainingBOSS} reports a relative error of about 15–25$\%$ for $E_G$ using the BOSS survey. It is also important to note that the observable estimate for $E_G$, based on the method in Ref.~\cite{Wenzl2024ConstrainingBOSS}, is biased low by approximately 1–3$\%$, as illustrated in Fig.~\ref{Fig:acuracy}. However, the error bars are not small enough to distinguish between the F5 and GR theory models clearly, as their average predictions differ by only about 5$\%$ at these scales. Observational estimates of the gravity estimator are also affected by complex real-world systematics that substantially increase the overall error budget, thus further compromising the power of the estimator to discriminate gravity theories.

We note that this work places particular emphasis on small scales, where deviations between F5 and GR models are expected to be most pronounced. Unfortunately, this is also the regime where fitting challenges are greatest, even under ideal simulation conditions. Specially at scales below $10 \ \rm{Mpc/h}$ where we seem to find the highest deviations in the RSD (see Fig.~\ref{fig:2Dplot}). Again, the large nonlinearities and limited amount of uncorrelated modes on these scales undermine the ability of the estimator to distinguish between competing models. Despite these challenges, the methodology underscores the potential for enhanced small-scale estimations with novel hybrid approaches that combine perturbative and fully non-linear (N-body) tools such as the BACCO emulator \cite{Pellejero_Iba_ez_2023}, and higher-quality observational data. This is particularly promising since much of the error budget at these small scales appears to originate from the growth rate-related parameter, $\beta$. 

Tests like the $E_G$ estimator do not currently provide sufficient constraints to distinguish between models. The results in Section \ref{sec:nulltest} suggest that, at least in simulations, the data is robust enough to achieve a significant detection (3$\sigma$ for the F5 model) of  deviations from GR at small scales across all the redshifts studied (for $z<1$). However, in observational contexts, this approach is more limited,  as we only have access to a single Universe with an unknown gravity model. This means that we cannot perform such model dependent comparisons, and most of the relevant information will be captured by nuisance parameters, e.g the linear galaxy bias being higher in GR than F5, that we cannot use to directly distinguish gravity models.

\subsection{\textbf{Comparison with alternative nonlinear modified gravity probes}}

The $E_G$ estimator provides a linear galaxy bias robust combination of galaxy clustering and weak lensing designed to test the consistency between the matter clustering and lensing potentials \cite{Zhang2007AScales}. Its principal advantage lies in its lack of sensitivity to linear galaxy bias and its ability to combine independent tracers of the large-scale structure. However, its construction also implies that it probes a specific combination of observables rather than directly isolating the velocity field or higher-order clustering information.

As mentioned in the introduction a number of studies have demonstrated that nonlinear and velocity-sensitive statistics can exhibit enhanced sensitivity to modified gravity effects, particularly in screened theories such as chameleon $f(R)$ gravity. 
Weak gravitational lensing also provides a powerful probe of Hu-Sawicki $f(R)$ gravity, particularly on nonlinear scales where screening transitions occur. Several studies have shown that both two-point and higher-order lensing statistics are sensitive to F5-like models. For example, \cite{Liu_2016} and \cite{Higuchi_2016} demonstrated that weak-lensing peak counts and cosmic shear statistics can constrain $f(R)$ gravity beyond linear scales, while \cite{Shirasaki_2016} quantified the additional information contained in nonlinear shear observables. Galaxy-galaxy lensing has also been explored in this context \cite{Li_2018}. Recent analyses \cite{davies2024constrainingmodifiedgravityweak} indicate, similarly to our analysis, that next-generation surveys can be used to distinguish GR from $f(R)$ and at the 2$\sigma$ level. Another approach, would be to directly measure model dependent parameters like $f_{R_0}$ as they do in \cite{Casas2023Euclid}, although a model of gravity, in this case $f(R)$, needs to be assumed.  

In chameleon scenarios, screening suppresses deviations from GR in dense regions while allowing enhanced growth in lower-density environments. As shown in \cite{Jennings_2012}, velocity-sensitive observables provide a larger  discriminatory power between $f(R)$ and GR on nonlinear scales . By contrast, the $E_G$ estimator combines projected clustering and lensing in a manner that partially cancels non-linear growth enhancements and remains dominated by large-scale contributions where deviations from GR are modest.

Our results are consistent with this broader picture: in the full-sky, idealized systematic free baseline considered here, the difference in $E_G$ between GR and F5 remains at the few-percent level, while redshift-space multipoles and the 2D correlation function exhibit substantially larger relative deviations. We therefore interpret $E_G$ not as a leading discriminator of chameleon-type modified gravity in the nonlinear regime, but rather as a complementary, largely galaxy-bias-independent consistency test that can be combined with velocity-based diagnostics to strengthen gravity constraints.

\subsection{\textbf{Limitations of the theory modeling and future directions}} 
\label{sec:Discussion4} 

The reliance on GR-based assumptions for estimating certain cosmological parameters (e.g, $\beta$) introduces biases into the $E_G$ calculation for F5 models, a limitation acknowledged in the analysis with the "GR theo" case (see sec.~\ref{sec:egresults}). These biases are inherent to the model-dependent nature of parameter fitting and underscore the difficulty of achieving truly model-independent estimates for $E_G$. In Section \ref{sec:Comet}, we highlighted that the emulator has a $k_{max} = 0.3502 \ \rm{Mpc^{-1}}$ (or $k_{max} = 0.5171 \ h/\rm{Mpc}$ for the mocks used, h=0.6774). Using Eq.~\eqref{eq:k trans}, this gives a $s_{min} \approx 6 \ \rm{Mpc}/$$h$. However, this may not be sufficient for our $s_{min} = 20 \ \rm{Mpc}/$$h$, as shown in Eq.~\eqref{eq: multipoles final comet}, where we need to integrate over the entire range of Fourier scales, $k$. Additionally, our implementation of \textsc{e-mantis}'s amplitude boost to the final multipoles, rather than applying it to the original matter power spectrum, could introduce a bias in the results at small scales since it is not a gravity-consistent MG RSD template. This means that the enhanced discriminatory power of small-scale RSD diagnostics identified here must be interpreted in light of modeling assumptions. We employ the GR-calibrated COMET-VDG template supplemented by a scale-dependent power-spectrum boost to approximate the F5 case which may not capture all MG-induced dynamical effects. Therefore absolute parameter constraints derived from RSD multipoles may therefore differ under a fully self-consistent MG likelihood framework. Dedicated MG-consistent RSD models demonstrate that modified gravity alters not only the matter power spectrum amplitude but also the velocity divergence field and nonlinear couplings \cite{Valogiannis_2020}. In particular, \cite{Bose_2017} investigated the impact of applying GR templates to MG data and found significant biases in the recovered cosmological parameters for volumes comparable to those of Stage IV surveys. However, they did not account for the additional freedom introduced by the linear galaxy bias, so we expect the approach to perform well as an approximation.

At small scales, the relationship between the correlation multipoles and the power spectrum is not necessarily linear, which may further affect the accuracy of our incomplete MG template. Nonetheless, the study demonstrates that, despite these limitations, the results remain consistent within acceptable error margins with the simulated MG data vectors, highlighting the robustness of the methodology. However, the potential impact of using GR theory (\emph{e.g.}, perturbation theory counterterms) to estimate cosmological parameters using RSD for the F5 model remains uncertain. In  \cite{alemanygotor2025testingmodifiedgravity3x2pt} a 3x2pt analysis in real space is conducted to constrain gravity using a GR template on the mocks presented in this work.

Moreover, the $\chi_\nu^2$ test when comparing data to best-fit models are unusually high, even for the full-scale case. We are confident that this is not an issue with the modeling done with COMET, as this emulator has been successfully used to fit other simulated catalogs with good agreement \cite{euclidcollaboration2026euclidpreparationgalaxypower}. At small scales, it is evident that the small error-bars (as compared to the large scales) make  the $\chi_\nu^2$ increase, specially considering that the perturbative model seems to start breaking down. On the other hand, for large scales, the off-diagonal terms of the covariance matrix seem to boost the value of $\chi_\nu^2$ since setting them to zero (\emph{i.e..}, taking only the diagonal part) tends to keep the value close to 1. Typically, these fits are performed at high redshifts to avoid large non-linear effects, but even at redshifts around $z=1$, the results do not improve in any significant way. 

The statistical significances  reported in this work are derived with covariance matrices estimated from a 100 Jackknife resampling using only 1 mock realizations. As is well known, such covariance estimates are subject to noise, and their inversion can introduce biases that affect $\chi^2$ statistics and derived $\sigma$-levels. We therefore regard that the quoted $\sigma$-levels as indicative of the information estimation within the adopted analysis framework, and caution that they should not be interpreted as robust forecasts representative of a full survey realization.

Another consideration is that the relationship between the wavenumber $k$ and the projected distance $s$ (as given by Eq.~\eqref{eq:k trans}) is based on several approximations that are only valid at small scales.  This could potentially affect the accuracy of the predictions for the growth rate and the $E_G$ in multipoles, although we have not investigated this in detail. An alternative approach would be to perform the fits in Fourier space, using the power spectrum multipoles, but as previously discussed, this introduces additional complications, such as extra noise parameters and noisy measurements of the multipoles, particularly when trying to split the measurements in small and large scales.

\subsection{\textbf{Observational and astrophysical systematics: scope and limitations}}

The analysis presented in this work is restricted to a gravity-only, idealized baseline constructed from dark-matter-only N-body simulations with full-sky coverage and no observational systematics. In this subsection we summarize the main neglected effects and comment qualitatively on their expected impact.

\begin{itemize}

\item \textbf{Baryonic physics}: Hydrodynamical simulations indicate that baryonic feedback modifies the matter density field on nonlinear scales, affecting the matter power spectrum and lensing convergence at the several-percent level for 
$k\geq 0.5$ $h \cdot $Mpc$^{-1}$. Comparisons between dark-matter-only (DMO) and hydrodynamical simulations, including EAGLE-based analyses, show that baryons can significantly alter projected mass statistics relevant for weak lensing, while having a comparatively smaller impact on halo velocities and redshift-space distortions (e.g. \cite{Hellwing_2016, collier2024galaxyclusteringmodifiedgravity}). Since the $E_G$ estimator explicitly involves galaxy–lensing correlations, it may therefore be more susceptible to baryonic uncertainties than purely velocity-based RSD diagnostics.

\item \textbf{Galaxy bias modeling}: Although $E_G$ is constructed to cancel linear galaxy bias, nonlinear and scale-dependent bias can enter in the nonlinear regime. Similarly, small-scale RSD modeling depends on assumptions regarding bias and velocity dispersion. In this work, identical HOD+SHAM prescriptions are applied to GR and F5 simulations to isolate gravitational effects; however, realistic galaxy–halo connections may differ across gravity models and introduce additional degeneracies.

\item \textbf{Survey geometry and observational systematics}: We assume full-sky coverage, neglect mask effects, photometric redshift uncertainties, and redshift failures. We also ignore shape noise and intrinsic alignments (not present in alternative analyses that use the CMB lensing instead of the galaxy lensing, see \emph{e.g.}, Ref.~\cite{Pullen2014ProbingLensing}). In realistic surveys, these effects degrade the signal-to-noise of angular power spectra entering $E_G$ and increase covariance between observables. Consequently, the statistical precision reported here should be interpreted as an upper bound under ideal conditions.

\end{itemize}

Taken together, these considerations imply that the present results represent a controlled gravity-only benchmark rather than a performance forecast for any specific survey. The qualitative comparison between $E_G$ and nonlinear velocity-sensitive diagnostics is expected to remain informative; however, more accurate constraints will depend on the treatment of baryons, survey systematics, and modeling uncertainties in future analyses.

\section{Conclusions}
\label{sec:Conclusions}

In this paper we have presented an end-to-end cosmological analysis pipeline to constrain gravity using one of the largest modified gravity simulations to date (see Ref.~\cite{Arnold2019RealisticGravity}). In particular, we use a comprehensive galaxy mock built out of a N-body simulation of the Hu $\&$ Sawicki $f(R)$ model, with amplitude ${f}_{R_{0}} = 10^{-5}$ (denoted as F5), that is still viable given current observational constraints, and a twin LCDM simulation that assumes General Relativity (denoted as GR) with the same cosmological parameters and initial conditions, to investigate whether future surveys (in the limit of an ideal noise-free full-sky survey) can detect deviations from standard gravity using the so-called $E_G$ estimator \cite{Zhang2007AScales}. This estimator combines 2-point statistics of galaxy clustering in real and redshift space, along with weak-lensing (galaxy-galaxy lensing). A key advantage of this estimator is that it is independent of the galaxy bias on large scales, and it is a direct test of gravity.

We have presented a consistent framework to compute this gravity estimator across a range of scales, incorporating a number of theoretical and modeling ingredients relevant for its accurate evaluation. In particular, we examine the assumption of scale independence on large scales and its limitations, especially in the context of non-standard gravity models where the estimation of the linear growth rate may be affected. By considering this assumption, we investigate the impact of scale dependence in the theoretical modeling of the $E_G$
estimator, focusing on its ability to probe deviations from General Relativity in scenarios with similar expansion histories and clustering properties, such as the F5 model.

Our main results can be summarized as follows:

\begin{itemize}
    
\item{Even for ideal all-sky galaxy surveys, the widely used $E_G$ estimator is unable to clearly distinguish between the currently viable, \emph{e.g.}, F5 and GR, gravity theories (see section \ref{sec:egresults}). This is mainly due to three reasons. First, the fact that our simulation only has one realization which can add a noise term affecting cosmic variance and stochastic components from the simulation associated to the galaxy assignment step. Secondly, the degeneracies between perturbative and cosmological parameters of current state-of-the-art non-linear models for galaxy clustering in redshift space (such as VDG or EFT), that bias the linear growth rate estimation at the low scales and redshifts explored. As shown in appendix \ref{annex:degeneracies} these degeneracies seem to be subdominant since we can recover the fiducial values at 1$\sigma$ using theoretical data vectors with our pipeline.  Lastly, the fact that such perturbative models can not accurately model RSD clustering on small-enough scales (typically $<10 \ \rm{Mpc/h}$) where most of the constraining power resides.}

\item{We have proposed a simple null-test, based on the correlation function multipoles, to quantify the optimal detection level for deviations with respect to standard gravity, that we illustrate for the working example of the F5 model. We find indicative detection levels of about 4$\sigma$ for all the low-redshift samples considered. However, these values should be interpreted with caution, as they do not account for uncertainties in the inverse covariance estimation and are based on a single galaxy mock realization. As such, they should be regarded as relative measures of the information content rather than as robust survey-level forecasts. This result holds when using either a purely data-based approach or a model-dependent one (see section \ref{sec:nulltest}).}

\end{itemize}

Finally, this work identifies several promising directions for future research. The use of Fourier-space fits, combined with advanced emulators such as BACCO, that effectively extend the range of (small) scales that are accurately modeled, could significantly enhance the precision and accuracy of the growth rate and $E_G$ estimations. These emulators offer improved modeling of the galaxy bias parameters in the non-linear regime and allow for direct estimation of noise parameters, potentially addressing many of the limitations identified in this study. Expanding the analysis to include alternative simulations and novel tools for modeling clustering in the non-linear regime could also provide new pathways to constrain gravity using the next generation of galaxy surveys.

In conclusion, this study presents an end-to-end cosmological analysis pipeline to constrain gravity using high-fidelity galaxy mocks for F5 and GR. Our results show the limitations of the $E_G$ statistic even for idealized next-generation surveys in the case where a fully modified-gravity-consistent modeling is not included. This motivates the exploration of alternative approaches that combine multiple observables to break degeneracies between nuisance (\emph{e.g.}, galaxy bias) and cosmological parameters. In particular, we envisage that using the now standard combination of photometric galaxy clustering and weak-lensing observables, known as the 3x2pt analysis, can provide a more optimal way of breaking the observed parameter degeneracies and provide more competitive constraints of gravity. We leave this study for future work \cite{alemanygotor2025testingmodifiedgravity3x2pt}.

\section*{Acknowledgements}
We acknowledge valuable discussions and feedback from Enrique Gaztañaga, Kazuya Koyama, and Elizabeth Gonzalez. We also thank Benjamín Camacho for his assistance with the COMET emulator and for providing access to the VDG model within it. Additionally, we thank David Alonso for his support with the developer version of \textit{pyCCL}, which allowed us to incorporate non-Limber calculations.

CV and PF acknowledge support form the Spanish Ministerio de Ciencia, Innovaci\'on y Universidades, projects PID2019-11317GB, PID2022-141079NB, PID2022-138896NB; the European Research Executive Agency HORIZON-MSCA-2021-SE-01 Research and Innovation programme under the Marie Skłodowska-Curie grant agreement number 101086388 (LACEGAL) and the programme Unidad de Excelencia Mar\'{\i}a de Maeztu, project CEX2020-001058-M. This work has made use of CosmoHub. CosmoHub has been developed by the Port d’Informació Científica (PIC), maintained through a collaboration of the Institut de Física d’Altes Energies (IFAE) and the Centro de Investigaciones Energéticas, Medioambientales y Tecnológicas (CIEMAT) and the Institute of Space Sciences (CSIC $\&$ IEEC).
CosmoHub was partially funded by the “Plan Estatal de Investigación Científica y Técnica y de Innovación” program of the Spanish government, has been supported by the call for grants for Scientific and Technical Equipment 2021 of the State Program for Knowledge Generation and Scientific and Technological Strengthening of the R+D+i System, financed by MCIN/AEI/ 10.13039/501100011033 and the EU NextGeneration/PRTR (Hadoop Cluster for the comprehensive management of massive scientific data, reference EQC2021-007479-P) and by MICIIN with funding from European Union NextGenerationEU(PRTR-C17.I1) and by Generalitat de Catalunya. 

\section*{Data Availability}

The mock galaxy catalogs used in this work can be accessed upon reasonable request to the authors from the web portal \textit{Cosmohub}\footnote{\url{https://cosmohub.pic.es/home}}.


\newpage
\clearpage
\FloatBarrier
\begin{widetext}

\appendix

\section{Additional plots for $E_G$ and $\beta$ PDF}
\label{annex:additional plots}

In this section we show the same plots for Fig.~\ref{fig:EG0.55} but for the other two redshift bins used in this work: z=0.35 (Fig.~\ref{fig:EG0.35}) and z=0.8 (Fig.~\ref{fig:EG0.8}).

\begin{figure}[h]
    \centering
    \includegraphics[width=0.8\textwidth]{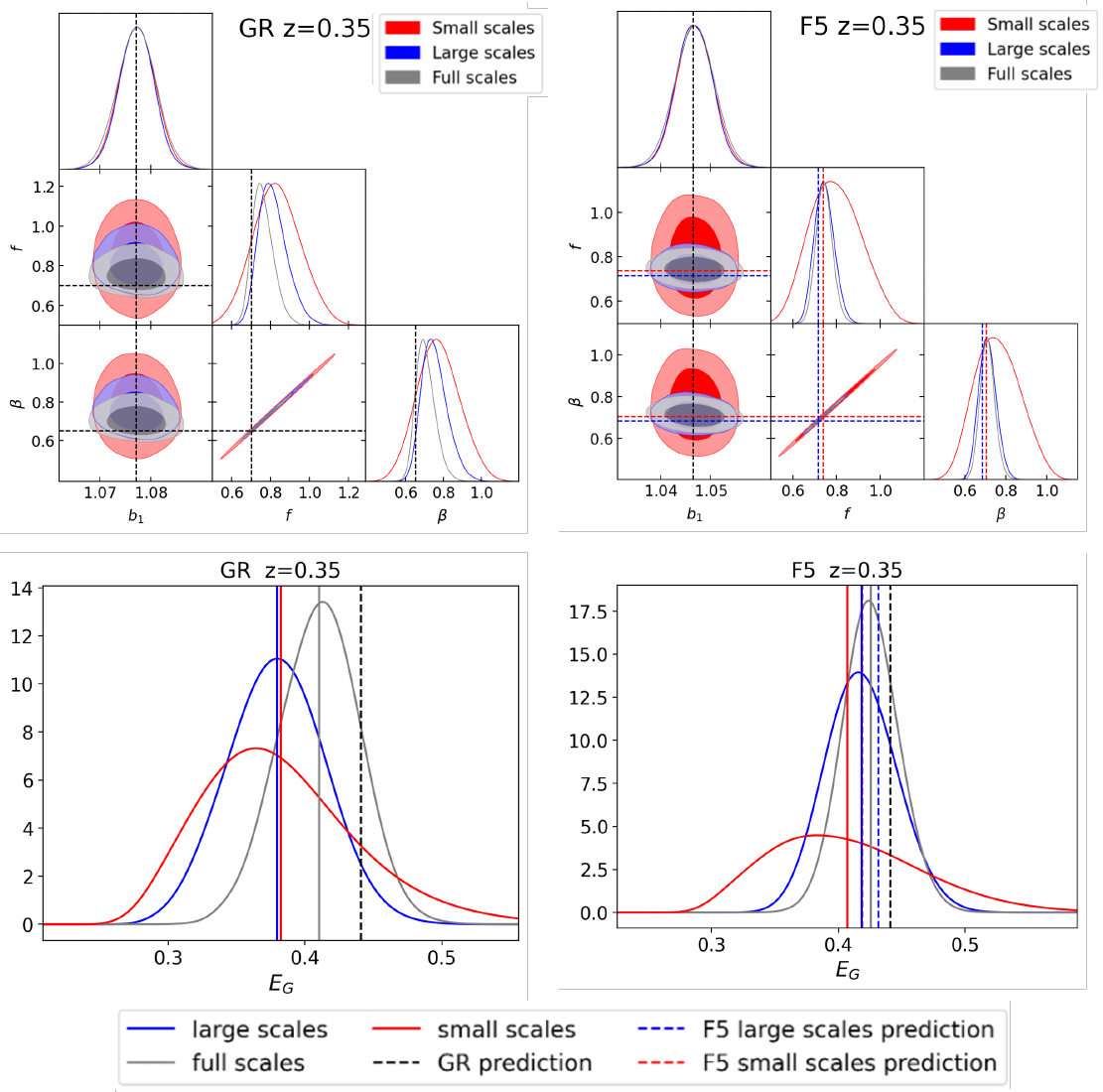}
    \caption{Same as Fig.~\ref{fig:EG0.55} but for the lowest redshift bin z=0.35. Top plots show the contours results for $b_1$ and $f$ and the corresponding derived $\beta$ parameter for small, large, and full scales for GR (Left) and F5 (Right). The dashed lines show the predictions for $f$ and $b_1$ (estimated from the $C_\ell$s) for GR (black), and F5 small (red) and large (blue) scales. The bottom plots show the PDF for the EG estimator for the respective scales, the solid lines show the mean of the respective same color PDF while the dashed lines represent the predictions for the same cases discussed earlier.}
    \label{fig:EG0.35}
\end{figure}

\newpage

\begin{figure}[h]
    \centering
    \includegraphics[width=0.8\textwidth]{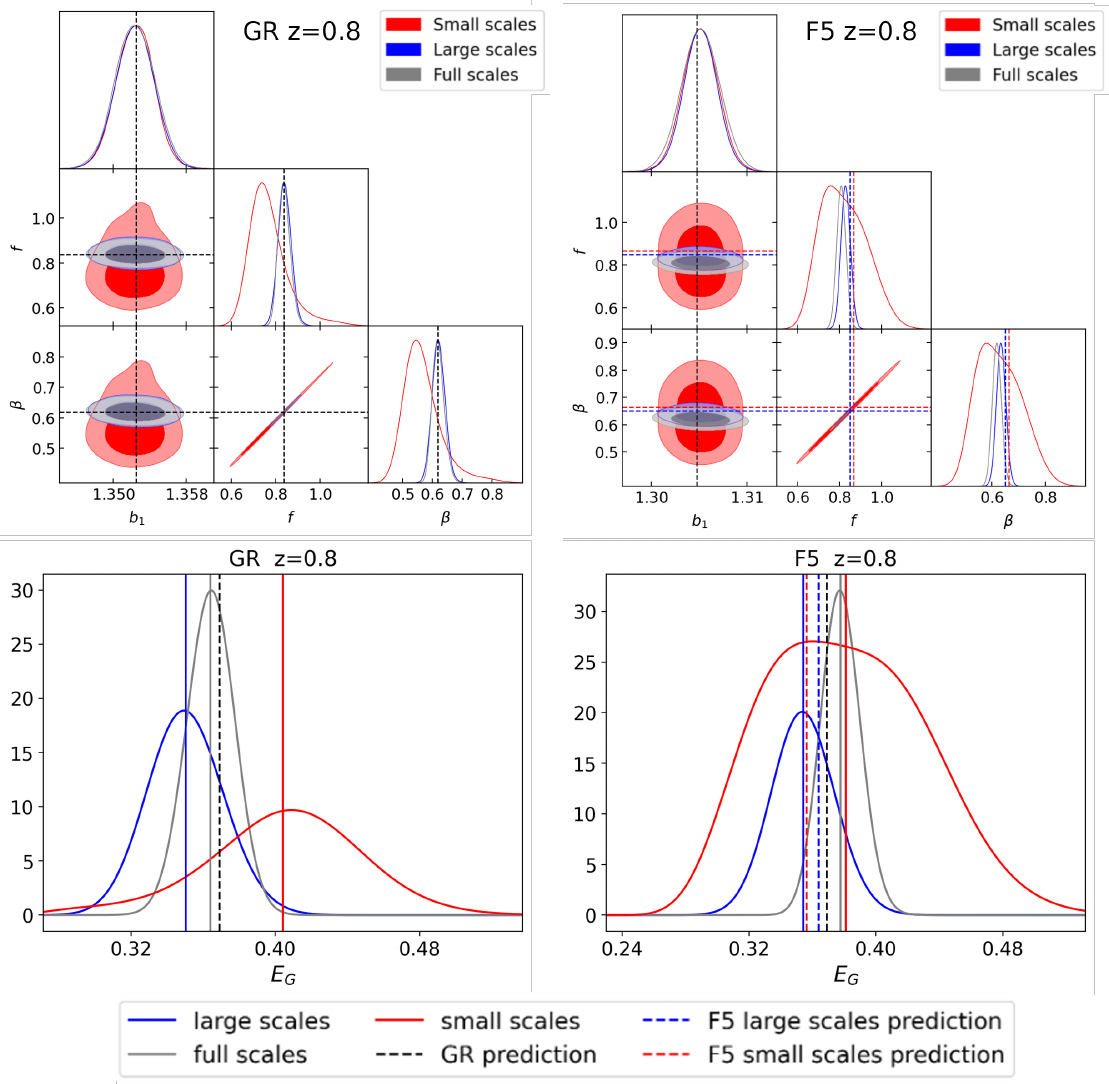}
    \caption{Same as Fig.~\ref{fig:EG0.35} but for the highest redshift bin z=0.8.}
    \label{fig:EG0.8}
\end{figure}

\newpage

\section{Accuracy of the $E_G$ estimator}
\label{annex:EG1}

Below we show how the different theory estimators for the $E_G$ estimator compare to the fiducial expression given by Eq.~\eqref{eq: EG general}.

\begin{figure}[h]
    \includegraphics[width=\textwidth]{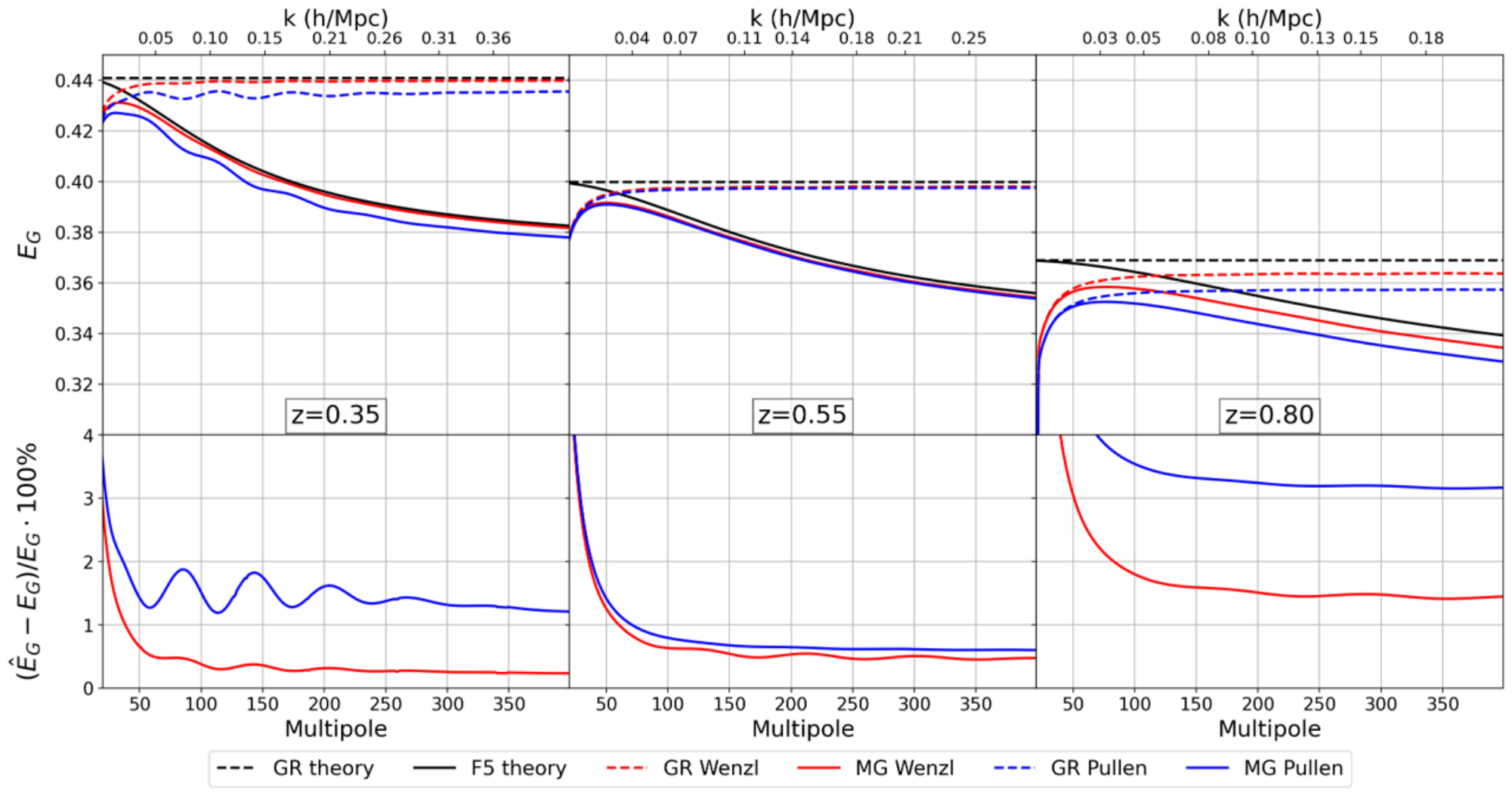}
\caption{Testing the accuracy of the estimator given by Ref.~\cite{Wenzl2024ConstrainingBOSS} (red lines) and Ref.~\cite{Pullen2016ConstrainingVelocities} (blue lines). Each column represents one of the redshift bins (from left to right: 0.35, 0.55, 0.8). The dashed (solid) lines represent the theoretical prediction ($E_G$ in Eq.~\eqref{eq: EG general}) for GR (F5). The black lines represent the observable quantities ($\hat{E}_G^{\text{Wenzl}}$ in Eq.~\eqref{eq:EG final}) with theoretical $C_\ell$s. The plots below show the respective relative difference of $\hat{E}_G^{\text{Wenzl}}$ and $\hat{E}_G^{\text{Pullen}}$ with respect to the theoretical prediction using the same color legend. The F5 relative difference is not included due to overlapping since it is the same difference as for the same model in GR.}
\label{Fig:acuracy}
\end{figure}

\section{Testing projection effects on the VDG model}
\label{annex:degeneracies}

In this section we analyze how much degeneracies or "projection effects" are present in our pipeline with the VDG model. We simply perform the MCMC chains using a theoretical data vector (DV) with the following values for the non-linear parameters: $c_0=8$, $c_2=5$, $b_2=0.34$ and $a_{vir}=2$. The value of $b_1$ (which is given as a Gaussian prior for the fit) is taken from the value of Table \ref{tab:b1} and $\gamma_2$, $\gamma_{21}$ follow the relation $\gamma_2=0.524 - 0.547b_1 + 0.046b_1^2$ and $\gamma_{21}=(2/21)(b_1-1)+ (6/7)\gamma_{2}$ from Ref.~\cite{Sanchez2016TheWedges}. The values of $f$ and $s_{12}$ are the fiducial values of the simulation at the respective $z$ with $q_{tr}=1$, $q_{lo}=1$. The chains are run using the same pipeline as in the analysis for the standard case based on GR theory. We use the corresponding corrected covariance matrix (Eq.~\eqref{eq:covfix}) from the respective GR redshift bin simulation data vector. As shown in Fig.~\ref{fig:GRtriangleteo}, we recover the fiducial values for each parameter within the \(1\sigma\) contours. This suggests that the pipeline and model used do not introduce significant projection effects. However, given the presence of irregular contours and some emerging bi-modalities, we still consider projection effects to be a subdominant factor in the final results.  

Additionally, using theoretical DV, we verified that directly fitting either small or large scales often fails to recover the fiducial values, significantly impacting the estimation of \( f \). This validates our two-step fitting approach, as the Gaussian (informative) priors from the full-scale fit play a crucial role in "guiding" the chains toward the fiducial value without being overly restrictive—allowing \( f \) in \( f(R) \) gravity to vary as expected.        

\begin{figure}[h]
\centering
\includegraphics[width=0.98\linewidth]{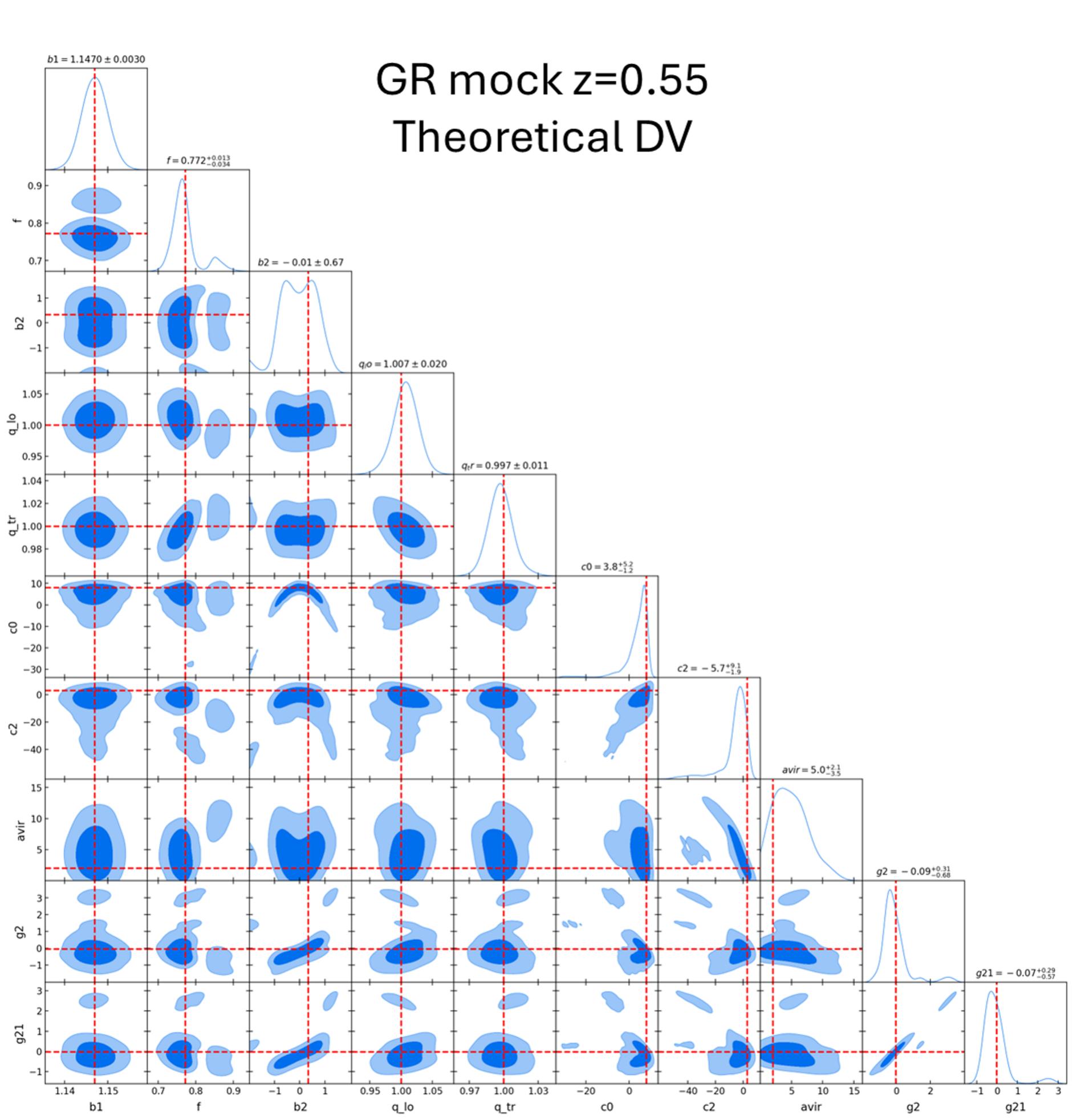}
\caption{Full triangle plot using a COMET VDG theoretical data vector with z=0.55. We use GR theory for the fits. The red dashed lines represent the fiducial values used to generate the data vector.}
\label{fig:GRtriangleteo}
\end{figure}

\newpage
\clearpage
\FloatBarrier

\section{Full parameter space triangle plots.}
\label{annex:triangle}

\begin{figure}[h]
\centering
\includegraphics[width=0.98\linewidth]{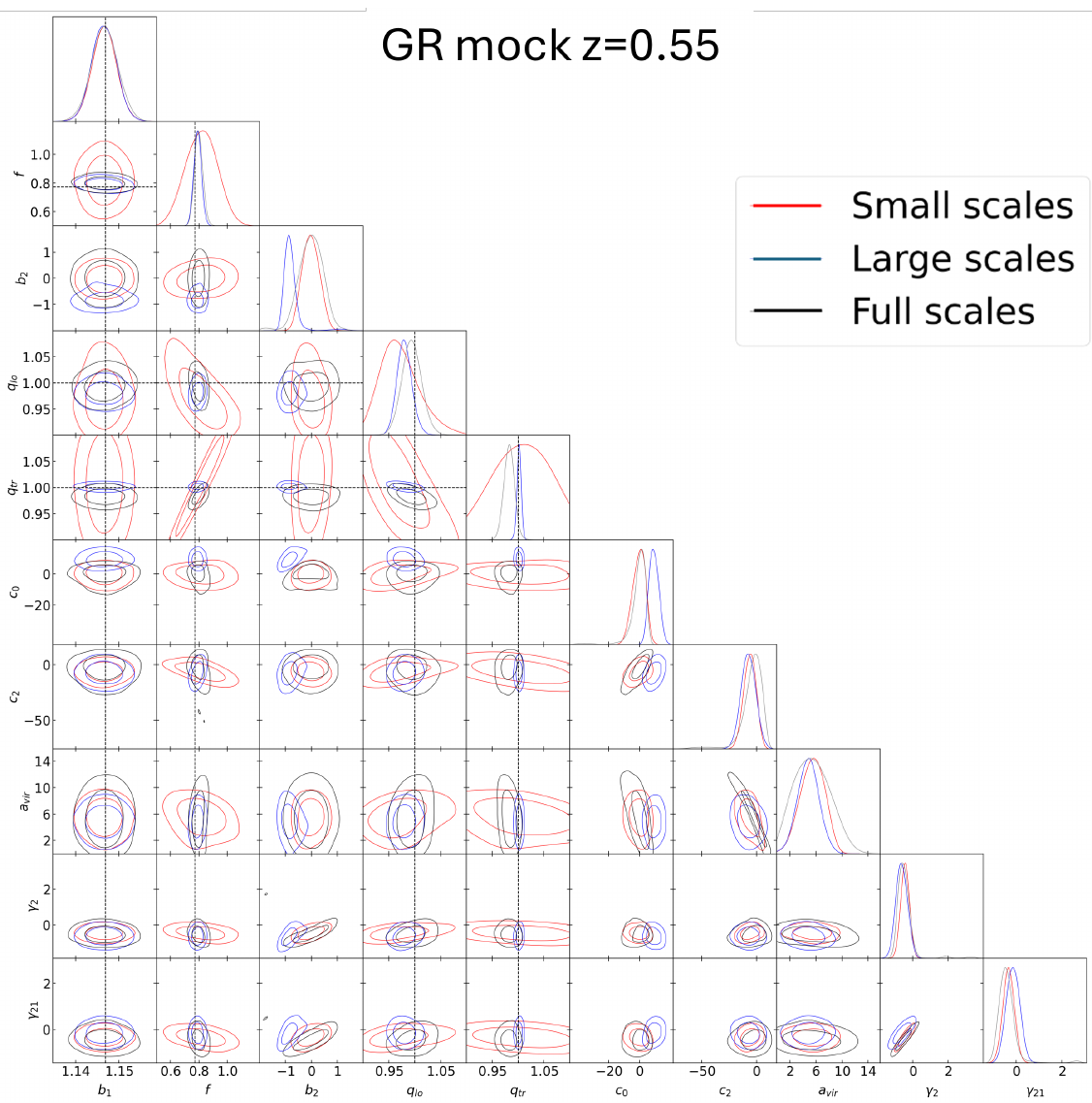}
\caption{Full triangle plot for the GR mock z=0.55 all galaxies COMET VDG fit. The contours for the small (red), large (blue) and full (black) scales are shown. While the black dashed lines represent the fiducial values of the cosmology of the simulation ($b_1$ is from the $C_\ell$s fit).}
\label{fig:GRtriangle}
\end{figure}

\newpage

\begin{figure}[h]
\centering
\includegraphics[width=0.98\linewidth]{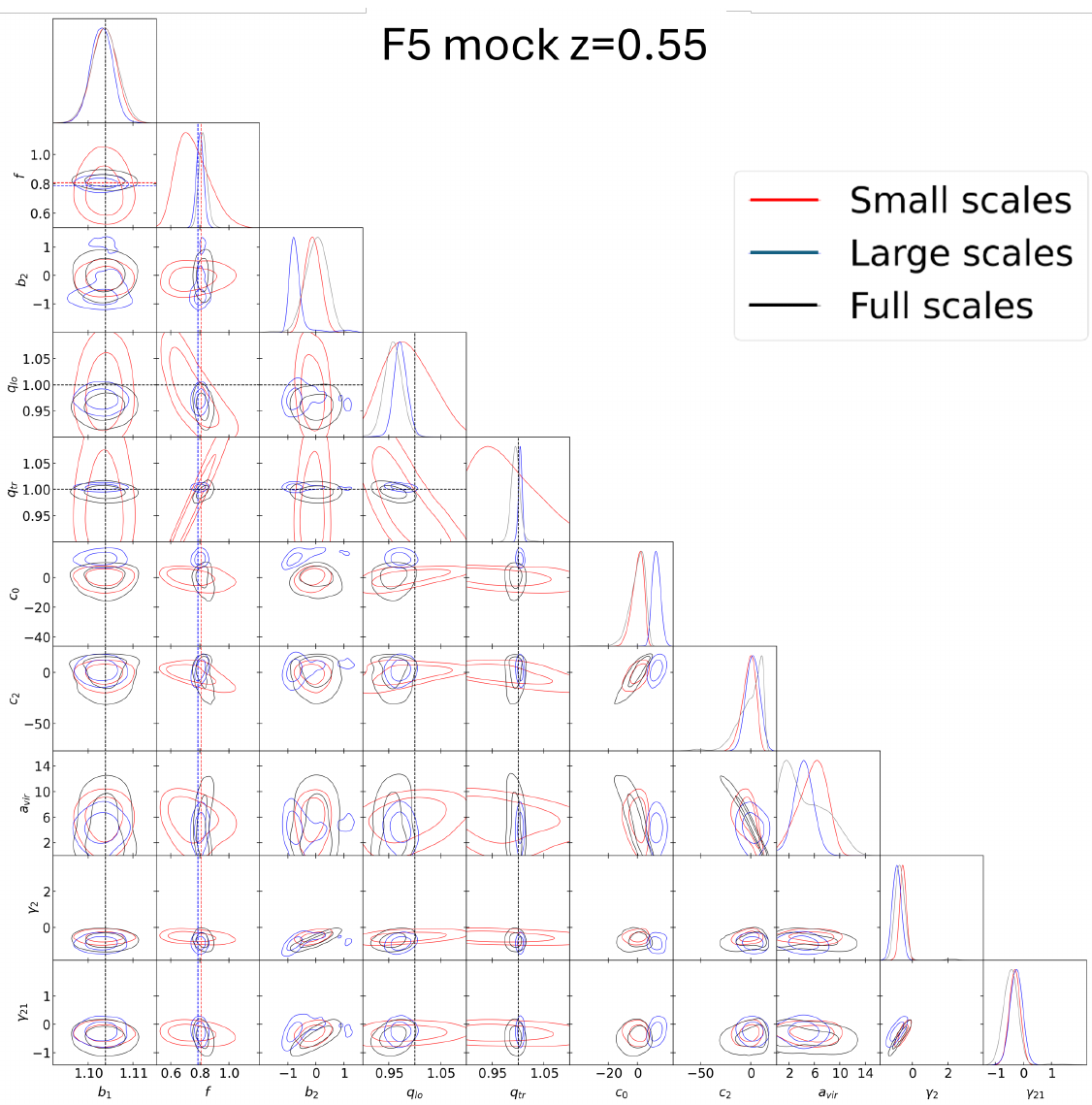}
\caption{Similar plot as Fig.~\ref{fig:GRtriangle} but for the F5 catalog using COMET VDG F5 theory fits. In this case the fiducial values for the growth rate are divided in small (red) and large (blue) scales.}
\label{fig:F5triangle}
\end{figure}

\newpage

\begin{figure}[h]
\centering
\includegraphics[width=0.98\linewidth]{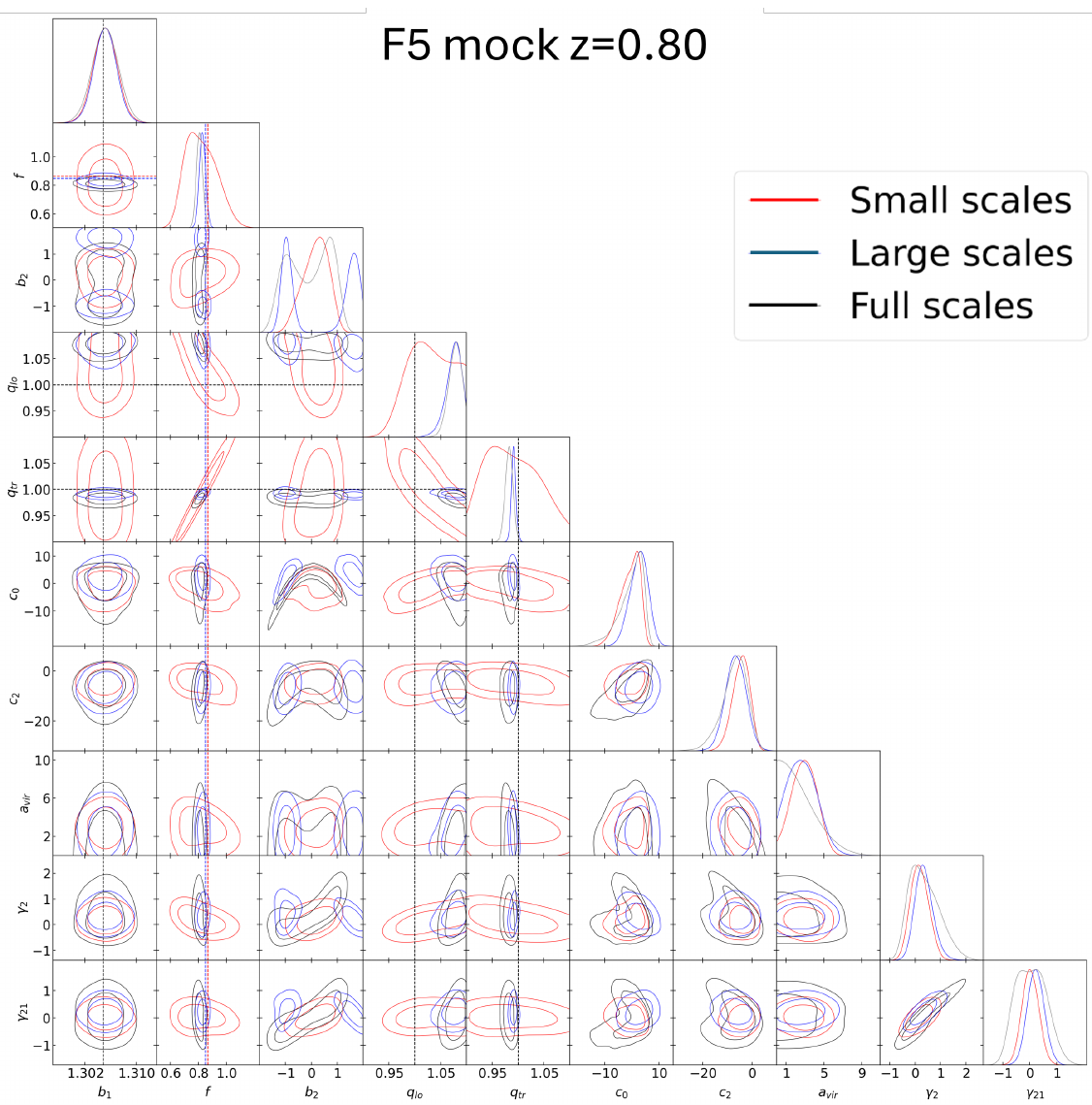}
\caption{Similar plot as Fig.~\ref{fig:F5triangle} but for z=0.8 catalog using COMET VDG F5 theory fits. In this case the fiducial values for the growth rate are divided in small (red) and large (blue) scales.}
\label{fig:F5triangle08}
\end{figure}

\newpage
\clearpage
\FloatBarrier

\section{COMET VDG model description}
\label{annex:VDG}

In sec.~\ref{sec:VDG} we introduced COMET implementation of the VDG model that we use throughout this work. Due to the complexity of the model we include a full summary of the model in this appendix. As also mentioned in sec.~\ref{sec:VDG} this description is just a summary of the key points presented in Ref.~\cite{Eggemeier2023COMETTheory} in order to cover the parameter space that we use to fit our data. In Ref.~\cite{eggemeier2025boostinggalaxyclusteringanalyses} we can find a even more detailed description of this model.

In redshift space the density perturbations can be expressed as\footnote{For the configuration and Fourier space integrals we use the short-hand notation:  $\int_{\vec{x}_{1}, \ldots, \vec{x}_{n}} \equiv\int d^{3} x_{1} \cdots\ d^{3} x_{n}$ and $\int_{{\vec{k}}_{1}, \ldots, {\vec{k}}_{n}} \ \equiv\ \int d^{3} k_{1} / ( 2 \pi)^{3} \cdots d^{3}k_{n} / ( 2 \pi)^{3}$}, respectively:
so that,
\begin{equation}
\delta_{s} ( \vec{k}, z )=\int_{\vec{x}} \mathrm{e}^{i \vec{k} \cdot \vec{x}} \mathrm{e}^{-i f k_{z} \text{v}_{z} ( \vec{x} )} D_{s} ( \vec{x} ) 
\end{equation}

where \( D_s(\vec{x}) \equiv \delta_g(\vec{x}) + f \nabla_z \text{v}_{z}(\vec{x}) \) is a combination of the galaxy density contrast \( \delta_g \) and the gradient of \( \text{v}_{\|} \) along the line-of-sight. The velocity $\text{v}$ is related to the normalized divergence field $\nu$:
\begin{equation}
    \vec{\text{v}} \equiv -f H a \vec{u} \rightarrow \nu = \nabla \vec{\text{v}} = \delta_0
\end{equation}

Expanding \( \delta \) and \( \vec{\text{v}} \) in perturbation theory yields the perturbation theory kernels in redshift-space \( Z_n \) \cite{Scoccimarro1998TheRedshift-Space}:

\begin{equation}
  \label{eq:Zn_expansion}
  \delta_s(\vec{k},z) = \sum_{n=1} D^n(z) \int_{\vec{k}_1,\ldots\,,\vec{k}_n} \delta_D(\vec{k} - \vec{k}_{1 \ldots n})\, \times\ Z_n(\vec{k}_1,\ldots\,,\vec{k}_n),\delta_L(\vec{k}_1) \cdots \delta_L(\vec{k}_n)\,,
\end{equation}

This \( Z_n \) kernels can be found on appendix A of Ref.~\cite{Eggemeier2023COMETTheory}. $Z_1$ is given by the expression relating the real space and redshift space power spectrum in linear theory:

\begin{equation}
    Z_1 (\vec{k}) = b_1(1 + \beta\mu^2),
\end{equation}

where the linear galaxy bias appears since these kernels are related to the galaxy power spectrum. In perturbation theory, predicting the clustering of biased tracers like galaxies involves relating their over-densities to various properties of the underlying dark matter field. This process, known as galaxy bias expansion, involves a series of operators that capture how the large-scale environment affects galaxy formation and evolution, with each operator associated with specific galaxy bias parameters. While these bias parameters cannot be calculated from first principles and depend on the selected tracer population, the relevant operators at each perturbative order can be determined based on symmetry considerations.

\subsection{\textbf{Tree and one-loop power spectrum}}

The galaxy bias expansion relevant for the power spectrum at next-to-leading order is given by:
\begin{equation}
\delta_g = b_1\,\delta + \frac{b_2}{2}\,\delta^2 + \gamma_2\,{\cal G}_2(\Phi_v) 
+ \gamma_{21}\,{\cal G}_{21}(\varphi_2,\varphi_1) 
+ b_{\nabla^2}\nabla^2\delta + \epsilon_g + \ldots,
\label{eq:bias expansion}
\end{equation}

where \(b_1, b_2\) are linear and quadratic bias parameters, \(\gamma_2, \gamma_{21}\) are parameters for Galileon operators, and \(b_{\nabla^2}\) represents a higher-derivative bias. The operators \({\cal G}_2(\Phi_v)\) and \({\cal G}_{21}(\varphi_2,\varphi_1)\) measure the effects of large-scale tides at different orders.

Different bases for galaxy bias exist, and their parameters can be transformed into the parameters in the above expansion. The higher-derivative term \(b_{\nabla^2}\nabla^2\delta\) accounts for the finite size of galaxy formation regions. This term is absorbed into a counterterm in some models.

If we want to expand the power spectrum to orders higher than linear theory (which is called the tree level), we usually start by the one-loop power  spectrum from Standard Perturbation Theory (SPT) which contains terms that are quadratic in the initial density perturbations:

\begin{equation}
    P_{gg,\rm SPT}^{\rm tree}(\vec{k}) = Z_1(\vec{k})^2\,P_L(k) 
    \label{eq:Ptree}
\end{equation}

\begin{equation}
    P_{gg,\rm SPT}^{\rm 1-loop}(\vec{k}) = 2 \int_{\vec{q}} \left[Z_2(\vec{k}-\vec{q},\vec{q})\right]^2 P_L(|\vec{k}-\vec{q}|) P_L(q)+6\,P_L(k) \int_{\vec{q}} Z_3(\vec{k},\vec{q},-\vec{q}) P_L(q)  
\label{eq:P1Lspt}
\end{equation}

we are using the plane-parallel approximation, so the wave vector \(\vec{k}\) is described by its magnitude \(k\) and its cosine \(\mu\) with respect to the line of sight (LOS).\(P_L(z)\) represents the linear matter power spectrum. 

\subsection{\textbf{Velocity generating function}}
\label{sec:Velocity generating function}

To connect the power spectrum in the VDG model with the Standard Perturbation Theory computation, we begin with the exact expression for the redshift-space galaxy power spectrum. This can be derived from mapping from real to redshift space and involves the pairwise velocity generating function \({\cal M}(\lambda,\vec{r})\) \cite{Scoccimarro1998TheRedshift-Space}:

\begin{equation}
    P_{gg}(\vec{k}) = \int_{\vec{r}} \mathrm{e}^{i \vec{k} \cdot \vec{r}}\,\left[\big(1 + \xi_{gg}(r)\big)\,{\cal M}(\lambda,\vec{r}) - 1\right],
\end{equation}

where \(\xi_{gg}(r)\) is the real-space galaxy correlation function at separation \(r\), and \(\lambda = -i\,f k \mu\). The pairwise velocity generating function can be broken down into components that include connected correlators, which are sensitive to small-scale modes. The VDG model treats the velocity difference generating function non-perturbatively by using an effective damping function \(W_{\infty}(\lambda)\), which accounts for small-scale velocity dispersion. This damping function is defined as:
\begin{equation}
W_{\infty}(\lambda) = \frac{1}{\sqrt{1 - \lambda^2\,a_{\rm vir}^2}}\,\exp{\left(\frac{\lambda^2\,\sigma_v^2}{1 - \lambda^2\,a_{\rm vir}^2}\right)},
\label{eq:velocity generating}
\end{equation}
where \(a_{\rm vir}\) is a free parameter that controls the non-Gaussianity of velocity differences. The remaining terms in the expression are treated perturbatively, expanding the exponentials to one-loop order. This involves galaxy bias, stochastic terms, and counterterms from small-scale modes, all evaluated using BAO-damped linear power spectra. The additional terms \(\Delta P(\vec{k})\) from this expansion, with respect to EFT, are given by:

\begin{equation}
\Delta P(k,\mu) = \lambda^2\,\sigma_v^2\,P_{D_s D_s}(k,\mu)-\lambda^2 \int_{\vec{q}} \frac{q_z^2}{q^4} P_{\nu \nu}(q)\ P_{D_s D_s}(\vec{k}-\vec{q}),
\end{equation}

where \(P_{D_s D_s}\) and \(P_{\nu \nu}\) are the power spectra of the density and velocity divergence fields, respectively, and \(\sigma_v^2\) is the linear velocity dispersion given by:

\begin{equation}
\label{eq:sigmav}
\sigma_v^2 = \frac{1}{3} \int_{\vec{k}} \frac{P_{\nu\nu}(k)}{k^2} = \frac{1}{3}\int_{\vec{k}} \frac{P_{L}(k)}{k^2}.
\end{equation}

\subsection{\textbf{The stochastic power spectrum}}
\label{sec: stochastic}

The stochastic field \(\epsilon_g\) captures highly non-linear effects in galaxy formation that are uncorrelated with large-scale fields and are considered stochastic at large scales. The contribution of this stochasticity to the galaxy power spectrum, \(P_{\epsilon_g\epsilon_g}(k)\), can be expanded as:
\begin{equation}
P_{\epsilon_g\epsilon_g}(k) = \frac{1}{\bar{n}}\left(N^P_{0} + N^P_{2,0}\,k^2 + \ldots\right),
\end{equation}
where \(\bar{n}\) is the mean number density of tracers, and \(N^P_{0}, N^P_{2,0}\) are stochastic bias parameters.

Additionally, redshift-space distortions introduce a stochastic term related to the gradient of the line-of-sight velocity field, \(P_{\epsilon_g\epsilon_{\nabla_z \text{v}_{\|}}}(k,\mu)\), which can be expressed as:
\begin{equation}
P_{\epsilon_g\epsilon_{\nabla_z \text{v}_{\|}}}(k,\mu) = \frac{N^P_{2,2}}{\bar{n}} {\cal L}_2(\mu)\,k^2,
\end{equation}
where \({\cal L}_2(\mu)\) is the second Legendre polynomial. The total stochastic contribution to the galaxy power spectrum is:
\begin{equation}
P_{gg}^{\rm stoch}(k,\mu) = P_{\epsilon_g\epsilon_g}(k) + P_{\epsilon_g\epsilon_{\nabla_z \text{v}_{\|}}}(k,\mu).
\end{equation}
In the VDG model, the impact of small-scale velocities is captured by the effective damping function \(W_{\infty}\), making the contribution from the stochastic velocity term less significant.

\subsection{\textbf{Counterterms power spectrum}}

In perturbation theory, the loop integrals involved in predicting the galaxy power spectrum extend over all scales, including those where the perturbative approach breaks down. To maintain a consistent theoretical framework, it becomes necessary to introduce counterterms. These counterterms have adjustable amplitudes (free parameters) that are designed to absorb any sensitivity to non-linear modes in the large-scale limit. Recent studies such as Ref.~\cite{Desjacques2018TheSpace} that have shown that the leading counterterms for the galaxy power spectrum in redshift space typically scale as \(\sim \mu^{2n}\,k^2\,P_L(k)\), with \(n = 0, 1, 2\). For simplicity, these are assumed to be local in time.

The first of these counterterms scales identically to the higher-derivative bias term \(\nabla^2\delta\), Eq.~\eqref{eq:bias expansion}, allowing us to absorb the coefficient \(b_{\nabla^2\delta}\) into the corresponding counterterm parameter. This same counterterm also captures the leading effect from deviations in the perfect fluid approximation for the matter field . The other two counterterms, corresponding to \(n = 1\) and \(n = 2\), can account for relevant velocity bias effects, which have been neglected so far. Then, three free parameters: \(c_0\), \(c_2\), and \(c_4\), are introduced to define the contribution of leading-order (LO) counterterms to the galaxy power spectrum as follows:
\begin{equation}
P_{gg}^{\rm ctr}(k,\mu) = -2 \sum_{n=0}^{2} c_{2n}\,{\cal L}_{2n}(\mu)\,k^2\,P_L(k)\,,
\end{equation}
where \({\cal L}_{2n}(\mu)\) are the Legendre polynomials of order \(2n\). This choice of polynomials instead of \(\mu^{2n}\) involves a linear transformation of the counterterm parameters, ensuring that each primarily contributes to a single power spectrum multipole. This means that $c_i$ will contribute mostly to $P_{g, l=i}$ for $i=0, 2, 4$.

\subsection{\textbf{Infrared Resummation}}

Considering all the parts derived in the previous subsections, the VDG model power spectrum is given by:
\begin{equation}
P_{gg, \mathrm{VDG}}(k) = W_{\infty}(k) \Big[ P_{gg, \mathrm{SPT}}^{\mathrm{tree}}(k) + P_{gg, \mathrm{SPT}}^{\mathrm{1-loop}}(k) 
+ P_{gg}^{\mathrm{stoch}}(k) + P_{gg}^{\mathrm{ctr}}(k) - \Delta P(k) \Big].
\label{eq:galaxy power spectrum}
\end{equation}

Although this model accurately describes the general behavior of the anisotropic galaxy power spectrum at mildly non-linear scales, it struggles to represent the amplitude of the Baryon Acoustic Oscillation (BAO) wiggles with precision. These wiggles are especially sensitive to large-scale bulk flows, which can blur the BAO signal due to extensive relative displacement fields.

In a perturbative approach, the corrections to the matter power spectrum can be resummed at each wavemode \( k \), accounting for the effects of fluctuations on larger scales. At leading order, this resummation manifests as a damping factor that primarily impacts the BAO wiggles. A common practice is to decompose the linear matter power spectrum into a smooth component and a wiggly component as follows:
\begin{equation}
P_L(k) = P_{nw}(k) + P_{w}(k),
\end{equation}

Following the leading order approximation, the infrared-resummed matter power spectrum is then expressed as the sum of the smooth component and the damped wiggly component:
\begin{equation}
P_{\rm{mm}}^{\,\rm{IR-LO}}(k) = P_{nw}(k) + e^{-k^2\Sigma^2}P_w(k),
\end{equation}
where the damping factor \(\Sigma^2\) is given by:
\begin{equation}
\Sigma^2 = \frac{1}{6\pi^2}\int_0^{k_{s}} P_{nw}(q)\left[1 - j_0\left(\frac{q}{k_{osc}}\right) + 2j_2\left(\frac{q}{k_{osc}}\right)\right]{\rm{d}}q
\end{equation}
Here, \(j_n\) represents the \(n\)-th order spherical Bessel function, \(k_{osc} = 1/\ell_{\rm{osc}}\) corresponds to the wavemode at the BAO scale \(\ell_{\rm{osc}} = 110\,h^{-1}{\rm{Mpc}}\), and \(k_{s}\) denotes the ultraviolet integration limit.

At next-to-leading order, the infrared-resummed matter power spectrum receives further contributions, including standard one-loop corrections sourced by higher powers of the density field. The full expression is then:
\begin{equation}
P_{\rm{mm}}^{\,\rm{IR-NLO}}(k) = \; P_{nw}(k) + \left(1+k^2\Sigma^2\right)e^{-k^2\Sigma^2}P_{w}(k) \,
 + P^{\rm{1-loop}}\left[P_{\rm{mm}}^{\,\rm{IR-LO}}\right](k),
\end{equation}
where the square brackets indicate that the one-loop integrals are computed using the leading order IR-resummed power spectrum instead of the linear power spectrum.

When extending this approach to the redshift-space galaxy power spectrum, the most significant change is that the damping factor now depends on the line-of-sight angle \(\mu\). At leading order, this can be written as:
\begin{equation}
P_{\rm{gg}}^{\,s,\rm{IR-LO}}(k,\mu) = \left(b_1 + f\mu^2\right)^2\left[P_{nw}(k) + e^{-k^2\Sigma_{\rm{tot}}^{\,2}(\mu)}P_{w}(k)\right].
\end{equation}
The angular dependence of the new damping factor \(\Sigma_{\rm{tot}}^{\,2}(\mu)\) is given by:
\begin{equation}
\Sigma_{\rm{tot}}^{\,2}(\mu) = \left[1 + f\mu^2(2 + f)\right]\Sigma^2 + f^2\mu^2(\mu^2 - 1){\rm{d}}\Sigma^2,
\end{equation}
where
\begin{equation}
{\rm{d}}\Sigma^2 = \frac{1}{2\pi^2}\int_0^{k_{s}}P_{nw}(q)\,j_2\left(\frac{q}{k_{osc}}\right){\rm{d}}q.
\end{equation}
At next-to-leading order, the expression becomes :
\begin{equation}
\begin{aligned}
    P_{\rm{gg}}^{\,s,\rm{IR-NLO}}(k,\mu) &= \left(b_1 + f\mu^2\right)^2 \Bigg[P_{nw}(k) + 
    \left(1 + k^2\Sigma_{\rm{tot}}^{\,2}(\mu)\right) e^{-k^2\Sigma_{\rm{tot}}^{\,2}(\mu)} P_{w}(k)\Bigg]
+ P_{\rm{gg}}^{\,s,{\rm{1-loop}}}\left[P_{nw}\right](k) \\
    &\quad + e^{-k^2\Sigma_{\rm{tot}}^{\,2}(\mu)} \Bigg(P_{\rm{gg}}^{\,s,{\rm{1-loop}}}\left[P_{nw} + P_{w}\right](k) 
    - P_{\rm{gg}}^{\,s,{\rm{1-loop}}}\left[P_{nw}\right](k)\Bigg),
\label{eq: wedges}
\end{aligned}
\end{equation}

where, as before, the square brackets indicate that the one-loop terms are evaluated using either the total linear matter power spectrum (\(P_{nw} + P_{w}\)) or just the smooth component (\(P_{nw}\)).

\subsection{\textbf{Modeling multipoles of the correlation function}}

When considering the redshift of our galaxy sample we need to consider a fiducial cosmology in order to convert from observed redshfits to scales in redshift space ($\vec{s}$ or $\vec{k}$). If the selected fiducial cosmology deviates from the true cosmology, this results in an incorrect rescaling of both the parallel and perpendicular components relative to the line of sight. Such discrepancies in rescaling can significantly affect the two-point statistics, which are central to the analysis presented in this work.

The anisotropic distortions that arise from an incorrect choice of fiducial cosmology are, to some extent, degenerate with anisotropies induced by the peculiar velocity field. Therefore, to accurately interpret the information contained within the galaxy power spectrum, it is imperative to account for these potential distortions. To correct for the effects of the chosen fiducial cosmology, a standard approach is to rescale the model power spectrum along the directions parallel and perpendicular to the line of sight. The rescaling is defined as follows:

\begin{equation}
k^{fid}_\perp = q_\perp k_\perp, \quad k^{fid}_\parallel = q_\parallel k_\parallel,
\end{equation}

where \( q_\perp \) and \( q_\parallel \) are known as the Alcock-Paczynski parameters. This parameters quantify the ratios of the angular diameter distance \( D_M(z) \) and the Hubble distance \( D_H(z) \) between the true and fiducial cosmologies, and are given by:

\begin{equation}
q_\perp(z) = \frac{D_M(z)}{D^{fid}_M(z)}, \quad q_\parallel(z) = \frac{D_H(z)}{D^{fid}_H(z)} = \frac{H^{fid}(z)}{H(z)}.
\end{equation}

With these definitions, the Alcock-Paczynski corrected wavenumber \( k(k^{fid}, \mu^{fid}) \) and the cosine of the angle to the line of sight \( \mu(k^{fid}, \mu^{fid}) \) can be expressed as:

\begin{align}
k(k^{fid}, \mu^{fid}) &= k^{fid} \left[ \frac{(\mu^{fid})^2}{q_\parallel^2} + \frac{1 - (\mu^{fid})^2}{q_\perp^2} \right]^{\frac{1}{2}}, \\
\mu(k^{fid}, \mu^{fid}) &= \frac{\mu^{fid}}{q_\parallel} \left[ \frac{(\mu^{fid})^2}{q_\parallel^2} + \frac{1 - (\mu^{fid})^2}{q_\perp^2} \right]^{-\frac{1}{2}}.
\end{align}

Finally, the corrected galaxy power spectrum multipoles \( P_\ell(k^{fid}) \) are evaluated as:

\begin{equation}
P_\ell(k^{\mathrm{fid}}) = \frac{2\ell + 1}{2q_\perp^2 q_\parallel} \int_{-1}^{+1} 
P_{gg}^{s} \left( k(k^{\mathrm{fid}},\mu^{\mathrm{fid}}), \mu(k^{\mathrm{fid}},\mu^{\mathrm{fid}}) \right)
\times \mathcal{L}_\ell \left( \mu(k^{\mathrm{fid}},\mu^{\mathrm{fid}}) \right) d\mu^{\mathrm{fid}},
\label{eq: multipoles final comet}
\end{equation}

where \( \mathcal{L}_\ell \) represents the Legendre polynomial of order \( \ell \). This approach ensures that the derived power spectrum is correctly interpreted within the context of the selected fiducial cosmology, thereby allowing for accurate cosmological inferences from observational data.

\section{An improved estimation of the linear galaxy bias}
\label{annex:bias}

In section \ref{sec:galaxy bias} we estimated the linear galaxy bias using \textit{pyCCL} up to linear theory. In most cases this should be enough since we are estimating the linear bias, but due to the small redshifts used in this work this left us with a small dataset to work with at large (linear) scales since things get nonlinear fast, \emph{e.g.} for $\ell_{max}=250$ at $z=0.35$. For this reason we used the non-linear models implemented in \textit{pyCCL} in order to extend the scales over which we can estimate the linear bias by also fitting the nonlinear bias $b_2$. We use the Eulerian perturbation theory and hybrid Lagrangian bias expansion correlations using the emulator baccoemu \cite{Angulo2021TheCosmology} to fit $b_1$ and $b_2$ over a maximum multipole of 700 independently of the redshift. 

The results using the BACCO simulation model are shown in Fig.~\ref{fig:nonlinear}. We can see that for the smallest redshift bin, \emph{i.e.} z=0.35, the models seems to break down as it is incapable of finding a good fit between data and theory despite that the parameters have converged, which results in a very high reduced $\chi_\nu$. The values are consistent with the ones from Table \ref{tab:b1}, although they are not within one sigma errors due to the small error-bars in both cases. 

Just as we did in section \ref{sec:galaxy bias} we use the values and standard deviation of $b_1$ obtained here as Gaussian priors for the MCMC. The value for $b_2$ obtained here are not used as a Gaussian prior since we cannot ensure that the $b_2$ from Eulerian and Bacco model corresponds to the $b_2$ from the VDG model. The fitted growth rate did not change much which makes sense since as stated earlier the value of $b_1$ is close to the one used in the standard case (Table \ref{tab:b1}). 

\begin{figure}[h]
\centering
\includegraphics[width=0.9\linewidth]{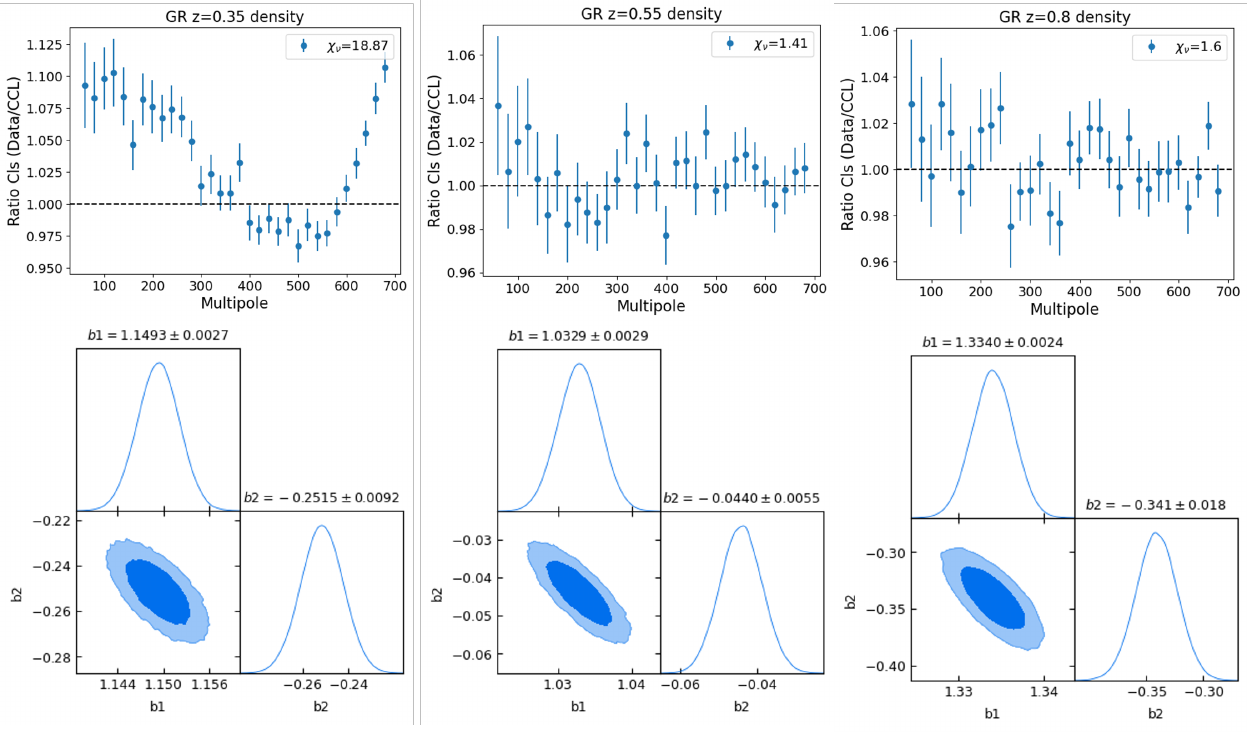}
\caption{Results for the fitting of the baccoemu nonlinear model to the angular power spectrum DV for $z=0.35, 0.55$ and $0.8$. The upper plots show the ratio of the DV $C_\ell$s over the best fit theoretical $C_\ell$ obtained with \textit{pyCCL}. The corresponding reduced chi squared is also shown. The lower plots show the triangular plots of the fits for $b_1$ and $b_2$.}
\label{fig:nonlinear}
\end{figure}

\newpage
\end{widetext}


\bibliographystyle{mnras}
\bibliography{updatedref}


\label{lastpage}

\end{document}